\documentclass[12pt]{article}
\usepackage{a4wide,epsfig,psfrag,amsmath,amssymb,cite,scalefnt}
\usepackage{color}

\graphicspath{ {figs_vasp/} }

\newcommand\eps{\epsilon}

\newcommand{\ep}{\epsilon}
\newcommand{\nc}{N_c}
\newcommand{\nl}{n_l}
\newcommand{\logx}{l_x}
\newcommand{\cR}{C_F}
\newcommand{\cA}{C_A}
\newcommand{\nh}{n_h}
\newcommand{\tf}{T_F}
\newcommand{\zthree}{\zeta(3)}
\newcommand{\logtwo}{l_2}
\newcommand{\afour}{a_4}

\allowdisplaybreaks

\begin{document}

\title{\vskip-3cm{\baselineskip14pt
    \begin{flushleft}
     \normalsize {TTP18-015}
    \end{flushleft}} \vskip1.5cm 
  Three-loop massive form factors: complete light-fermion and large-$N_c$
  corrections for vector, axial-vector, scalar and pseudo-scalar currents
  }

\author{
  Roman N. Lee$^{a}$,
  Alexander V. Smirnov$^{b,d}$,
  \\
  Vladimir A. Smirnov$^{c}$,
  Matthias Steinhauser$^{d}$,
  \\[1em]
  {\small\it (a) Budker Institute of Nuclear Physics}\\
  {\small\it  630090 Novosibirsk, Russia}
  \\
  {\small\it (b) Research Computing Center, Moscow State University}\\
  {\small\it 119991, Moscow, Russia}
  \\  
  {\small\it (c) Skobeltsyn Institute of Nuclear Physics of Moscow State University}\\
  {\small\it 119991, Moscow, Russia}
  \\
  {\small\it (d) Institut f{\"u}r Theoretische Teilchenphysik,
    Karlsruhe Institute of Technology (KIT)}\\
  {\small\it 76128 Karlsruhe, Germany}  
}
  
\date{}

\maketitle

\thispagestyle{empty}

\begin{abstract}
  We compute the three-loop QCD corrections to the massive quark form
  factors with external vector, axial-vector, scalar and pseudo-scalar
  currents. All corrections with closed loops of massless fermions are
  included. The non-fermionic part is computed in the large-$N_c$ limit, where
  only planar Feynman diagrams contribute.

%

\end{abstract}

\thispagestyle{empty}



\newpage


\section{Introduction}

Vertex form factors with external fermions play a crucial role in a number of
phenomenologically interesting processes. Among them are the massive fermion
production in electron positron collisions, and in particular the
forward-backward asymmetry, where form factor contributions induced by vector
and axial-vector currents are needed. Furthermore, building blocks to the
decay rates of scalar and pseudo-scalar Higgs bosons are provided by the
corresponding form factors. Last but not least, form factors constitute
important toys, which help to investigate the structure of high-order quantum
corrections.

In this work we consider vertex form factors where the external current is of
vector, axial-vector, scalar or pseudo-scalar type. They
are given by
\begin{eqnarray}
  j_\mu^v &=& \bar{\psi}\gamma_\mu\psi\,,\nonumber\\
  j_\mu^a &=& \bar{\psi}\gamma_\mu\gamma_5\psi\,,\nonumber\\
  j^s &=& m \,\bar{\psi}\psi\,,\nonumber\\
  j^p &=& i m \,\bar{\psi}\gamma_5\psi\,,
  \label{eq::currents}
\end{eqnarray}
where for convenience the heavy quark mass $m$ has been introduced in
the scalar and pseudo-scalar currents such that no additional overall
(ultraviolet) renormalization constants have to be introduced (as for the
vector and axial-vector\footnote{In this paper we do not consider Feynman
  diagrams which contribute to the axial anomaly.}
cases)~\cite{Chetyrkin:1994js}. If $j^s$ or $j^p$ are used to compute
properties of the Higgs boson there is a one-to-one relation of $m$ to the
corresponding Yukawa coupling.

We consider the three-point functions of the currents in
Eq.~(\ref{eq::currents}) and a quark-anti-quark pair. The
corresponding vertex functions can be decomposed into scalar
form factors which are defined as
\begin{eqnarray}
  \Gamma_\mu^v(q_1,q_2) &=& 
  F_1^v(q^2)\gamma_\mu - \frac{i}{2m}F_2^v(q^2) \sigma_{\mu\nu}q^\nu
  \,, \nonumber\\
  \Gamma_\mu^a(q_1,q_2) &=& 
  F_1^a(q^2)\gamma_\mu\gamma_5 {- \frac{1}{2m}F_2^a(q^2) q_\mu }\gamma_5
  \,, \nonumber\\
  \Gamma^s(q_1,q_2) &=& {m} F^s(q^2)
  \,, \nonumber\\
  \Gamma^p(q_1,q_2) &=& {i m} F^p(q^2) {\gamma_5}
  \,,
  \label{eq::Gamma}
\end{eqnarray}
with incoming momentum $q_1$, outgoing momentum
$q_2$ and $q=q_1-q_2$ being the outgoing momentum at $j^\delta$.  The external
quarks are on-shell, i.e., $q_1^2=q_2^2=m^2$ and we have $\sigma^{\mu\nu} =
i[\gamma^\mu,\gamma^\nu]/2$.  We note that in all cases the colour structure
is a simple Kronecker delta in the fundamental colour indices of the external
quarks and not written out explicitly.

For later convenience we define the perturbative expansion of the scalar form
factors as
\begin{eqnarray}
  F = \sum_{n\ge0} F^{(n)}
  \left(\frac{\alpha_s(\mu)}{4\pi}\right)^n
  \,,
\end{eqnarray}
with $F_1^{v,(0)}=F_1^{a,(0)}=F^{a,(0)}=F^{p,(0)}=1$ and
$F_2^{v,(0)}=F_2^{a,(0)}=0$.

The two-loop corrections to the vector current contributions $F_1^v$ and
$F_2^v$ have been computed for the first time in
Ref.~\cite{Bernreuther:2004ih} (see also Ref.~\cite{Hoang:1997ca} for the
fermionic contributions) and have been cross checked by several
groups~\cite{Gluza:2009yy,Henn:2016tyf,Ahmed:2017gyt,Ablinger:2017hst,Lee:2018nxa}. In
some cases higher order terms in $\epsilon$ have been added.  Two-loop
axial-vector, scalar and pseudo-scalar contributions have been computed in
Refs.~\cite{Bernreuther:2004th,Bernreuther:2005rw,Bernreuther:2005gw} and
recently been confirmed in Ref.~\cite{Ablinger:2017hst}
where ${\cal O}(\epsilon)$ and ${\cal O}(\epsilon^2)$ terms have been added.
Three-loop corrections are only known for well-defined subsets of the vector
form factor: The large-$N_c$ limit has been computed in
Ref.~\cite{Henn:2016tyf} using the master integrals
of~\cite{Henn:2016kjz}. This involves only planar integrals.  The complete
(planar and non-planar) light-fermion contributions to $F_1^v$ and $F_2^v$
have been obtained in Ref.~\cite{Lee:2018nxa}. In this reference also the
results of the relevant master integrals are given.  Let us mention that
all-order corrections to the massive vector form factor in the large-$\beta_0$
limit have been considered in Ref.~\cite{Grozin:2017aty}.

\begin{figure}[t] 
  \begin{center}
    \begin{tabular}{cccc}
      \includegraphics[width=0.2\textwidth]{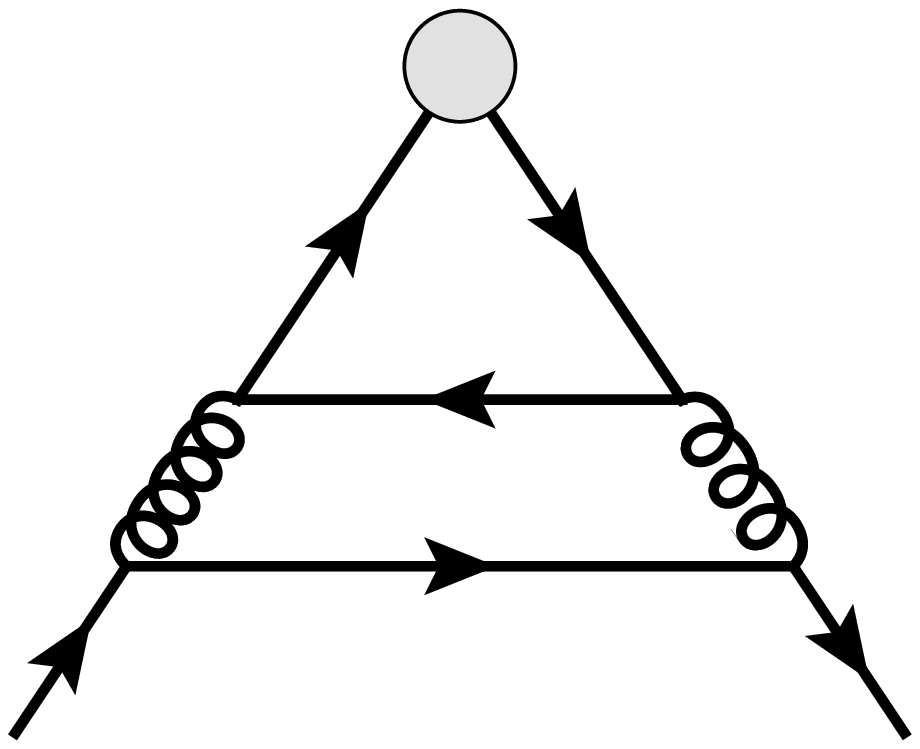} &
      \includegraphics[width=0.2\textwidth]{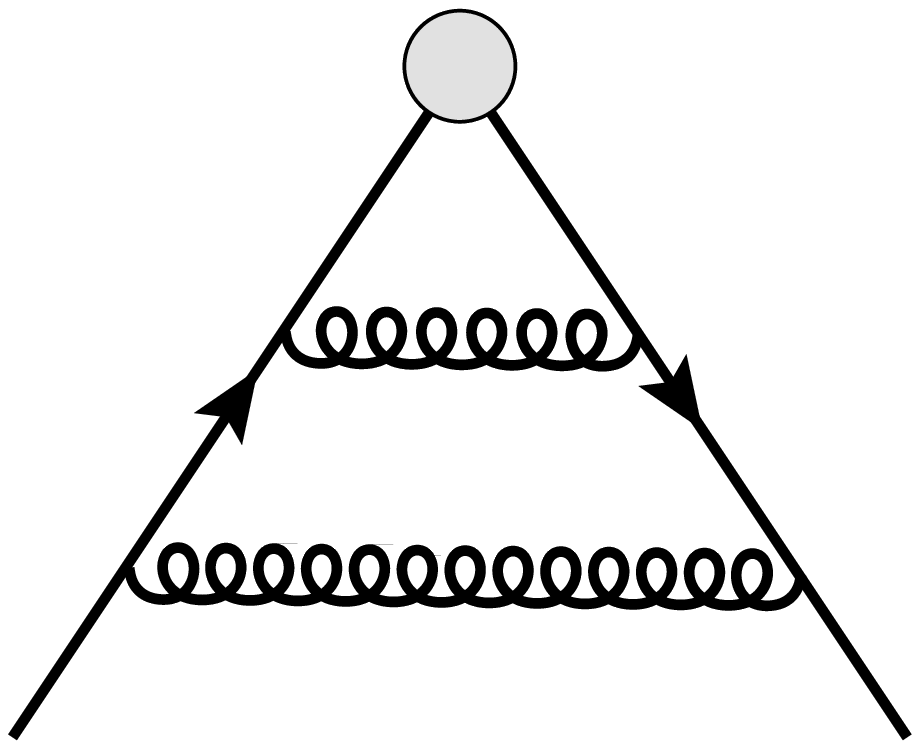} &
      \includegraphics[width=0.2\textwidth]{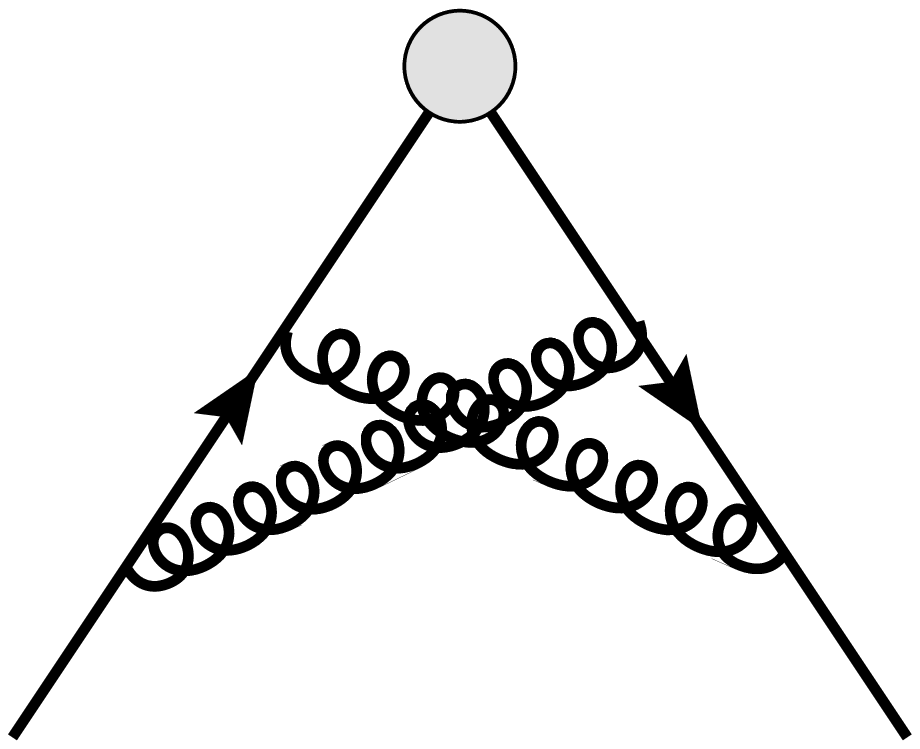} &
      \includegraphics[width=0.2\textwidth]{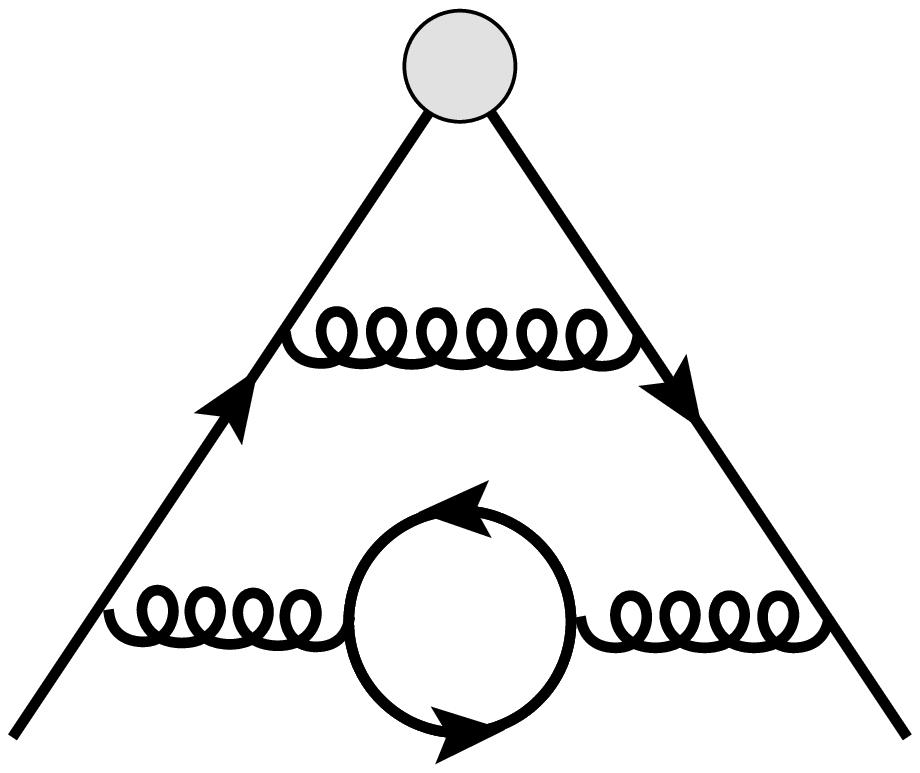}
      \\
      (a) & (b) & (c) & (d)
      \\[.3em]
      \includegraphics[width=0.2\textwidth]{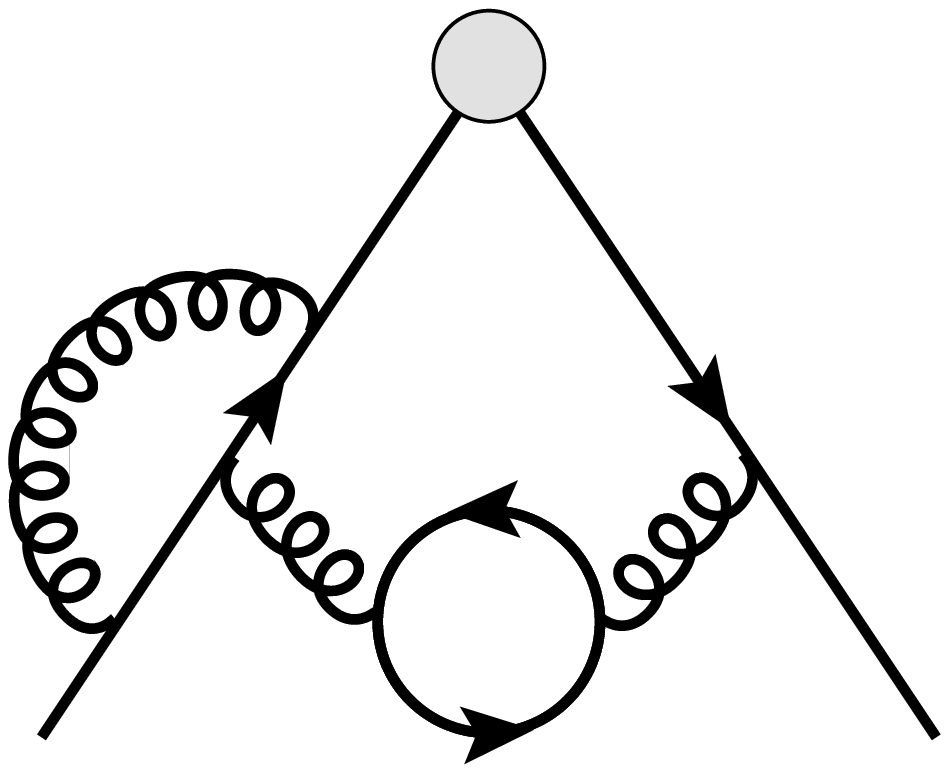} &
      \includegraphics[width=0.2\textwidth]{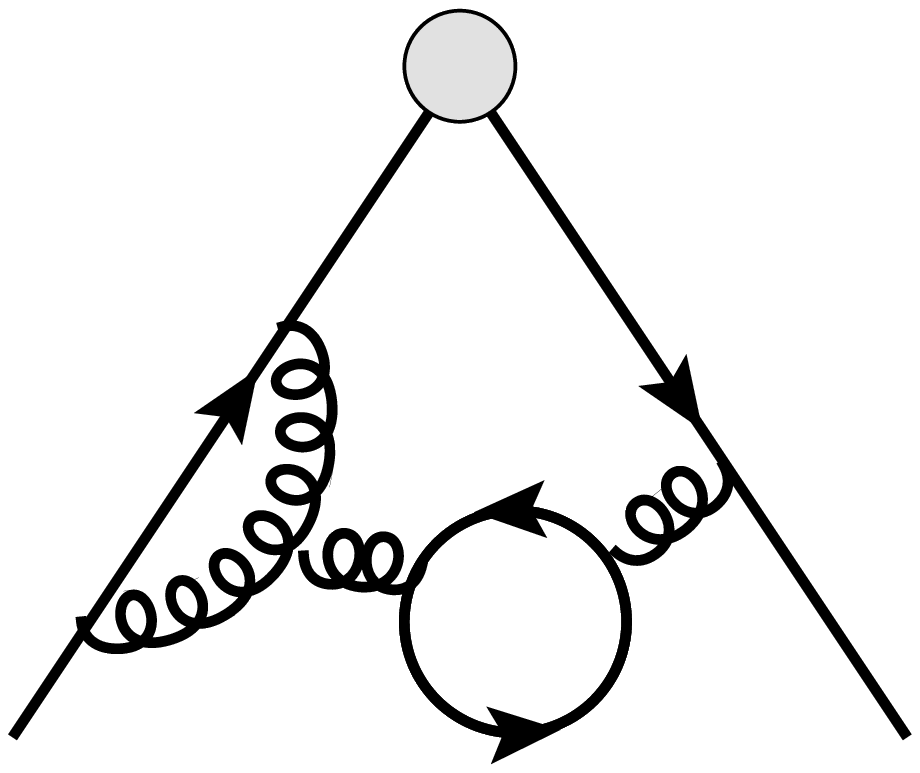} &
      \includegraphics[width=0.2\textwidth]{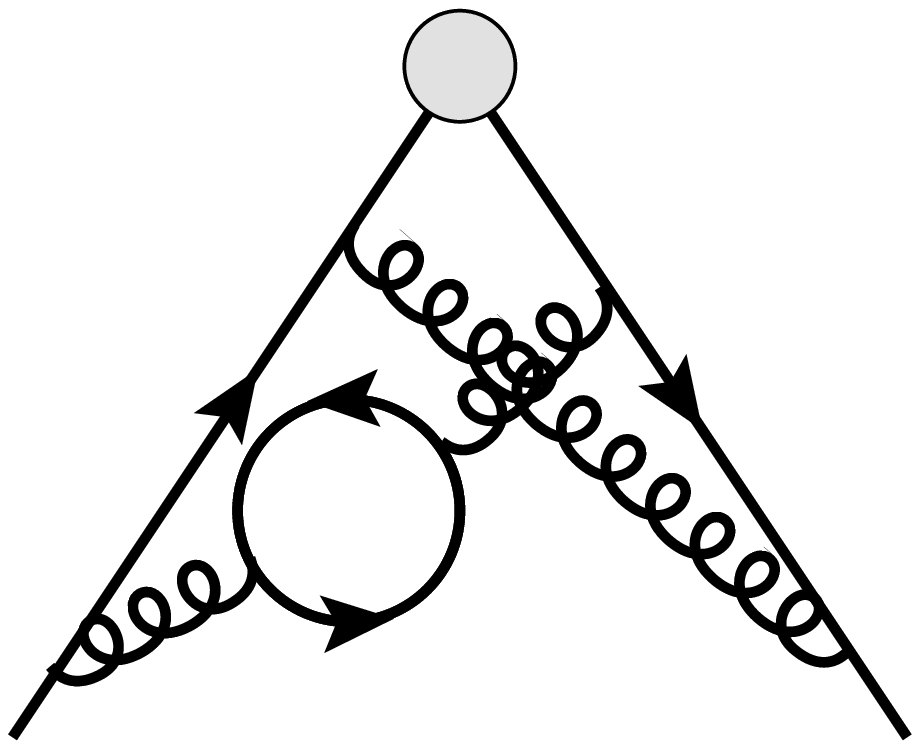} &
      \includegraphics[width=0.2\textwidth]{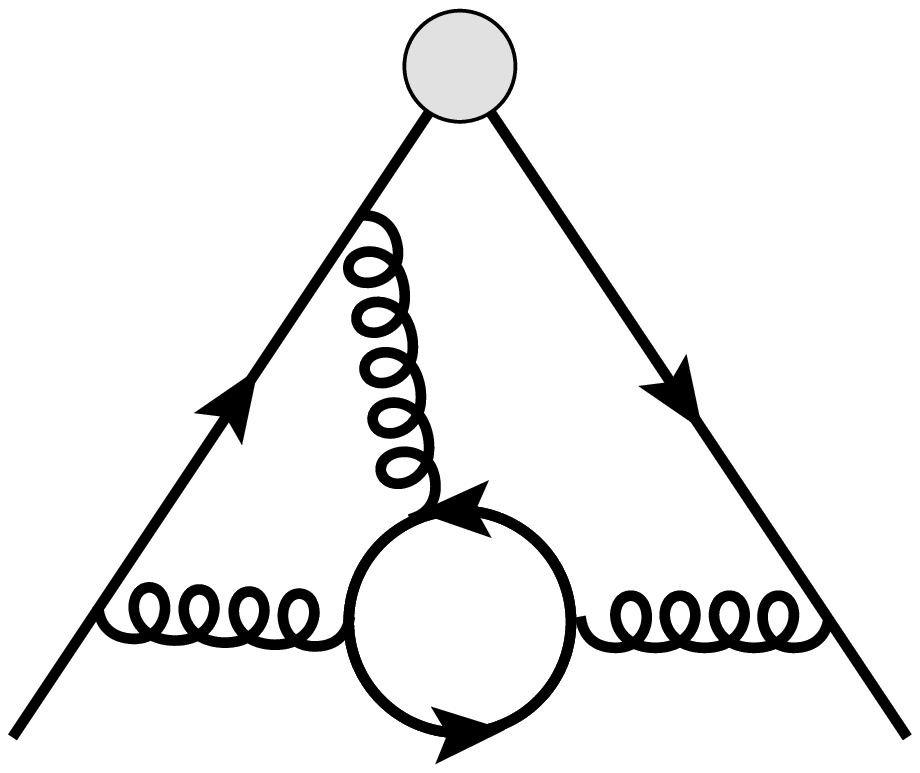}
      \\
      (e) & (f) & (g) & (h)
      \\[.3em]
      \includegraphics[width=0.2\textwidth]{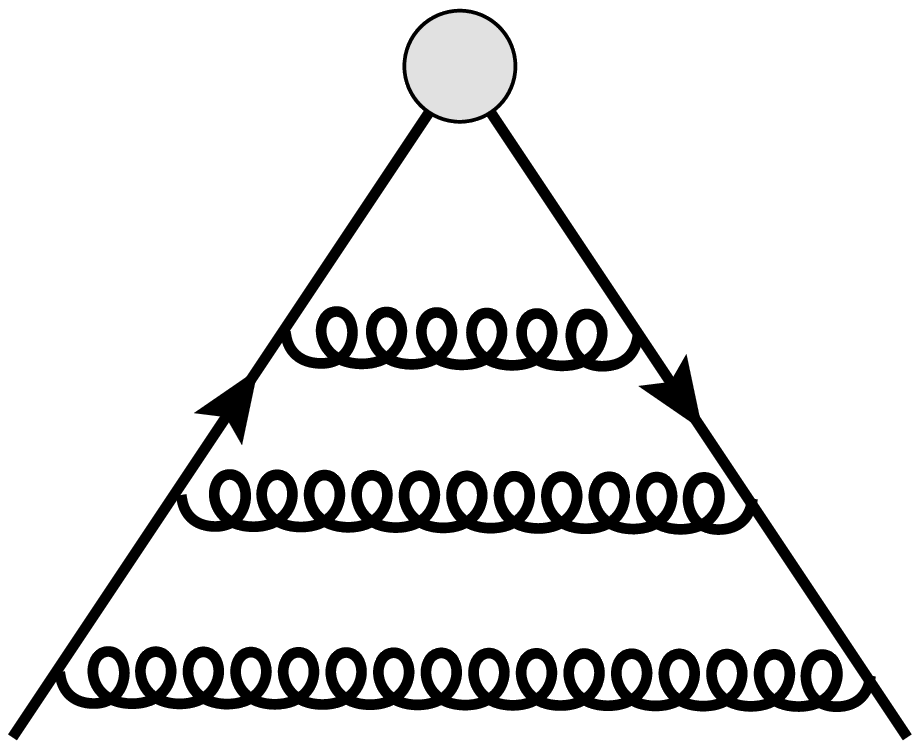} &
      \includegraphics[width=0.2\textwidth]{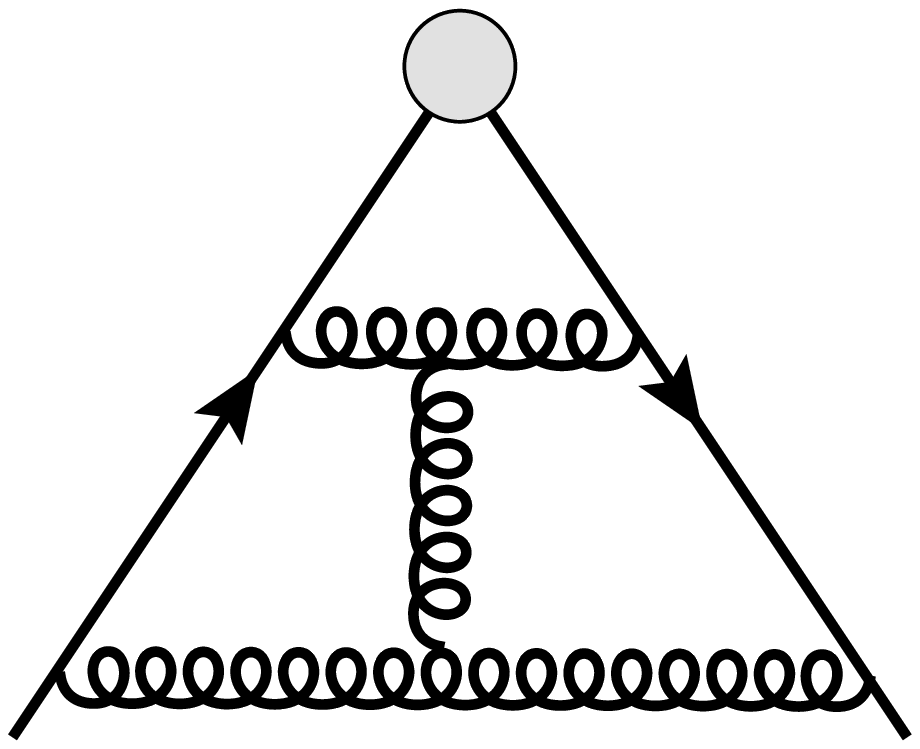} &
      \includegraphics[width=0.2\textwidth]{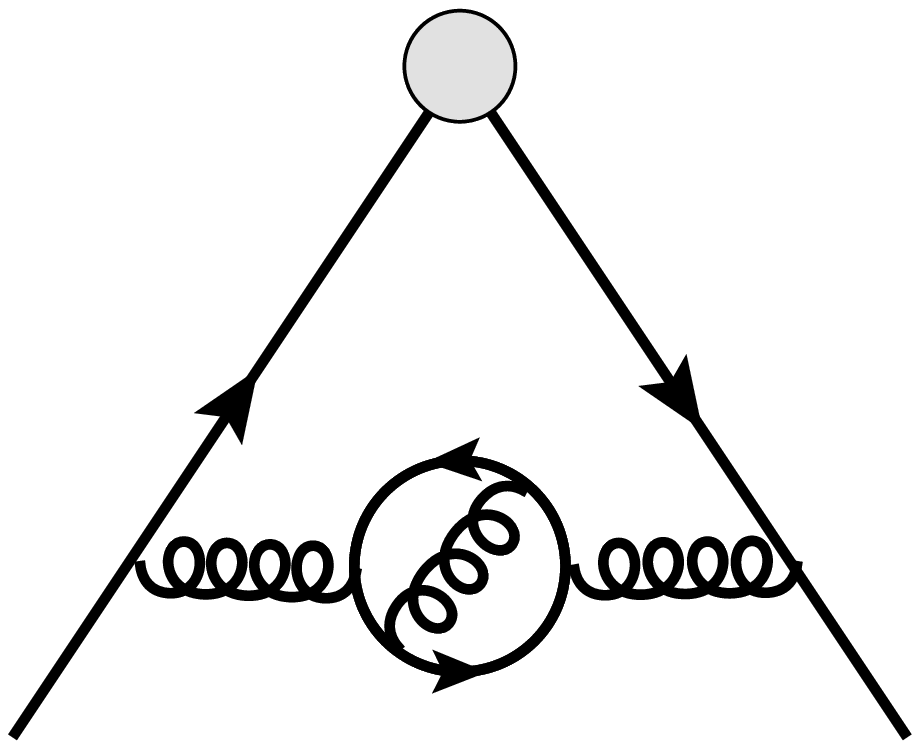} &
      \includegraphics[width=0.2\textwidth]{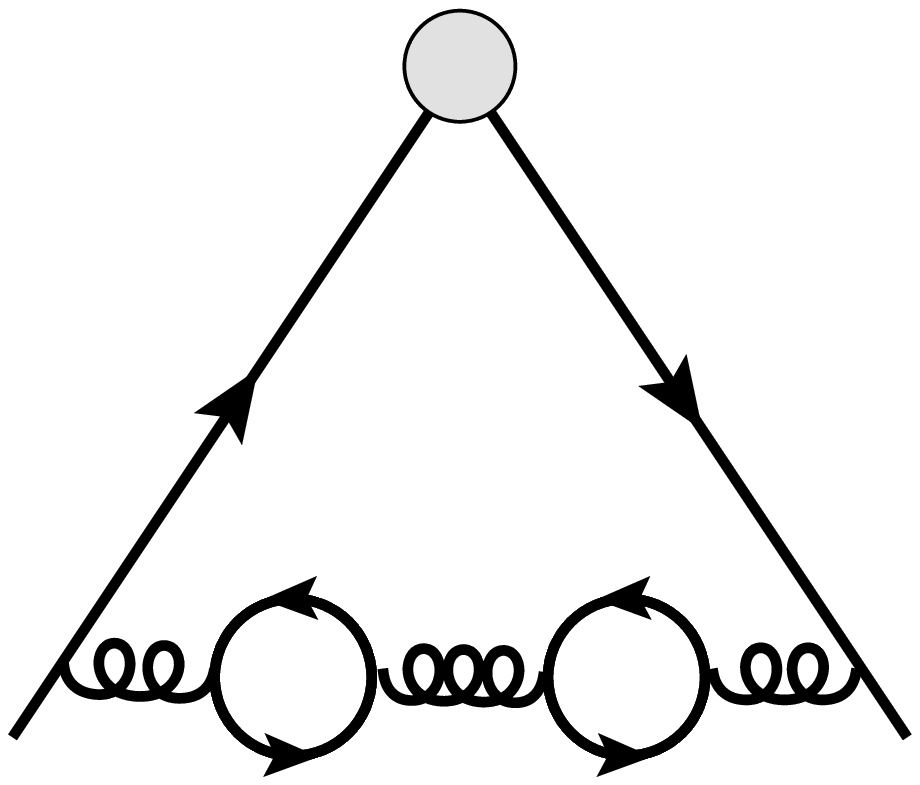}
      \\
      (i) & (j) & (k) & (l)
    \end{tabular}
    \caption{\label{fig::diags}Sample diagrams contributing to the form
      factors.  Solid and curly lines represent quarks and gluons,
      respectively. The grey blob refers to one of the external currents given
      in Eq.~(\ref{eq::currents}). Singlet contributions, as shown in (a), are
      not considered in this paper.}
  \end{center}
\end{figure}

For the three-point functions one in general distinguishes singlet and
non-singlet contributions.  The former includes a closed fermion loop which
contains the coupling to the external current. It is connected to the fermions
in the final state via gluons as is shown in Fig.~\ref{fig::diags}(a).  In
case the external current contains $\gamma_5$ singlet contributions need
special attention since the anti-commuting definition for $\gamma_5$ can not
be used.  Instead prescriptions like the one introduced in
Ref.~\cite{Larin:1993tq} have to be applied.

If the external current does not contain $\gamma_5$ the singlet contributions
can be treated along the same lines as the non-singlet
part. However, in contrast to the latter the singlet contributions have
massless cuts which requires modifications of the technique described
in~\cite{Henn:2016kjz,Henn:2016tyf,Lee:2018nxa} to compute the master
integrals. Thus, in this paper we restrict ourselves to non-singlet
contributions (cf. Figs.~\ref{fig::diags}(b)--(l)), i.e., the external
current couples to the fermions in the final state.  At three loops we compute
the complete light-fermion contributions and consider the large-$N_c$
expansion of the remaining part.  At one- and two-loop order all
colour factors are computed and agreement with the
literature~\cite{Ablinger:2017hst} is found.

In the next section we introduce the notation and briefly mention some
techniques used for the calculation. Afterwards analytical and numerical
results are presented in Sections~\ref{sec::ana} and~\ref{sec::num}.
We close with a brief summary in Section~\ref{sec::con}.


\section{Technicalities}

The techniques and the setup of the programs, which are used to obtain the
results of this paper, are straightforward extensions of the
works~\cite{Henn:2016tyf,Lee:2018nxa} and thus we refrain from repeating in
detail the technical descriptions. Note, however, that in contrast to
Ref.~\cite{Henn:2016tyf} we do not define a ``super family'', which includes
the eight relevant planar families as sub-cases.  Rather, we generated the
input files for {\tt FIRE}~\cite{Smirnov:2014hma} from scratch and computed
separate tables for each individual family.  Let us mention that for the
reduction to master integrals and the minimization of the latter it is 
useful to combine {\tt FIRE}~\cite{Smirnov:2014hma} with {\tt
  LiteRed}~\cite{Lee:2012cn,Lee:2013mka}, which provides important symmetry
information. In fact, for the most complicated integral family the
reduction took about a day of CPU time on a computer with 18 cores, even
for general gauge parameter.

\begin{figure}[t] 
  \begin{center}
    \includegraphics[width=\textwidth]{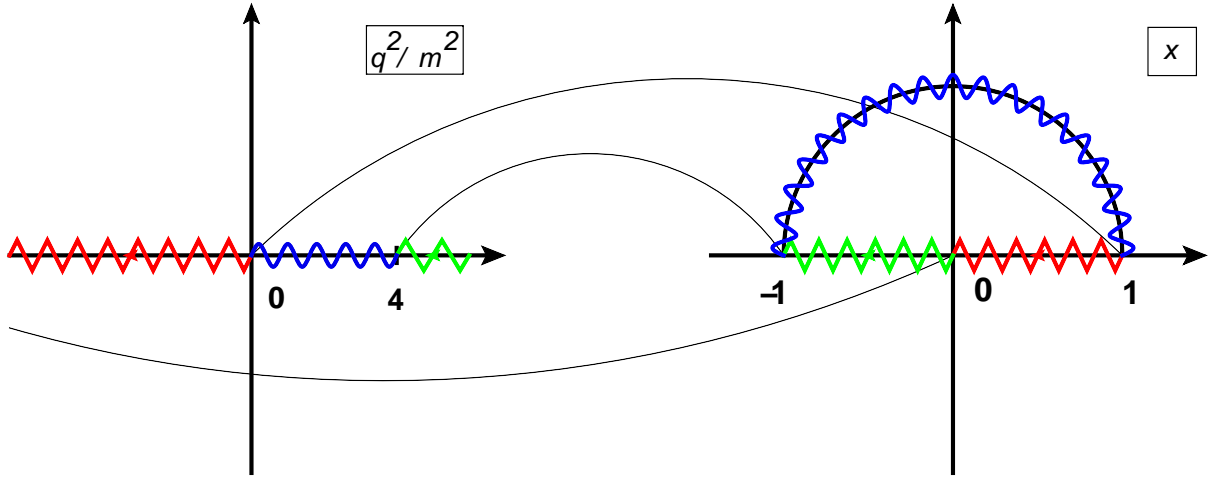}
    \caption{\label{fig::trans_x_q}Illustration of the variable transformation
      between $q^2/m^2$ and $x$ as given in Eq.~(\ref{eq::trans_x_q}).  The
      left graph represents the $q^2/m^2$ plane and on the right the
      complex $x$ plane is shown.  The (coloured) wiggled and zigzag lines
      show the mapping of the various intervals, whereas the straight lines
      indicate the mapping for special values of $q^2/m^2$ and $x$.  }
  \end{center}
\end{figure}

For the form factors it is useful to introduce the following variable
\begin{eqnarray}
  \frac{q^2}{m^2} &=& - \frac{(1-x)^2}{x}
  \,,
  \label{eq::trans_x_q}
\end{eqnarray}
which maps the complex $q^2/m^2$ plane into the unit circle, as illustrated in
Fig.~\ref{fig::trans_x_q}.  The low-energy ($q^2\to0$), high-energy
($q^2\to\infty$) and threshold ($q^2\to4m^2$) limits correspond to $x\to 1$,
$x\to 0$ and $x\to -1$, respectively.  Furthermore, the interval $q^2<0$ is
mapped to $x\in(0,1)$ and $q^2\in[0,4m^2]$ to the upper semi-circle.  Note
that for $x\in(0,1)$ and $x=e^{i\phi}$ with $\phi\in[0,\pi]$ the form factors
have to be real-valued since the corresponding Feynman diagrams do not have
cuts. This is different for the region $q^2>4m^2$, which corresponds to
$x\in(-1,0)$, where the form factors are complex-valued.  Note that for
negative $x$ we interpret $\log(x)$ as $\log(x+i0)=\log(-x)+i\pi$.

For the threshold limit ($q^2\to 4 m^2$, $x\to-1$) it is convenient to
introduce the velocity of the produced quarks
\begin{eqnarray}
  \beta &=& \sqrt{1 - \frac{4 m^2}{s}}
  \,,
\end{eqnarray}
which is related to $x$ via
\begin{eqnarray}
  x &=& \frac{2\beta}{1+\beta} - 1
  \,.
\end{eqnarray}

For the analytic three-loop expressions we furthermore define
\begin{eqnarray}
  r_{1/2}=e^{\pm i\pi/3} = (1\pm i \sqrt{3})/2 \,,\nonumber\\
  r_{3/4}=e^{\pm i2\pi/3} = (-1\pm i \sqrt{3})/2 \,.
\end{eqnarray}

In the practical calculation it is convenient to apply projectors
in order to extract the scalar form factors. We refrain to provide
them explicitly but refer to Ref.~\cite{Ablinger:2017hst}
where projectors for the four currents in Eq.~(\ref{eq::currents})
can be found.

All one- and two-loop Feynman integrals can be expressed as a linear
combination of the $2+17$ master integrals discussed in
Ref.~\cite{Lee:2018nxa}. Note that our two-loop basis is smaller than the one
of Ref.~\cite{Ablinger:2017hst} where 23 non-singlet master integrals are
given. After inserting the $\epsilon$-expanded results for the master
integrals into the expressions for the form factors we obtain the one- and
two-loop expressions expanded up to order $\epsilon^4$ and $\epsilon^2$,
respectively. Our two-loop results agree with~\cite{Ablinger:2017hst}.  Let us
repeat that we do not consider singlet contributions which occur for the first
time at two loops. Note that they vanish for the vector current but give
non-vanishing contributions for the other three currents.

At three-loop order we have 89 planar master integrals entering the
large-$N_c$ expressions and 15 additional master integrals for the complete
light-fermion $n_l$ and $n_l^2$ contributions, only two of the them
are non-planar.

To obtain the renormalized form factors we use the $\overline{\rm MS}$ scheme
for the strong coupling constant and the on-shell scheme for the heavy quark
mass and wave function of the external quarks. In all cases the counterterm
contributions are simply obtained by re-scaling the bare parameters with the
corresponding renormalization constants, $Z_{\alpha_s}$, $Z_m^{\rm OS}$ and
$Z_2^{\rm OS}$. The latter is needed to three loops whereas two-loop
corrections for $Z_{\alpha_s}$ and $Z_m^{\rm OS}$ are required.  For the
scalar and pseudo-scalar form factors also the overall overall factor $m$ has
to be renormalized (to three-loop order), which we choose to do in the
$\overline{\rm MS}$ scheme. Note that this is the natural choice if $F^s$ or
$F^p$ are used for Higgs boson production or decay since then $m$ takes over
the role of the Yukawa coupling.  The $\overline{\rm MS}$ renormalization
constants, of course, only contain pole parts. However, for the on-shell
quantities also higher order $\epsilon$ coefficients are needed since the one-
and two-loop form factors develop $1/\epsilon$ and $1/\epsilon^2$ poles,
respectively. Note that in our case the overall renormalization constants
of all currents in Eq.~(\ref{eq::currents}) are equal to unity.


\section{\label{sec::ana}Analytic results}


The analytic results for the form factors are expressed in terms of 
Goncharov polylogarithms (GPLs)~\cite{Goncharov:1998kja} with letters $-1, 0,
+1$ and $r_1$. They are quite long and we refrain
from presenting them in the paper. Rather we collect all relevant expressions
in a computer-readable format; the corresponding
file can be downloaded from~\cite{progdata}.
To fix the notation we provide one-loop results for
the six scalar form factors introduced in Eq.~(\ref{eq::Gamma})
up to the constant term in $\epsilon$.
For $\mu^2=m^2$ they are given by
\begin{eqnarray}
  F_1^{v,(1)} &=& 
\cR\left[
  \frac{1}{\epsilon}\left(\frac{2 \left(x^2+1\right) G(0|x)}{(x-1)
    (x+1)}-2\right)+\frac{\left(x^2+1\right) [G(0|x)]^2}{(x-1)
    (x+1)}
  \right.\nonumber\\&&\left.\mbox{}
  +\frac{\left(3 x^2+2 x+3\right) G(0|x)}{(x-1) (x+1)}-\frac{4
    \left(x^2+1\right) G(-1,0|x)}{(x-1) (x+1)}
  \right.\nonumber\\&&\left.\mbox{}
  -\frac{\pi ^2 \left(x^2+1\right)}{3 (x-1) (x+1)}-4
  \right] 
  \,,\nonumber\\
  F_2^{v,(1)} &=& \cR \frac{4 x G(0|x)}{(x-1) (x+1)}
  \,,\nonumber\\
  F_1^{a,(1)} &=& F_1^{v,(1)} - \cR \frac{4 x G(0|x)}{(x-1) (x+1)}
  \,,\nonumber\\
  F_2^{a,(1)} &=& \cR\left[ 
      \frac{4 x \left(3 x^2-2 x+3\right) G(0|x)}{(x-1)^3 (x+1)}-\frac{8
    x}{(x-1)^2}
  \right] \,,\nonumber\\
  F^{s,(1)} &=& F_1^{v,(1)} 
  + \cR\left[ 6-\frac{(x+3) (3 x+1) G(0|x)}{(x-1) (x+1)} \right]
  \,,\nonumber\\
  F^{p,(1)} &=& F_1^{v,(1)}
  + \cR\left[ 6-\frac{\left(3 x^2+2 x+3\right) G(0|x)}{(x-1) (x+1)} \right]
  \,,
  \label{eq::FF_1l}
\end{eqnarray}
where the two GPLs can be written as
\begin{eqnarray}
  G(0| x)     &=& \log(x)\,,\nonumber\\
  G(-1, 0| x) &=& \mbox{Li}_2(-x) + \log(x)\log(1+x) \,.
\end{eqnarray}
In Eq.~(\ref{eq::FF_1l}) we have the colour factor $C_F=(N_c^2-1)/(2N_c)$.  At
two-loop order one has $C_F^2$, $C_FC_A$, $C_FT_Fn_l$ and $C_FT_Fn_h$ where
$C_A=N_c$, $n_l$ counts the massless quark loops and $n_h=1$ the quark loops
with mass $m$. At three-loop order there are the colour factors
$C_FT_F^2n_l^2$, $C_F^2T_Fn_l$, $C_FC_AT_Fn_l$, $C_FT_F^2n_hn_l$ and
$N_c^3$. The latter is obtained by the replacements $C_F\to N_c/2$ and $C_A\to
N_c$ in the non-$n_l$ terms and taking only the leading contribution for large $N_c$.
This limit removes in particular all (pure) $n_h$ terms.

In the following three subsections we discuss the analytic structure of the
form factors in three important kinematical regions where the external
momentum $q^2$ is either small, large or close to the threshold for producing
the heavy quarks on-shell.  In these limits the expressions are compact and
analytic results can be reproduced in this paper.  We obtain the expansions of
the full result by expanding the GPLs in the respective region.  We restrict
ourselves to the choice $\mu^2=m^2$ and refer to~\cite{progdata} for the
general results. Subsection~\ref{sub::cusp} contains a brief discussion
on the infrared structure of the form factors and mentions several checks on
our calculation.



\subsection{Static limit}

After expanding the GPLs in the low-energy limit we obtain the expansion of
the form factor up to order $(1-x)^6$. In the following we present the results
for the first two terms for the axial-vector, scalar and pseudo-scalar
cases. The results for the vector currents can be found in
Refs.~\cite{Henn:2016kjz,Lee:2018nxa} (see Sections 4.2.1 and 5.2,
respectively).  Since the interval $q^2/m^2\in[0,4]$ is mapped to the upper
semi-unit circle in the complex $x$ plane we use $x=e^{i\phi}$ and parametrize
our results as a function of $\phi$ which is real-valued.  Our results read
\begin{eqnarray}
  F_1^{a,(1)} &=&
-2 \cR
+ \phi^2
\cR
\Bigg[
-\frac{2}{3 \ep}-\frac{5}{6}
\Bigg]
  \,,\nonumber\\
  F_1^{a,(2)} &=&
\cR^2
\Bigg[
-\frac{16 \pi ^2 \logtwo}{3}+8 \zthree+\frac{16 \pi ^2}{3}-\frac{29}{3}
\Bigg]
+\cA \cR
\Bigg[
\frac{8 \pi ^2 \logtwo}{3}-4 \zthree
\nonumber\\&&\mbox{}
-\frac{4 \pi ^2}{3}-\frac{143}{9}
\Bigg]
+\cR \tf \nl
\Bigg[
\frac{28}{9}
\Bigg]
+\cR \tf \nh
\Bigg[
\frac{460}{9}-\frac{16 \pi ^2}{3}
\Bigg]
\nonumber\\&&\mbox{}
+ \phi^2\Bigg\{
\cR^2
\Bigg[
\frac{4}{3 \ep}+\frac{64 \pi ^2 \logtwo}{15}-\frac{32 \zthree}{5}-\frac{217 \pi ^2}{270}-\frac{1121}{180}
\Bigg]
\nonumber\\&&\mbox{}
+\cA \cR
\Bigg[
\frac{11}{9 \ep^2}+\frac{\frac{2 \pi ^2}{9}-\frac{94}{27}}{\ep}-\frac{32 \pi ^2 \logtwo}{15}+\frac{88 \zthree}{15}+\frac{317 \pi ^2}{540}-\frac{19813}{1620}
\Bigg]
\nonumber\\&&\mbox{}
+\cR \tf \nl
\Bigg[
-\frac{4}{9 \ep^2}+\frac{20}{27 \ep}+\frac{8 \pi ^2}{27}+\frac{361}{81}
\Bigg]
+\cR \tf \nh
\Bigg[
\frac{491}{81}-\frac{\pi ^2}{2}
\Bigg]
\Bigg\}
  \,,\nonumber\\
  F_1^{a,(3)} &=&
\nc^3
\Bigg[
8 \pi ^2 \zthree+\frac{92 \zthree}{3}-20 \zeta (5)-\frac{2 \pi ^4}{3}+\frac{16 \pi ^2}{3}-\frac{107909}{648}
\Bigg]
\nonumber\\&&\mbox{}
+\cR^2 \tf \nl
\Bigg[
-\frac{1024 \afour}{9}-\frac{128 \logtwo^4}{27}-\frac{256 \pi ^2
  \logtwo^2}{27}+\frac{640 \pi ^2 \logtwo}{27}-\frac{736 \zthree}{9}+\frac{176
  \pi ^4}{81}
\nonumber\\&&\mbox{}
-\frac{896 \pi ^2}{27}+\frac{4586}{27}
\Bigg]
+\cA \cR \tf \nl
\Bigg[
\frac{512 \afour}{9}+\frac{64 \logtwo^4}{27}+\frac{128 \pi ^2
  \logtwo^2}{27}-\frac{320 \pi ^2 \logtwo}{27}
\nonumber\\&&\mbox{}
+\frac{608 \zthree}{9}-\frac{88 \pi ^4}{81}+\frac{392 \pi ^2}{27}+\frac{3752}{81}
\Bigg]
+\cR \tf^2 \nl^2
\Bigg[
\frac{400}{81}-\frac{64 \pi ^2}{27}
\Bigg]
\nonumber\\&&\mbox{}
+\cR \tf^2 \nh \nl
\Bigg[
\frac{448 \pi ^2}{27}-\frac{13600}{81}
{\Bigg]}
\nonumber\\&&\mbox{} 
  + \phi^2\Bigg\{
\nc^3
\Bigg[
-\frac{121}{81 \ep^3}+\frac{2383}{486 \ep^2}-\frac{22 \pi ^2}{81
  \ep^2}-\frac{10 \zthree}{27 \ep}-\frac{385}{108 \ep}-\frac{2 \pi ^4}{27
  \ep}+\frac{259 \pi ^2}{243 \ep}
\nonumber\\&&\mbox{} 
-\frac{122 \pi ^2 \zthree}{45}+\frac{226541 \zthree}{8100}-11 \zeta (5)+\frac{55 \pi ^4}{81}-\frac{262627 \pi ^2}{729000}-\frac{226620157}{1749600}
\Bigg]
\nonumber\\&&\mbox{}
+\cR^2 \tf \nl
\Bigg[
{\frac{\frac{14}{27}-\frac{32    \zthree}{9}}{\ep}
+\frac{4096 \afour}{45}}
+\frac{512 \logtwo^4}{135}+\frac{1024 \pi ^2
  \logtwo^2}{135}-\frac{1528 \pi ^2 \logtwo}{45}+\frac{6932 \zthree}{135}
\nonumber\\&&\mbox{}
-\frac{728 \pi ^4}{405}+\frac{64196 \pi ^2}{6075}+\frac{16453}{270}
\Bigg]
+\cA \cR \tf \nl
\Bigg[
\frac{176}{81 \ep^3}+\frac{\frac{16 \pi ^2}{81}-\frac{1552}{243}}{\ep^2}
\nonumber\\&&\mbox{}
+\frac{\frac{112 \zthree}{27}-\frac{160 \pi
    ^2}{243}+\frac{1556}{243}}{\ep}-\frac{256 \logtwo^4}{135}-\frac{512 \pi ^2
  \logtwo^2}{135}+\frac{764 \pi ^2 \logtwo}{45}-\frac{754
  \zthree}{45}+\frac{44 \pi ^4}{405}
\nonumber\\&&\mbox{}
+\frac{15443 \pi ^2}{3645}+\frac{292648}{2187}
{ -\frac{2048 \afour}{45}}
\Bigg]
+\cR \tf^2 \nl^2
\Bigg[
-\frac{32}{81 \ep^3}+\frac{160}{243 \ep^2}+\frac{32}{243 \ep}
\nonumber\\&&\mbox{}
{ -\frac{448  \zthree}{81}}
-\frac{560 \pi ^2}{243}-\frac{30748}{2187}
\Bigg]
+\cR \tf^2 \nh \nl
\Bigg[
\frac{8 \pi ^2}{81 \ep}+\frac{8 \pi ^2 \logtwo}{9}-\frac{284
  \zthree}{81}+\frac{980 \pi ^2}{243}
\nonumber\\&&\mbox{}
-\frac{3544}{81} 
\Bigg]
\Bigg\}
  \,,
\end{eqnarray}

\begin{eqnarray}
  F_2^{a,(1)} &=&
\frac{14}{3}\cR
+\phi^2
\frac{11}{15}\cR
  \,,\nonumber\\
  F_2^{a,(2)} &=&
\cR^2
\Bigg[
-\frac{88 \pi ^2 \logtwo}{15}+\frac{44 \zthree}{5}+\frac{88 \pi ^2}{15}-\frac{23}{5}
\Bigg]
+\cA \cR
\Bigg[
\frac{44 \pi ^2 \logtwo}{15}-\frac{22 \zthree}{5}-\frac{376 \pi ^2}{135}
\nonumber\\&&\mbox{}
+\frac{7663}{135}
\Bigg]
+\cR \tf \nl
\Bigg[
-\frac{412}{27}
\Bigg]
+\cR \tf \nh
\Bigg[
\frac{16 \pi ^2}{9}-\frac{412}{27}
\Bigg]
\nonumber\\&&\mbox{}
+ \phi^2\Bigg\{
\cR^2
\Bigg[
-\frac{28}{9 \ep}-\frac{44 \pi ^2 \logtwo}{21}+\frac{22 \zthree}{7}+\frac{296 \pi ^2}{135}-\frac{11111}{945}
\Bigg]
+\cA \cR
\Bigg[
\frac{22 \pi ^2 \logtwo}{21}
\nonumber\\&&\mbox{}
-\frac{11 \zthree}{7}
-\frac{211 \pi ^2}{225}+\frac{5039}{378}
\Bigg]
+\cR \tf \nl
\Bigg[
-\frac{466}{135}
\Bigg]
+\cR \tf \nh
\Bigg[
\frac{\pi ^2}{15}-\frac{14}{27}
\Bigg]
\Bigg\}
  \,,\nonumber\\
  F_2^{a,(3)} &=&
\nc^3
\Bigg[
-\frac{24 \pi ^2 \zthree}{5}+\frac{1198 \zthree}{225}+12 \zeta (5)-\frac{2 \pi ^4}{27}+\frac{486416 \pi ^2}{30375}+\frac{20858429}{48600}
\Bigg]
\nonumber\\&&\mbox{}
+\cR^2 \tf \nl
\Bigg[
-\frac{5632 \afour}{45}-\frac{704 \logtwo^4}{135}-\frac{1408 \pi ^2
  \logtwo^2}{135}+\frac{10688 \pi ^2 \logtwo}{135}-\frac{2176
  \zthree}{15}+\frac{968 \pi ^4}{405}
\nonumber\\&&\mbox{}
-\frac{94144 \pi ^2}{2025}-\frac{5606}{45}
\Bigg]
+\cA \cR \tf \nl
\Bigg[
\frac{2816 \afour}{45}+\frac{352 \logtwo^4}{135}+\frac{704 \pi ^2
  \logtwo^2}{135}-\frac{5344 \pi ^2 \logtwo}{135}
\nonumber\\&&\mbox{}
+\frac{4304 \zthree}{135}-\frac{484 \pi ^4}{405}+\frac{4568 \pi ^2}{1215}-\frac{118496}{243}
\Bigg]
+\cR \tf^2 \nl^2
\Bigg[
\frac{13616}{243}+\frac{448 \pi ^2}{81}
\Bigg]
\nonumber\\&&\mbox{}
+\cR \tf^2 \nh \nl
\Bigg[
\frac{48544}{243}-\frac{1600 \pi ^2}{81}
{\Bigg]}
\nonumber\\&&\mbox{}
+ \phi^2\Bigg\{
\nc^3
\Bigg[
\frac{77}{54 \ep^2}-\frac{385}{36 \ep}+\frac{19 \pi ^2}{81 \ep}-\frac{36 \pi
  ^2 \zthree}{35}+\frac{16754723 \zthree}{661500}+\frac{18 \zeta (5)}{7}
\nonumber\\&&\mbox{}
-\frac{107 \pi ^4}{1350}+\frac{247172251 \pi ^2}{61740000}+\frac{5650924217}{190512000}
\Bigg]
+\cR^2 \tf \nl
\Bigg[
-\frac{56}{27 \ep^2}+\frac{272}{27 \ep}
\nonumber\\&&\mbox{}
-\frac{352 \logtwo^4}{189}-\frac{704 \pi ^2 \logtwo^2}{189}+\frac{108896 \pi
  ^2 \logtwo}{4725}-\frac{278864 \zthree}{4725}+\frac{484 \pi
  ^4}{567}-\frac{3041144 \pi ^2}{297675}
\nonumber\\&&\mbox{}
+\frac{887293}{28350}
{ -\frac{2816 \afour}{63}}
\Bigg]
+\cA \cR \tf \nl
\Bigg[
\frac{1408 \afour}{63}+\frac{176 \logtwo^4}{189}+\frac{352 \pi ^2
  \logtwo^2}{189}-\frac{54448 \pi ^2 \logtwo}{4725}
\nonumber\\&&\mbox{}
+\frac{34024 \zthree}{1575}-\frac{242 \pi ^4}{567}+\frac{92494 \pi ^2}{23625}-\frac{5303386}{42525}
\Bigg]
+\cR \tf^2 \nl^2
\Bigg[
\frac{19496}{1215}+\frac{352 \pi ^2}{405}
\Bigg]
\nonumber\\&&\mbox{}
+\cR \tf^2 \nh \nl
\Bigg[
-\frac{{ 112} \pi ^2 \logtwo}{45}+\frac{392 \zthree}{45}-\frac{24 \pi ^2}{5}+\frac{13040}{243}
\Bigg]
\Bigg\}
  \,,
\end{eqnarray}

\begin{eqnarray}
  F^{s,(1)} &=&
-2\cR
+\phi^2
\cR
\Bigg[
-\frac{2}{3 \ep}-\frac{1}{3}
\Bigg]
  \,,\nonumber\\
  F^{s,(2)} &=&
\cR^2
\Bigg[
-8 \pi ^2 \logtwo+12 \zthree+5 \pi ^2+\frac{193}{8}
\Bigg]
+\cA \cR
\Bigg[
4 \pi ^2 \logtwo-6 \zthree
\nonumber\\&&\mbox{}
-\frac{4 \pi ^2}{3}-\frac{123}{8}
\Bigg]
+\cR \tf \nl
\Bigg[
\frac{11}{2}-\frac{4 \pi ^2}{3}
\Bigg]
+\cR \tf \nh
\Bigg[
\frac{51}{2}-\frac{8 \pi ^2}{3}
\Bigg]
\nonumber\\&&\mbox{}
+\phi^2\Bigg\{
\cR^2
\Bigg[
\frac{4}{3 \ep}+\frac{14 \pi ^2 \logtwo}{3}-7 \zthree-\frac{40 \pi ^2}{27}+\frac{2}{9}
\Bigg]
\nonumber\\&&\mbox{}
+\cA \cR
\Bigg[
\frac{11}{9 \ep^2}+\frac{\frac{2 \pi ^2}{9}-\frac{94}{27}}{\ep}-\frac{7 \pi ^2 \logtwo}{3}+\frac{37 \zthree}{6}+\frac{47 \pi ^2}{108}-\frac{650}{81}
\Bigg]
\nonumber\\&&\mbox{}
+\cR \tf \nl
\Bigg[
-\frac{4}{9 \ep^2}+\frac{20}{27 \ep}+\frac{8 \pi ^2}{27}+\frac{316}{81}
\Bigg]
+\cR \tf \nh
\Bigg[
\frac{5 \pi ^2}{18}-\frac{94}{81}
\Bigg]
\Bigg\}
  \,,\nonumber\\
  F^{s,(3)} &=&
\nc^3
\Bigg[
-6 \pi ^2 \zthree+\frac{181 \zthree}{9}+15 \zeta (5)-\frac{7 \pi ^4}{12}+\frac{4651 \pi ^2}{216}-\frac{428095}{7776}
\Bigg]
\nonumber\\&&\mbox{}
+\cR^2 \tf \nl
\Bigg[
-\frac{512 \afour}{3}-\frac{64 \logtwo^4}{9}-\frac{128 \pi ^2
  \logtwo^2}{9}+\frac{320 \pi ^2 \logtwo}{9}-\frac{536 \zthree}{3}+\frac{476
  \pi ^4}{135}
\nonumber\\&&\mbox{}
-\frac{200 \pi ^2}{9}-\frac{1286}{9}
\Bigg]
+\cA \cR \tf \nl
\Bigg[
\frac{256 \afour}{3}+\frac{32 \logtwo^4}{9}+\frac{64 \pi ^2
  \logtwo^2}{9}-\frac{160 \pi ^2 \logtwo}{9}
\nonumber\\&&\mbox{}
+\frac{220 \zthree}{9}-\frac{76
  \pi ^4}{135}-\frac{364 \pi ^2}{27}+\frac{54373}{243} 
\Bigg]
+\cR \tf^2 \nl^2
\Bigg[
\frac{224 \zthree}{9}+\frac{16 \pi ^2}{27}-\frac{8110}{243}
\Bigg]
\nonumber\\&&\mbox{}
+\cR \tf^2 \nh \nl
\Bigg[
-\frac{128 \zthree}{9}+\frac{208 \pi ^2}{9}-\frac{52076}{243}
\Bigg]
\nonumber\\&&\mbox{}
+ \phi^2\Bigg\{
\nc^3
\Bigg[
-\frac{121}{81 \ep^3}+\frac{2383}{486 \ep^2}-\frac{22 \pi ^2}{81
  \ep^2}-\frac{10 \zthree}{27 \ep}-\frac{5587}{864 \ep}-\frac{2 \pi ^4}{27
  \ep}+\frac{4549 \pi ^2}{3888 \ep}
\nonumber\\&&\mbox{}
-\frac{64 \pi ^2 \zthree}{9}+\frac{2089 \zthree}{162}+\frac{581 \pi ^4}{648}-\frac{4157 \pi ^2}{23328}-\frac{1444729}{17496}
\Bigg]
+\cR^2 \tf \nl
\Bigg[
\frac{896 \afour}{9}
\nonumber\\&&\mbox{}
+\frac{-\frac{32 \zthree}{9}+\frac{7 \pi ^2}{9}-\frac{29}{27}}{\ep}+\frac{112
  \logtwo^4}{27}+\frac{224 \pi ^2 \logtwo^2}{27}-\frac{904 \pi ^2
  \logtwo}{27}+\frac{1432 \zthree}{27}-\frac{794 \pi ^4}{405}
\nonumber\\&&\mbox{}
+\frac{4442 \pi
  ^2}{243}-\frac{620}{81} 
\Bigg]
+\cA \cR \tf \nl
\Bigg[
\frac{176}{81 \ep^3}+\frac{\frac{16 \pi ^2}{81}-\frac{1552}{243}}{\ep^2}
\nonumber\\&&\mbox{}
+\frac{\frac{112 \zthree}{27}-\frac{160 \pi
    ^2}{243}+\frac{1556}{243}}{\ep}-\frac{56 \logtwo^4}{27}-\frac{112 \pi ^2
  \logtwo^2}{27}+\frac{452 \pi ^2 \logtwo}{27}-\frac{494 \zthree}{27}+\frac{77
  \pi ^4}{405}
\nonumber\\&&\mbox{}
+\frac{1927 \pi ^2}{729}+\frac{264568}{2187}
{ -\frac{448 \afour}{9}}
\Bigg]
+\cR \tf^2 \nl^2
\Bigg[
-\frac{32}{81 \ep^3}+\frac{160}{243 \ep^2}+\frac{32}{243 \ep}-\frac{448
  \zthree}{81}
\nonumber\\&&\mbox{}
-\frac{416 \pi ^2}{243}-\frac{34960}{2187}
\Bigg]
+\cR \tf^2 \nh \nl
\Bigg[
\frac{8 \pi ^2}{81 \ep}+\frac{8 \pi ^2 \logtwo}{3}-\frac{788
  \zthree}{81}+\frac{572 \pi ^2}{243}
\nonumber\\&&\mbox{}
-\frac{2464}{81} 
\Bigg]
\Bigg\}
  \,,
\end{eqnarray}

\begin{eqnarray}
  F^{p,(1)} &=&
2\cR + \phi^2\cR
\Bigg[
\frac{1}{3}-\frac{2}{3 \ep}
\Bigg]
  \,,\nonumber\\
  F^{p,(2)} &=&
\cR^2
\Bigg[
-\frac{40 \pi ^2 \logtwo}{3}+20 \zthree+\frac{31 \pi ^2}{3}-\frac{61}{24}
\Bigg]
+\cA \cR
\Bigg[
\frac{20 \pi ^2 \logtwo}{3}-10 \zthree
\nonumber\\&&\mbox{}
-\frac{8 \pi ^2}{3}+\frac{2189}{72}
\Bigg]
+\cR \tf \nl
\Bigg[
-\frac{157}{18}-\frac{4 \pi ^2}{3}
\Bigg]
+\cR \tf \nh
\Bigg[
\frac{491}{18}-\frac{8 \pi ^2}{3}
\Bigg]
\nonumber\\&&\mbox{}
+\phi^2\Bigg\{
\cR^2
\Bigg[
-\frac{4}{3 \ep}+\frac{14 \pi ^2 \logtwo}{5}-\frac{21 \zthree}{5}+\frac{67 \pi ^2}{135}-\frac{512}{45}
\Bigg]
\nonumber\\&&\mbox{}
+\cA \cR
\Bigg[
\frac{11}{9 \ep^2}+\frac{\frac{2 \pi ^2}{9}-\frac{94}{27}}{\ep}-\frac{7 \pi ^2 \logtwo}{5}+\frac{143 \zthree}{30}-\frac{59 \pi ^2}{540}+\frac{794}{405}
\Bigg]
\nonumber\\&&\mbox{}
+\cR \tf \nl
\Bigg[
-\frac{4}{9 \ep^2}+\frac{20}{27 \ep}+\frac{8 \pi ^2}{27}+\frac{52}{81}
\Bigg]
+\cR \tf \nh
\Bigg[
\frac{182}{81}-\frac{\pi ^2}{18}
\Bigg]
\Bigg\}
  \,,\nonumber\\
  F^{p,(3)} &=&
\nc^3
\Bigg[
2 \pi ^2 \zthree+\frac{457 \zthree}{9}-5 \zeta (5)-\frac{5 \pi ^4}{4}+\frac{8191 \pi ^2}{216}+\frac{1639379}{7776}
\Bigg]
\nonumber\\&&\mbox{}
+\cR^2 \tf \nl
\Bigg[
-\frac{2560 \afour}{9}-\frac{320 \logtwo^4}{27}-\frac{640 \pi ^2
  \logtwo^2}{27}+\frac{2752 \pi ^2 \logtwo}{27}-\frac{2056
  \zthree}{9}+\frac{2308 \pi ^4}{405}
\nonumber\\&&\mbox{}
-\frac{2360 \pi ^2}{27}-\frac{844}{27}
\Bigg]
+\cA \cR \tf \nl
\Bigg[
\frac{1280 \afour}{9}+\frac{160 \logtwo^4}{27}+\frac{320 \pi ^2
  \logtwo^2}{27}-\frac{1376 \pi ^2 \logtwo}{27}
\nonumber\\&&\mbox{}
+28 \zthree-\frac{668 \pi ^4}{405}-\frac{308 \pi ^2}{27}-\frac{59507}{243}
\Bigg]
+\cR \tf^2 \nl^2
\Bigg[
\frac{224 \zthree}{9}+\frac{16 \pi ^2}{3}+\frac{5906}{243}
\Bigg]
\nonumber\\&&\mbox{}
+\cR \tf^2 \nh \nl
\Bigg[
-\frac{128 \zthree}{9}+\frac{80 \pi ^2}{9}-\frac{17132}{243}
\Bigg]
\nonumber\\&&\mbox{}
+ \phi^2\Bigg\{
\nc^3
\Bigg[
-\frac{121}{81 \ep^3}+\frac{2977}{486 \ep^2}-\frac{22 \pi ^2}{81
  \ep^2}-\frac{10 \zthree}{27 \ep}-\frac{11155}{864 \ep}-\frac{2 \pi ^4}{27
  \ep}+\frac{4549 \pi ^2}{3888 \ep}
\nonumber\\&&\mbox{}
-\frac{176 \pi ^2 \zthree}{45}+\frac{59387 \zthree}{2025}-8 \zeta (5)+\frac{437 \pi ^4}{648}+\frac{7508351 \pi ^2}{2916000}-\frac{97652371}{1749600}
\Bigg]
\nonumber\\&&\mbox{}
+\cR^2 \tf \nl
\Bigg[
-\frac{16}{9 \ep^2}+\frac{-\frac{32 \zthree}{9}+\frac{7
    \pi ^2}{9}+\frac{211}{27}}{\ep}+\frac{112 \logtwo^4}{45}+\frac{224 \pi ^2
  \logtwo^2}{45}-\frac{1912 \pi ^2 \logtwo}{135}
\nonumber\\&&\mbox{}
+\frac{3536 \zthree}{135}-\frac{6 \pi ^4}{5}+\frac{20138 \pi ^2}{6075}+\frac{24101}{405}
{ +\frac{896 \afour}{15}}
\Bigg]
+\cA \cR \tf \nl
\Bigg[
\frac{176}{81 \ep^3}
\nonumber\\&&\mbox{}
{ +\frac{\frac{16 \pi ^2}{81}-\frac{1552}{243}}{\ep^2}}
+\frac{\frac{112
    \zthree}{27}-\frac{160 \pi ^2}{243}+\frac{1556}{243}}{\ep}-\frac{56
  \logtwo^4}{45}-\frac{112 \pi ^2 \logtwo^2}{45}+\frac{956 \pi ^2
  \logtwo}{135}-\frac{1186 \zthree}{135}
\nonumber\\&&\mbox{}
-\frac{77 \pi ^4}{405}
+\frac{18869 \pi ^2}{3645}+\frac{26032}{2187}
{ -\frac{448 \afour}{15}}
\Bigg]
+\cR \tf^2 \nl^2
\Bigg[
-\frac{32}{81 \ep^3}+\frac{160}{243 \ep^2}+\frac{32}{243 \ep}
\nonumber\\&&\mbox{}
{ -\frac{448 \zthree}{81}}
-\frac{224 \pi ^2}{243}-\frac{112}{2187}
\Bigg]
+\cR \tf^2 \nh \nl
\Bigg[
\frac{8 \pi ^2}{81 \ep}+\frac{8 \pi ^2 \logtwo}{9}-\frac{284
  \zthree}{81}-\frac{220 \pi ^2}{243}
\nonumber\\&&\mbox{}
+\frac{1504}{243}
\Bigg]
\Bigg\}
  \,,
\end{eqnarray}

In all cases the limit $q^2\to0$ exists (i.e., there are no logarithmic terms
in $q^2$) and all form factors become infrared finte.  The infrared
divergences are present starting from the $\phi^2$ term.  Note that for the
vector case all quantum corrections vanish for $q^2=0$ and we have
$F^v_1(0)=1$ whereas for all other form form factors this is not the case. The
static form factors for the vector and axial-vector case have been discussed in
Ref.~\cite{Bernreuther:2005gq} up to two-loop order and the physical
interpretations have nicely been summarized.  Once the complete non-fermionic
pieces and all singlet contributions are available the analysis of
Ref.~\cite{Bernreuther:2005gq} can be extended to three loops.  Three-loop
corrections to $F_2^v(0)$ have been computed in~\cite{Grozin:2007fh}.


\subsection{\label{sub::high}High-energy limit}

In the high-energy limit, i.e. for $x\to0$
it is convenient to introduce
\begin{eqnarray}
  F^{\delta,(n)} &=& \sum_{k\ge0} {f_{\rm lar}^{\delta,(n,k)} } x^k
  \,,
  \label{eq::F_i_lar}
\end{eqnarray}
where we have computed seven expansion terms, i.e., up to ${\cal O}(x^6)$, for
all six scalar form factors. Note that the leading terms are identical both for
$F_1^v$ and $F_1^a$ and for $F^s$ and $F^p$ since in this limit the quark masses
in the numerator can be neglected and $\gamma_5$ is anti-commuted through an
even number of $\gamma$ matrices to one end of the fermion string.
As a consequence we also have that $f_2^{a,(n,0)}=0$ since $f_2^{v,(n,0)}=0$.
We illustrate the structure of the analytic
expressions by showing the terms of order $x^0$ and $x^1$
which are given by
\begin{eqnarray}
  f_{1,\rm lar}^{a,(1,0)} &=& f_{1,\rm lar}^{v,(1,0)}
  \,,\nonumber\\
  f_{1,\rm lar}^{a,(1,1)} &=&
\cR
\Bigg[
6 \logx-4
\Bigg]
  \,,\nonumber\\
  f_{1,\rm lar}^{a,(2,0)} &=& f_{1,\rm lar}^{v,(2,0)}
  \,,\nonumber\\
  f_{1,\rm lar}^{a,(2,1)} &=&
\cR^2
\Bigg[
\left(-\frac{12}{\ep}+\frac{4 \pi ^2}{3}-19\right)
 \logx^2+\logx \left(-\frac{4}{\ep}-80 \zthree+\frac{22 \pi ^2}{3}+19\right)
\nonumber\\&&\mbox{}
+\frac{8}{\ep}+48 \pi ^2 \logtwo+\frac{\logx^4}{3}-12 \logx^3-104 \zthree+\frac{4 \pi ^4}{5}-11 \pi ^2-22
\Bigg]
\nonumber\\&&\mbox{}
+\cA \cR
\Bigg[
-24 \pi ^2 \logtwo-\frac{\logx^4}{6}+\left(\frac{4 \pi ^2}{3}-9\right)
 \logx^2+\logx \left(-56 \zthree+6 \pi ^2+\frac{241}{3}\right)
\nonumber\\&&\mbox{}
-168 \zthree+\frac{11 \pi ^4}{15}+\frac{79 \pi ^2}{3}-\frac{796}{9}
\Bigg]
+\cR \tf \nl
\Bigg[
-4 \logx^2-\frac{68 \logx}{3}+\frac{4 \pi ^2}{3}+\frac{200}{9}
\Bigg]
\nonumber\\&&\mbox{}
+\cR \tf \nh
\Bigg[
-20 \logx^2-\frac{164 \logx}{3}-\frac{68 \pi ^2}{3}-\frac{160}{9}
\Bigg]
  \,,\nonumber\\
  f_{1,\rm lar}^{a,(3,0)} &=& f_{1,\rm lar}^{v,(3,0)}
  \,,\nonumber\\
  f_{1,\rm lar}^{a,(3,1)} &=&
\nc^3
\Bigg[
\frac{3 \logx^3}{2 \ep^2}+\frac{15 \logx^2}{2 \ep^2}+\frac{4 \logx}{3
  \ep^2}-\frac{14}{3 \ep^2}+\frac{9 \logx^4}{4 \ep}+\frac{12
  \logx^3}{\ep}-\frac{\pi ^2 \logx^3}{\ep}+\frac{48 \logx^2
  \zthree}{\ep}-\frac{239 \logx^2}{6 \ep} 
\nonumber\\&&\mbox{}
-\frac{61 \pi ^2 \logx^2}{12 \ep}+\frac{155 \logx \zthree}{\ep}+\frac{47
  \logx}{12 \ep}-\frac{17 \pi ^4 \logx}{30 \ep}
-\frac{31 \pi ^2 \logx}{2 \ep}+\frac{112 \zthree}{\ep}+\frac{97}{2
  \ep}-\frac{17 \pi ^4}{30 \ep}
\nonumber\\&&\mbox{}
-\frac{137 \pi ^2}{12 \ep}+\frac{15
  \logx^5}{8}-\frac{3 \pi ^2 \logx^4}{2}+\frac{89 \logx^4}{8}
+40 \logx^3
\zthree-\frac{119 \pi ^2 \logx^3}{72}
-\frac{283 \logx^3}{4}-90 \logx^2 \zthree
\nonumber\\&&\mbox{}
+\frac{17 \pi ^4
  \logx^2}{10}
+\frac{11 \pi ^2 \logx^2}{18}-\frac{3797 \logx^2}{36}-6 \pi ^2
\logx \zthree-573 \logx \zthree-384 \logx \zeta (5)+\frac{1579 \pi ^4
  \logx}{360}
\nonumber\\&&\mbox{}
+\frac{22309 \pi ^2 \logx}{216}
+\frac{3211 \logx}{9}+426 \zthree^2-\frac{121 \pi ^2 \zthree}{3}-\frac{20134
  \zthree}{9}+678 \zeta (5)
\nonumber\\&&\mbox{}
-\frac{353 \pi ^6}{3780}+\frac{15593 \pi ^4}{2160}+\frac{18815 \pi ^2}{216}-\frac{89909}{324}
\Bigg]  
+\cR^2 \tf \nl
\Bigg[
1024 \afour
+\logx^2 \left(-\frac{8}{\ep^2}
\right.\nonumber\\&&\left.\mbox{}
+\frac{164}{3 \ep}+\frac{304 \zthree}{3}+\frac{122 \pi ^2}{27}+\frac{1856}{9}\right)
+\logx \left(-\frac{8}{3 \ep^2}+\frac{\frac{32}{3}-\frac{4 \pi
      ^2}{3}}{\ep}+\frac{3296 \zthree}{9}
\right.\nonumber\\&&\left.\mbox{}
+\frac{40 \pi ^4}{27}-\frac{2290 \pi ^2}{27}-\frac{346}{3}\right)
+\frac{16}{3 \ep^2}+\left(\frac{4}{\ep}-\frac{8 \pi ^2}{9}+\frac{868}{9}\right)
 \logx^3+\frac{-40-\frac{4 \pi ^2}{3}}{\ep}
\nonumber\\&&\mbox{}
+\frac{128 \logtwo^4}{3}+\frac{256 \pi ^2 \logtwo^2}{3}-\frac{1664 \pi ^2
  \logtwo}{3}-\frac{4 \logx^5}{9}+\frac{320 \logx^4}{27}-\frac{64 \pi ^2
  \zthree}{3}+\frac{10352 \zthree}{9}
\nonumber\\&&\mbox{}
-672 \zeta (5)-\frac{764 \pi ^4}{135}+\frac{1832 \pi ^2}{9}-\frac{1628}{9}
\Bigg]
+\cA \cR \tf \nl
\Bigg[
-512 \afour-\frac{64 \logtwo^4}{3}
\nonumber\\&&\mbox{}
-\frac{128 \pi ^2 \logtwo^2}{3}+\frac{832 \pi ^2 \logtwo}{3}+\frac{2 \logx^5}{9}+\frac{20 \logx^4}{27}+\left(\frac{32}{3}-\frac{20 \pi ^2}{9}\right)
 \logx^3+\logx^2 \left(\frac{88 \zthree}{3}-\frac{532 \pi ^2}{27}
\right.\nonumber\\&&\left.\mbox{}
-\frac{112}{3}\right)
+\logx \left(\frac{2960 \zthree}{9}+\frac{308 \pi ^4}{135}-\frac{724 \pi ^2}{9}-\frac{2120}{3}\right)
+\frac{32 \pi ^2 \zthree}{3}+\frac{3448 \zthree}{3}
\nonumber\\&&\mbox{}
-496 \zeta (5)+\frac{242 \pi ^4}{135}-\frac{4364 \pi ^2}{27}+\frac{83992}{81}
\Bigg]
+\cR \tf^2 \nl^2
\Bigg[
\frac{32 \logx^3}{9}+\frac{272 \logx^2}{9}+\left(\frac{880}{9}
\right.\nonumber\\&&\left.\mbox{}
+\frac{32 \pi ^2}{9}\right)
 \logx-\frac{128 \zthree}{3}-\frac{400 \pi ^2}{27}-\frac{11296}{81}
\Bigg]
+\cR \tf^2 \nh \nl
\Bigg[
\frac{64 \logx^3}{3}+\frac{1184 \logx^2}{9}
\nonumber\\&&\mbox{}
+\left(\frac{1504}{9}+\frac{64 \pi ^2}{3}\right)
 \logx+\frac{640 \zthree}{3}
+\frac{704 \pi ^2}{9}-\frac{18272}{81}
\Bigg]
  \,,
\end{eqnarray}
\begin{eqnarray}
  f_{2,\rm lar}^{a,(1,1)} &=&
\cR
\Bigg[
-12 \logx-8
\Bigg]
  \,,\nonumber\\
  f_{2,\rm lar}^{a,(2,1)} &=&
\cR^2
\Bigg[
\left(\frac{24}{\ep}+86\right)
 \logx^2+\left(\frac{40}{\ep}-8 \pi ^2+122\right)
 \logx+\frac{16}{\ep}-32 \pi ^2 \logtwo+24 \logx^3+48 \zthree
\nonumber\\&&\mbox{}
+22 \pi ^2+68
\Bigg]
+\cA \cR
\Bigg[
16 \pi ^2 \logtwo-22 \logx^2-\frac{458 \logx}{3}+48 \zthree-14 \pi ^2-\frac{968}{9}
\Bigg]
\nonumber\\&&\mbox{}
+\cR \tf \nl
\Bigg[
8 \logx^2+\frac{136 \logx}{3}-\frac{8 \pi ^2}{3}+\frac{304}{9}
\Bigg]
+\cR \tf \nh
\Bigg[
8 \logx^2+\frac{136 \logx}{3}+\frac{40 \pi ^2}{3}
\nonumber\\&&\mbox{}
-\frac{128}{9}
\Bigg]
  \,,\nonumber\\
  f_{2,\rm lar}^{a,(3,1)} &=&
\nc^3
\Bigg[
-\frac{3 \logx^3}{\ep^2}-\frac{19 \logx^2}{\ep^2}-\frac{76 \logx}{3
  \ep^2}-\frac{28}{3 \ep^2}-\frac{9 \logx^4}{2 \ep}-\frac{13
  \logx^3}{\ep}+\frac{143 \logx^2}{3 \ep}+\frac{\pi ^2 \logx^2}{2
  \ep}-\frac{30 \logx \zthree}{\ep} 
\nonumber\\&&\mbox{}
+\frac{547 \logx}{6 \ep}+\frac{8 \pi ^2 \logx}{3 \ep}-\frac{32
  \zthree}{\ep}+\frac{37}{\ep}+\frac{13 \pi ^2}{6 \ep}-\frac{15
  \logx^5}{4}-\frac{25 \logx^4}{12}+\frac{3 \pi ^2 \logx^3}{4}
+\frac{815  \logx^3}{9}
\nonumber\\&&\mbox{}
-111 \logx^2 \zthree 
+\frac{25 \pi ^2 \logx^2}{4}+\frac{1441 \logx^2}{6}-30 \logx \zthree+\frac{13
  \pi ^4 \logx}{30}-\frac{799 \pi ^2 \logx}{18}-\frac{5545 \logx}{9} 
\nonumber\\&&\mbox{}
-19 \pi ^2 \zthree+\frac{2333 \zthree}{3} 
-60 \zeta (5)-\frac{27 \pi ^4}{20}+\frac{3509 \pi ^2}{108}-\frac{46702}{81}
\Bigg] 
\nonumber\\&&\mbox{}
+\cR^2 \tf \nl
\Bigg[
-\frac{2048 \afour}{3}+\left(\frac{16}{\ep^2}-\frac{280}{3 \ep}-\frac{28 \pi ^2}{3}-\frac{7192}{9}\right)
 \logx^2+\logx \left(\frac{80}{3 \ep^2}
\right.\nonumber\\&&\left.\mbox{}
+\frac{\frac{8 \pi ^2}{3}-\frac{448}{3}}{\ep}+32 \zthree+\frac{92 \pi ^2}{9}-\frac{7732}{9}\right)
+\frac{32}{3 \ep^2}+\left(-\frac{8}{\ep}-\frac{2096}{9}\right)
 \logx^3
\nonumber\\&&\mbox{}
+\frac{\frac{8 \pi ^2}{3}-64}{\ep}-\frac{256 \logtwo^4}{9}-\frac{512 \pi ^2
  \logtwo^2}{9}+\frac{2816 \pi ^2 \logtwo}{9}-\frac{80 \logx^4}{3}-256
\zthree+\frac{1424 \pi ^4}{135}
\nonumber\\&&\mbox{}
-\frac{2056 \pi ^2}{9}-\frac{4328}{9}
\Bigg]
+\cA \cR \tf \nl
\Bigg[
\frac{1024 \afour}{3}+\frac{128 \logtwo^4}{9}+\frac{256 \pi ^2
  \logtwo^2}{9}-\frac{1408 \pi ^2 \logtwo}{9}
\nonumber\\&&\mbox{}
+\frac{352 \logx^3}{9}+\frac{3688 \logx^2}{9}+\left(\frac{14560}{9}+\frac{128 \pi ^2}{3}\right)
 \logx-\frac{1744 \zthree}{3}-\frac{196 \pi ^4}{135}+\frac{1264 \pi ^2}{27}
\nonumber\\&&\mbox{}
+\frac{99344}{81}
\Bigg]
+\cR \tf^2 \nl^2
\Bigg[
-\frac{64 \logx^3}{9}-\frac{544 \logx^2}{9}+\left(-\frac{1760}{9}-\frac{64 \pi ^2}{9}\right)
 \logx
+\frac{256 \zthree}{3}
\nonumber\\&&\mbox{}
+\frac{32 \pi ^2}{3}-\frac{12992}{81}
\Bigg]
+\cR \tf^2 \nh \nl
\Bigg[
-\frac{128 \logx^3}{9}-\frac{1088 \logx^2}{9}
\nonumber\\&&\mbox{}
+\left(-\frac{3520}{9}-\frac{128 \pi ^2}{9}\right)
 \logx-\frac{256 \zthree}{3}-\frac{640 \pi ^2}{9}-\frac{6976}{81}
\Bigg]
  \,,
\end{eqnarray}
\begin{eqnarray}
  f_{\rm lar}^{s,(1,0)} &=&
\cR
\Bigg[
-\frac{2 \logx}{\ep}-\frac{2}{\ep}-\logx^2+\frac{\pi ^2}{3}+2
\Bigg]
  \,,\nonumber\\
  f_{\rm lar}^{s,(1,1)} &=&
\cR
\Bigg[
12 \logx-4
\Bigg]
  \,,\nonumber\\
  f_{\rm lar}^{s,(2,0)} &=&
\cR^2
\Bigg[
\left(\frac{2}{\ep^2}+\frac{2}{\ep}-\frac{2 \pi ^2}{3}+2\right)
 \logx^2+\logx \left(\frac{4}{\ep^2}+\frac{-4-\frac{2 \pi ^2}{3}}{\ep}-32 \zthree+\frac{3 \pi ^2}{2}-4\right)
\nonumber\\&&\mbox{}
+\frac{2}{\ep^2}+\left(\frac{2}{\ep}+\frac{2}{3}\right)
 \logx^3+\frac{-4-\frac{2 \pi ^2}{3}}{\ep}-8 \pi ^2 \logtwo+\frac{7
   \logx^4}{6}-44 \zthree-\frac{59 \pi ^4}{90}+\frac{41 \pi ^2}{6}
\nonumber\\&&\mbox{}
+\frac{161}{8}
\Bigg]
+\cA \cR
\Bigg[
\logx \left(\frac{11}{3 \ep^2}+\frac{\frac{\pi ^2}{3}-\frac{67}{9}}{\ep}+26 \zthree-\frac{11 \pi ^2}{9}-\frac{242}{27}\right)
+\frac{11}{3 \ep^2}
\nonumber\\&&\mbox{}
+\frac{-2 \zthree+\frac{\pi ^2}{3}-\frac{49}{9}}{\ep}+4 \pi ^2 \logtwo-\frac{11 \logx^3}{9}+\left(\frac{\pi ^2}{3}-\frac{67}{9}\right)
 \logx^2+\frac{80 \zthree}{3}-\frac{\pi ^4}{60}
\nonumber\\&&\mbox{}
+\frac{55 \pi ^2}{27}+\frac{3047}{216}
\Bigg]
+\cR \tf \nl
\Bigg[
\left(-\frac{4}{3 \ep^2}+\frac{20}{9 \ep}+\frac{4 \pi ^2}{9}+\frac{112}{27}\right)
 \logx-\frac{4}{3 \ep^2}+\frac{20}{9 \ep}
\nonumber\\&&\mbox{}
+\frac{4 \logx^3}{9}+\frac{20 \logx^2}{9}-\frac{16 \zthree}{3}-\frac{32 \pi ^2}{27}-\frac{247}{54}
\Bigg]
+\cR \tf \nh
\Bigg[
\frac{4 \logx^3}{9}+\frac{20 \logx^2}{9}
\nonumber\\&&\mbox{}
+\left(\frac{224}{27}+\frac{2 \pi ^2}{3}\right)
 \logx-\frac{10 \pi ^2}{9}+\frac{1969}{54}
\Bigg]
  \,,\nonumber\\
  f_{\rm lar}^{s,(2,1)} &=&
\cR^2
\Bigg[
\left(-\frac{24}{\ep}+\frac{4 \pi ^2}{3}+4\right)
 \logx^2+\logx \left(-\frac{16}{\ep}-64 \zthree+\frac{32 \pi ^2}{3}-24\right)
\nonumber\\&&\mbox{}
+\frac{8}{\ep}+48 \pi ^2 \logtwo-24 \logx^3-120 \zthree+\frac{34 \pi ^4}{45}-\frac{56 \pi ^2}{3}+16
\Bigg]
+\cA \cR
\Bigg[
-24 \pi ^2 \logtwo
\nonumber\\&&\mbox{}
+12 \logx^2+\frac{416 \logx}{3}-44 \zthree+\frac{46 \pi ^2}{3}-\frac{580}{9}
\Bigg]
+\cR \tf \nl
\Bigg[
-8 \logx^2-\frac{112 \logx}{3}
\nonumber\\&&\mbox{}
+\frac{8 \pi ^2}{3}+\frac{128}{9}
\Bigg]
+\cR \tf \nh
\Bigg[
-8 \logx^2-\frac{64 \logx}{3}-8 \pi ^2-\frac{160}{9}
\Bigg]
  \,,\nonumber\\
  f_{\rm lar}^{s,(3,0)} &=&
\nc^3
\Bigg[
-\frac{\logx^3}{6 \ep^3}-\frac{7 \logx^2}{3 \ep^3}-\frac{467 \logx}{54
  \ep^3}-\frac{175}{27 \ep^3}-\frac{\logx^4}{4 \ep^2}-\frac{17 \logx^3}{12
  \ep^2}+\frac{55 \logx^2}{18 \ep^2}-\frac{\pi ^2 \logx^2}{12
  \ep^2}+\frac{\logx \zthree}{\ep^2} 
\nonumber\\&&\mbox{}
+\frac{3589 \logx}{162 \ep^2}-\frac{29 \pi ^2 \logx}{108 \ep^2}+\frac{31
  \zthree}{9 \ep^2}+\frac{2509}{162 \ep^2}-\frac{5 \pi ^2}{27 \ep^2}-\frac{5
  \logx^5}{24 \ep}+\frac{\logx^4}{72 \ep}+\frac{209 \logx^3}{36 \ep}-\frac{\pi
  ^2  \logx^3}{8 \ep}
\nonumber\\&&\mbox{}
-\frac{11 \logx^2 \zthree}{2 \ep}+\frac{889 \logx^2}{108
  \ep}
+\frac{29 \pi ^2 \logx^2}{144 \ep}-\frac{73 \logx \zthree}{9
  \ep}-\frac{20393 \logx}{864 \ep}+\frac{16 \pi ^4 \logx}{135 \ep}-\frac{1483
  \pi ^2 \logx}{1296 \ep}
\nonumber\\&&\mbox{}
-\frac{388 \zthree}{27 \ep}-\frac{7 \pi ^2
  \zthree}{18 \ep}+\frac{6 \zeta (5)}{\ep}-\frac{15977}{864 \ep}+\frac{16 \pi
  ^4}{135 \ep}
-\frac{701 \pi ^2}{648 \ep}-\frac{\logx^6}{8}+\frac{9
  \logx^5}{16}
\nonumber\\&&\mbox{}
-\frac{\pi ^2 \logx^4}{9}+\frac{809 \logx^4}{216}-\frac{49
  \logx^3 \zthree}{6}+\frac{7 \pi ^2 \logx^3}{18}-\frac{301
  \logx^3}{162}+\frac{361 \logx^2 \zthree}{36}-\frac{\pi ^4
  \logx^2}{9}
\nonumber\\&&\mbox{}
-\frac{145 \pi ^2 \logx^2}{432}-\frac{16883
  \logx^2}{576}-\frac{13}{36} \pi ^2 \logx \zthree+\frac{9575 \logx
  \zthree}{108}+15 \logx \zeta (5)
-\frac{11 \pi ^4 \logx}{36}
\nonumber\\&&\mbox{}
-\frac{81859 \pi
  ^2 \logx}{7776}-\frac{2738633 \logx}{46656}-\frac{16 \zthree^2}{3}+\frac{353
  \pi ^2 \zthree}{72}+\frac{7489 \zthree}{324}
\nonumber\\&&\mbox{}
-\frac{785 \zeta
  (5)}{3}+\frac{2039 \pi ^6}{17010}-\frac{4519 \pi ^4}{2160}+\frac{231895 \pi
  ^2}{5184}+\frac{11170537}{46656}  
\Bigg]
\nonumber\\&&\mbox{}
+\cR^2 \tf \nl
\Bigg[
-\frac{512 \afour}{3}+\frac{8}{3 \ep^3}+\left(\frac{4}{3 \ep^2}-\frac{64}{9 \ep}-\frac{29 \pi ^2}{27}-\frac{412}{27}\right)
 \logx^3+\frac{-\frac{88}{9}-\frac{4 \pi ^2}{9}}{\ep^2}
\nonumber\\&&\mbox{}
+\logx^2 \left(\frac{8}{3 \ep^3}-\frac{28}{9 \ep^2}+\frac{-\frac{332}{27}-\frac{10 \pi ^2}{9}}{\ep}+\frac{232 \zthree}{9}-\frac{17 \pi ^2}{9}-\frac{1913}{162}\right)
+\logx \left(\frac{16}{3 \ep^3}
\right.\nonumber\\&&\left.\mbox{}
+\frac{-\frac{128}{9}-\frac{4 \pi ^2}{9}}{\ep^2}+\frac{-\frac{16 \zthree}{3}+\frac{17 \pi ^2}{9}+\frac{401}{27}}{\ep}+\frac{2204 \zthree}{9}+\frac{98 \pi ^4}{135}-\frac{68 \pi ^2}{27}+\frac{14969}{162}\right)
\nonumber\\&&\mbox{}
+\left(-\frac{4}{9 \ep}-\frac{175}{27}\right)
 \logx^4+\frac{-\frac{16 \zthree}{3}+3 \pi ^2+\frac{553}{27}}{\ep}-\frac{64
   \logtwo^4}{9}-\frac{128 \pi ^2 \logtwo^2}{9}+\frac{224 \pi ^2
   \logtwo}{9}-\logx^5
\nonumber\\&&\mbox{}
-8 \pi ^2 \zthree+\frac{2116 \zthree}{9}+40 \zeta (5)+\frac{3238 \pi ^4}{405}-\frac{2323 \pi ^2}{54}+\frac{673}{162}
\Bigg]
\nonumber\\&&\mbox{}
+\cA \cR \tf \nl
\Bigg[
\frac{256 \afour}{3}+\frac{176}{27 \ep^3}
+\frac{-\frac{16 \zthree}{9}+\frac{8 \pi ^2}{27}-\frac{1192}{81}}{\ep^2}+\logx
\left(\frac{176}{27 \ep^3}+\frac{\frac{8 \pi ^2}{27}-\frac{1336}{81}}{\ep^2}
\right.\nonumber\\&&\left.\mbox{}
+\frac{\frac{112 \zthree}{9}-\frac{80 \pi ^2}{81}+\frac{836}{81}}{\ep}-\frac{1448 \zthree}{9}-\frac{22 \pi ^4}{135}+\frac{3272 \pi ^2}{243}+\frac{14998}{729}\right)
\nonumber\\&&\mbox{}
+\frac{\frac{496 \zthree}{27}-\frac{80 \pi ^2}{81}+\frac{356}{81}}{\ep}+\frac{32 \logtwo^4}{9}+\frac{64 \pi ^2 \logtwo^2}{9}-\frac{112 \pi ^2 \logtwo}{9}+\frac{44 \logx^4}{27}+\left(\frac{1156}{81}-\frac{8 \pi ^2}{27}\right)
 \logx^3
\nonumber\\&&\mbox{}
+\logx^2 \left(-16 \zthree+\frac{16 \pi ^2}{9}+\frac{3454}{81}\right)
+\frac{4 \pi ^2 \zthree}{9}-\frac{22552 \zthree}{81}+\frac{596 \zeta (5)}{3}
\nonumber\\&&\mbox{}
-\frac{1903 \pi ^4}{405}-\frac{7640 \pi ^2}{243}-\frac{176663}{729}
\Bigg]
+\cR \tf^2 \nl^2
\Bigg[
-\frac{32}{27 \ep^3}
+\frac{160}{81 \ep^2}
\nonumber\\&&\mbox{}
+\logx \left(-\frac{32}{27 \ep^3}+\frac{160}{81 \ep^2}+\frac{32}{81 \ep}-\frac{64 \zthree}{27}-\frac{160 \pi ^2}{81}-\frac{3712}{729}\right)
+\frac{32}{81 \ep}-\frac{8 \logx^4}{27}-\frac{160 \logx^3}{81}
\nonumber\\&&\mbox{}
+\left(-\frac{400}{81}-\frac{16 \pi ^2}{27}\right)
 \logx^2+32 \zthree+\frac{232 \pi ^4}{405}+\frac{736 \pi ^2}{243}+\frac{14006}{729}
\Bigg]
\nonumber\\&&\mbox{}
+\cR \tf^2 \nh \nl
\Bigg[
\logx \left(\frac{8 \pi ^2}{27 \ep}-\frac{416 \zthree}{27}-\frac{40 \pi ^2}{9}+\frac{512}{81}\right)
+\frac{8 \pi ^2}{27 \ep}-\frac{16 \logx^4}{27}-\frac{320 \logx^3}{81}
\nonumber\\&&\mbox{}
+\left(-\frac{800}{81}-\frac{32 \pi ^2}{27}\right)
 \logx^2-\frac{352 \zthree}{9}-\frac{16 \pi ^4}{27}+\frac{3224 \pi ^2}{243}-\frac{16748}{243}
\Bigg]
  \,,\nonumber\\
  f_{\rm lar}^{s,(3,1)} &=&
\nc^3
\Bigg[
\frac{3 \logx^3}{\ep^2}+\frac{16 \logx^2}{\ep^2}+\frac{25 \logx}{3
  \ep^2}-\frac{14}{3 \ep^2}+\frac{9 \logx^4}{2 \ep}-\frac{3 \logx^3}{2
  \ep}-\frac{\pi ^2 \logx^3}{3 \ep}+\frac{16 \logx^2 \zthree}{\ep}-\frac{248
  \logx^2}{3 \ep}
\nonumber\\&&\mbox{}
-\frac{3 \pi ^2 \logx^2}{2 \ep}+\frac{62 \logx
  \zthree}{\ep}-\frac{113 \logx}{3 \ep}-\frac{17 \pi ^4 \logx}{90
  \ep}-\frac{35 \pi ^2 \logx}{6 \ep}+\frac{54
  \zthree}{\ep}
+\frac{31}{\ep}
-\frac{17 \pi ^4}{90 \ep}
\nonumber\\&&\mbox{}
-\frac{14 \pi ^2}{3
  \ep}+\frac{15 \logx^5}{4}-\frac{\pi ^2 \logx^4}{2}-\frac{203
  \logx^4}{12}+\frac{40 \logx^3 \zthree}{3}
+\frac{85 \pi ^2
  \logx^3}{36}-\frac{1043 \logx^3}{9}
\nonumber\\&&\mbox{}
-\frac{239 \logx^2 \zthree}{3}+\frac{17
  \pi ^4 \logx^2}{30}+\frac{248 \pi ^2 \logx^2}{27}+\frac{1070
  \logx^2}{9}-\frac{2}{3} \pi ^2 \logx \zthree-\frac{6428 \logx \zthree}{9}
\nonumber\\&&\mbox{}
-48\logx \zeta (5)+\frac{631 \pi ^4 \logx}{270}+\frac{6263 \pi ^2
  \logx}{54}+\frac{95897 \logx}{144}+74 \zthree^2-\frac{89 \pi ^2
  \zthree}{3}
\nonumber\\&&\mbox{}
-\frac{31657 \zthree}{18}+\frac{592 \zeta (5)}{3}-\frac{275 \pi
  ^6}{2268}
+\frac{24373 \pi ^4}{3240}+\frac{4975 \pi
  ^2}{108}+\frac{22981}{1296} 
\Bigg]
\nonumber\\&&\mbox{}
+\cR^2 \tf \nl
\Bigg[
1024 \afour+\logx^2 \left(-\frac{16}{\ep^2}+\frac{280}{3 \ep}+\frac{160 \zthree}{3}+\frac{308 \pi ^2}{27}+\frac{1768}{9}\right)
\nonumber\\&&\mbox{}
+\logx \left(-\frac{32}{3 \ep^2}+\frac{\frac{160}{3}-\frac{8 \pi ^2}{3}}{\ep}+\frac{1504 \zthree}{9}+\frac{272 \pi ^4}{135}-\frac{2120 \pi ^2}{27}+\frac{278}{9}\right)
\nonumber\\&&\mbox{}
+\frac{16}{3 \ep^2}+\left(\frac{8}{\ep}-\frac{16 \pi ^2}{9}+120\right)
 \logx^3+\frac{-32-\frac{8 \pi ^2}{3}}{\ep}+\frac{128 \logtwo^4}{3}+\frac{256
   \pi ^2 \logtwo^2}{3}-\frac{1408 \pi ^2 \logtwo}{3}
\nonumber\\&&\mbox{}
+\frac{80 \logx^4}{3}+1040 \zthree-\frac{1664 \zeta (5)}{3}-\frac{5368 \pi ^4}{405}+\frac{6404 \pi ^2}{27}-\frac{742}{3}
\Bigg]
\nonumber\\&&\mbox{}
+\cA \cR \tf \nl
\Bigg[
-512 \afour-\frac{64 \logtwo^4}{3}-\frac{128 \pi ^2 \logtwo^2}{3}+\frac{704
  \pi ^2 \logtwo}{3}-24 \logx^3
\nonumber\\&&\mbox{}
+\left(-\frac{896}{3}-\frac{38 \pi ^2}{9}\right)
 \logx^2+\logx \left(\frac{32 \zthree}{3}-\frac{352 \pi ^2}{9}-\frac{10400}{9}\right)
+\frac{5624 \zthree}{9}+\frac{373 \pi ^4}{135}
\nonumber\\&&\mbox{}
-\frac{956 \pi ^2}{27}+\frac{38632}{81}
\Bigg]
+\cR \tf^2 \nl^2
\Bigg[
\frac{64 \logx^3}{9}+\frac{448 \logx^2}{9}+\left(\frac{1216}{9}+\frac{64 \pi ^2}{9}\right)
 \logx
\nonumber\\&&\mbox{}
-\frac{256 \zthree}{3}-\frac{64 \pi ^2}{3}-\frac{4672}{81}
\Bigg]
+\cR \tf^2 \nh \nl
\Bigg[
\frac{128 \logx^3}{9}+\frac{704 \logx^2}{9}
\nonumber\\&&\mbox{}
+\left(\frac{2176}{9}+\frac{128 \pi ^2}{9}\right)
 \logx
+\frac{256 \zthree}{3}+\frac{1856 \pi ^2}{27}-\frac{11648}{81}
\Bigg]
  \,,
\end{eqnarray}
\begin{eqnarray}
  f_{\rm lar}^{p,(1,0)} &=& f_{\rm lar}^{s,(1,0)}
  \,,\nonumber\\
  f_{\rm lar}^{p,(1,1)} &=&
\cR
\Bigg[
4 \logx-4
\Bigg]
  \,,\nonumber\\
  f_{\rm lar}^{p,(2,0)} &=& f_{\rm lar}^{s,(2,0)}
  \,,\nonumber\\
  f_{\rm lar}^{p,(2,1)} &=&
\cR^2
\Bigg[
\left(-\frac{8}{\ep}+\frac{4 \pi ^2}{3}+12\right)
 \logx^2+\frac{8}{\ep}+16 \pi ^2 \logtwo-8 \logx^3+\logx \left(-64 \zthree+\frac{8 \pi ^2}{3}-8\right)
\nonumber\\&&\mbox{}
-40 \zthree+\frac{34 \pi ^4}{45}+16
\Bigg]
+\cA \cR
\Bigg[
-8 \pi ^2 \logtwo+\frac{16 \logx^2}{3}+\frac{520 \logx}{9}-4 \zthree
\nonumber\\&&\mbox{}
+\frac{26 \pi ^2}{9}-\frac{580}{9}
\Bigg]
+\cR \tf \nl
\Bigg[
-\frac{8 \logx^2}{3}-\frac{80 \logx}{9}+\frac{8 \pi ^2}{9}+\frac{128}{9}
\Bigg]
\nonumber\\&&\mbox{}
+\cR \tf \nh
\Bigg[
-\frac{8 \logx^2}{3}+\frac{64 \logx}{9}-\frac{8 \pi ^2}{3}+\frac{32}{9}
\Bigg]
  \,,\nonumber\\
  f_{\rm lar}^{p,(3,0)} &=& f_{\rm lar}^{s,(3,0)}
  \,,\nonumber\\
  f_{\rm lar}^{p,(3,1)} &=&
\nc^3
\Bigg[
\frac{\logx^3}{\ep^2}+\frac{14 \logx^2}{3
  \ep^2}-\frac{\logx}{\ep^2}-\frac{14}{3 \ep^2}+\frac{3 \logx^4}{2
  \ep}-\frac{11 \logx^3}{6 \ep}-\frac{\pi ^2 \logx^3}{3 \ep}+\frac{16 \logx^2
  \zthree}{\ep}-\frac{118 \logx^2}{3 \ep}-\frac{\pi ^2 \logx^2}{2
  \ep}
\nonumber\\&&\mbox{} 
+\frac{26 \logx \zthree}{\ep}-\frac{\logx}{\ep}-\frac{17 \pi ^4
  \logx}{90 \ep}-\frac{43 \pi ^2 \logx}{18 \ep}+\frac{14
  \zthree}{\ep}+\frac{31}{\ep}-\frac{17 \pi ^4}{90 \ep}-\frac{20 \pi ^2}{9
  \ep}
\nonumber\\&&\mbox{} 
+\frac{5 \logx^5}{4}-\frac{\pi ^2 \logx^4}{2}-\frac{253
  \logx^4}{36}+\frac{40 \logx^3 \zthree}{3}+\frac{127 \pi ^2
  \logx^3}{36}-\frac{1459 \logx^3}{27}-\frac{425 \logx^2 \zthree}{3}
\nonumber\\&&\mbox{} 
+\frac{17
  \pi ^4 \logx^2}{30}+\frac{373 \pi ^2 \logx^2}{54}+\frac{7217
  \logx^2}{54}-\frac{2}{3} \pi ^2 \logx \zthree-\frac{4190 \logx
  \zthree}{9}
-48 \logx \zeta (5)
\nonumber\\&&\mbox{} 
+\frac{101 \pi ^4 \logx}{54}
+\frac{1175 \pi ^2
  \logx}{27}+\frac{186587 \logx}{1296}+74 \zthree^2-\frac{67 \pi ^2
  \zthree}{3}-\frac{11815 \zthree}{18}
\nonumber\\&&\mbox{} 
+\frac{124 \zeta (5)}{3}
-\frac{275 \pi
  ^6}{2268}+\frac{4109 \pi ^4}{648}+\frac{9527 \pi ^2}{324}+\frac{17797}{1296} 
\Bigg]
\nonumber\\&&\mbox{}
+\cR^2 \tf \nl
\Bigg[
\frac{1024 \afour}{3}+\logx^2 \left(-\frac{16}{3 \ep^2}+\frac{104}{3 \ep}+\frac{160 \zthree}{3}+\frac{140 \pi ^2}{27}+\frac{664}{9}\right)
\nonumber\\&&\mbox{}
+\frac{16}{3 \ep^2}+\left(\frac{8}{3 \ep}-\frac{16 \pi ^2}{9}+\frac{280}{9}\right)
 \logx^3+\logx \left(-\frac{8 \pi ^2}{9 \ep}+\frac{1120 \zthree}{9}+\frac{272
     \pi ^4}{135}-\frac{400 \pi ^2}{9}
\right.\nonumber\\&&\left.\mbox{}
-\frac{110}{9}\right)
+\frac{-32-\frac{8 \pi ^2}{9}}{\ep}+\frac{128 \logtwo^4}{9}+\frac{256 \pi ^2
  \logtwo^2}{9}-\frac{1408 \pi ^2 \logtwo}{9}+\frac{80 \logx^4}{9}+\frac{3184
  \zthree}{9}
\nonumber\\&&\mbox{}
-\frac{1664 \zeta (5)}{3}-\frac{2296 \pi ^4}{405}+\frac{1508 \pi ^2}{27}-\frac{742}{3}
\Bigg]
+\cA \cR \tf \nl
\Bigg[
-\frac{512 \afour}{3}-\frac{64 \logtwo^4}{9}
\nonumber\\&&\mbox{} 
-\frac{128 \pi ^2 \logtwo^2}{9}+\frac{704 \pi ^2 \logtwo}{9}-\frac{232 \logx^3}{27}+\left(-\frac{3800}{27}-2 \pi ^2\right)
 \logx^2+\logx \left(-\frac{32 \zthree}{3}-\frac{64 \pi ^2}{27}
\right.\nonumber\\&&\left.\mbox{}
-\frac{27136}{81}\right)
+\frac{2024 \zthree}{9}+\frac{103 \pi ^4}{135}+\frac{308 \pi ^2}{81}+\frac{38632}{81}
\Bigg]
+\cR \tf^2 \nl^2
\Bigg[
\frac{64 \logx^3}{27}
\nonumber\\&&\mbox{}
+\frac{320 \logx^2}{27}+\left(\frac{1600}{81}+\frac{64 \pi ^2}{27}\right)
 \logx-\frac{256 \zthree}{9}-\frac{704 \pi ^2}{81}-\frac{4672}{81}
\Bigg]
\nonumber\\&&\mbox{}
+\cR \tf^2 \nh \nl
\Bigg[
\frac{128 \logx^3}{27}+\frac{64 \logx^2}{27}+\left(\frac{896}{81}+\frac{128 \pi ^2}{27}\right)
 \logx+\frac{256 \zthree}{9}+\frac{1216 \pi ^2}{81}
\nonumber\\&&\mbox{}
-\frac{5504}{81}
\Bigg]
  \,.
\end{eqnarray}
Results for $f_{i,\rm lar}^{v,(n,k)}$ can be found in Eq.~(13) of
Ref.~\cite{Lee:2018nxa} and Eq.~(16) of Ref.~\cite{Henn:2016kjz}.

In general one has two powers of $\logx$ for each loop-order.
At one and two loops we indeed observe $\logx^2$ and $\logx^4$ terms,
respectively. However, for the shown colour structures
we have at three-loop order at most $\logx^5$ terms in the above expressions.
Note that for the vector form factor one has $\logx^6$ terms in
the $N_c^3$ term of $f_{1,\rm lar}^{v,(3,0)}$~\cite{Sudakov:1954sw,Frenkel:1976bj}.
For a dedicated analysis of the leading 
logarithmically enhanced terms in power-suppressed
contributions we refer to~\cite{Penin:2014msa,Liu:2017axv,Liu:2017vkm}.


\subsection{Threshold expansion}

In the threshold region we can use our expressions for the form factors to
obtain physical results for decay rates and productions cross sections since
the corresponding real radiation is suppressed by a relative order
$\beta^3$. In the expressions we present in this subsection, the factor $m$ in
the definition of the scalar and pseudo-scalar currents, see
Eq.~(\ref{eq::currents}), has been transformed from the $\overline{\rm MS}$ to
the on-shell scheme which allows for a more straightforward comparison with
results present in the literature.

In Refs.~\cite{Henn:2016tyf,Lee:2018nxa} the vector form factors $F_1^v$ and
$F_2^v$ have been used to obtain results for the cross section
$\sigma(e^+e^-\to Q\bar{Q})$ in the limit of small quark velocities.
In principle these results can be extended in order to incorporate
the $Z$-boson contribution with vector and axial-vector couplings.
We prefer to represent the results in a slightly different way, namely
as the decay rate of a (hypotetical) boson with
either vector, axial-vector, scalar or pseudo-scalar couplings
which is related to the form factors via
\begin{eqnarray}
R^v &=& \beta \left( |F_1^v+F_2^v|^2 + \frac{|(1-\beta^2)F_1+F_2|^2}{2(1-\beta^2)}\right)
\,,\nonumber\\
R^a &=& \beta^3 |F_1^a|^2
\,,\nonumber\\
R^s &=& \beta^3 |F^s|^2
\,,\nonumber\\
R^p &=& \beta |F^p|^2
\label{eq::R_F}
\,,
\end{eqnarray}
$R^\delta$ are defined such that the exact result at tree level reads
\begin{eqnarray}
  R^{(0),v} &=& \frac{3\beta}{2}\left(1-\frac{\beta^2}{3}\right)\,,\nonumber\\
  R^{(0),a} &=& \beta^3\,,\nonumber\\
  R^{(0),s} &=& \beta^3\,,\nonumber\\
  R^{(0),p} &=& \beta\,.
\end{eqnarray}
Note that the $R^\delta$ enter physical quantities as building blocks.
For example, we have
\begin{eqnarray}
  \sigma(e^+e^-\to Q \bar{Q}) &=& \sigma_0 R^v + \ldots \,,\nonumber\\
  \Gamma(H\to Q\bar{Q}) &=& \frac{3 G_F M_H M_Q^2}{4\sqrt{2}\pi} R^s +\ldots \,,\nonumber\\
  \Gamma(A\to Q\bar{Q}) &=& \frac{3 G_F M_A M_Q^2}{4\sqrt{2}\pi} R^p + \ldots \,,
  \label{eq::R_phys}
\end{eqnarray}
where $H$ and $A$ are scalar and pseudo-scalar Higgs bosons with masses $M_H$
and $M_A$, respectively,
and the ellipses indicate quantum corrections from real radiation.  In
Eq.~(\ref{eq::R_phys}) $G_F$ is Fermi's constant, $M_Q$ is the heavy quark
on-shell mass, $\sigma_0 = 4\pi \alpha^2 Q_Q^2 /(3s)$, $\alpha$ is the fine
structure constant and $Q_Q$ is the electric charge of the quark $Q$.
The decay rates in~(\ref{eq::R_phys}) can be obtained from an effective
Lagrangian of the form
\begin{eqnarray}
  {\cal L}_{\rm eff} &=& -\frac{1}{v}\left( j^s H + j^p A \right)
  \,,
\end{eqnarray}
where $v=1/\sqrt{\sqrt{2}G_F}$ is the vacuum expectation value.

We cast $R^\delta$ in the form
\begin{eqnarray}
  R^\delta &=& R^{(0),\delta} 
  + K_\delta \beta^{n_\delta}
  \sum_{i\ge1} \left(\frac{\alpha_s(M)}{4\pi}\right)^i \Delta^{(i),\delta}
  \,,
  \label{eq::Rdelta}
\end{eqnarray}
with $K_v=3/2$, $K_a=K_s=K_p=1$, $n_v=n_p=1$ and $n_a=n_s=3$. Then the threshold expansion of 
$\Delta^{(i),\delta}$ starts at $1/\beta^i$ and the leading term of the
real radiation contribution to $\Delta^{(i),\delta}$ is of order $\beta^{3-i}$.
For the three-loop fermionic results we can provide results including $\beta^0$.
In the following we only show results for $\delta=a,s$ and $p$ since
the expressions for the vector current can be found in Refs.~\cite{Henn:2016tyf,Lee:2018nxa}.
The one- and two-loop results for $\Delta^{(i),\delta}$ are given by
\begin{eqnarray}
\Delta^{(1),a} &=&
\cR
\Bigg[
 \frac{2 \pi ^2}{\beta} 
-8
+ 2 \pi ^2\beta 
\Bigg]
\,,\nonumber\\
\Delta^{(1),s} &=&
\cR
\Bigg[
 \frac{2 \pi ^2}{\beta} 
-4
+ 2 \pi ^2\beta 
\Bigg]
\,,\nonumber\\
\Delta^{(1),p} &=&
\cR
\Bigg[
 \frac{2 \pi ^2}{\beta}
-12
+ 2 \pi ^2\beta 
\Bigg]
\,,
\end{eqnarray}
\begin{eqnarray}
\Delta^{(2),a} &=&
\cR^2
\Bigg[
 \frac{1}{\beta^2} \Bigg(4 \pi ^2+\frac{4 \pi ^4}{3}\Bigg)
- \frac{16\pi^2}{\beta}
+ \Bigg(-\frac{40}{3} \pi ^2 \log (2 \beta )-54 \zeta (3)+\frac{8 \pi
  ^4}{3}+\frac{4 \pi ^2}{3}
\nonumber\\&&\mbox{}
+\frac{140}{3}+\frac{76}{3} \pi ^2 { l_2}\Bigg)
\Bigg]
+\cA \cR
\Bigg[
 \frac{1}{\beta} \Bigg(\frac{194 \pi ^2}{9}-\frac{44}{3} \pi ^2 \log (2 \beta )\Bigg)
\nonumber\\&&\mbox{}
+ \Bigg(-\frac{16}{3} \pi ^2 \log (2 \beta )-36 \zeta (3)+\frac{178 \pi ^2}{9}-\frac{404}{9}-\frac{56}{3} \pi ^2 { l_2}\Bigg)
\Bigg]
\nonumber\\&&\mbox{}
+\cR \tf \nl
\Bigg[
 \frac{1}{\beta} \Bigg(\frac{16}{3} \pi ^2 \log (2 \beta )-\frac{88 \pi ^2}{9}\Bigg)
+ \frac{112}{9}
\Bigg]
+\cR \tf \nh
 \Bigg(\frac{640}{9}-\frac{64 \pi ^2}{9}\Bigg)
\,,\nonumber\\
\Delta^{(2),s} &=&
\cR^2
\Bigg[
 \frac{1}{\beta^2} \Bigg(4 \pi ^2+\frac{4 \pi ^4}{3}\Bigg)
- \frac{8 \pi ^2}{\beta}
+ \Bigg(-\frac{64}{3} \pi ^2 \log (2 \beta )-88 \zeta (3)+\frac{8 \pi
  ^4}{3}+\frac{62 \pi ^2}{3}
\nonumber\\&&\mbox{}
+14+16 \pi ^2 { l_2}\Bigg)
\Bigg]
+ \cA \cR
\Bigg[
 \frac{1}{\beta} \Bigg(\frac{194 \pi ^2}{9}-\frac{44}{3} \pi ^2 \log (2 \beta )\Bigg)
\nonumber\\&&\mbox{}
+ \Bigg(-\frac{16}{3} \pi ^2 \log (2 \beta )-40 \zeta (3)+\frac{38 \pi ^2}{3}+\frac{98}{9}-16 \pi ^2 { l_2}\Bigg)
\Bigg]
\nonumber\\&&\mbox{}
+\cR \tf \nl
\Bigg[
 \frac{1}{\beta} \Bigg(\frac{16}{3} \pi ^2 \log (2 \beta )-\frac{88 \pi ^2}{9}\Bigg)
-\frac{40}{9}
\Bigg]
+
\cR \tf \nh
 \Bigg(\frac{968}{9}-\frac{32 \pi ^2}{3}\Bigg)
\,,\nonumber\\
\Delta^{(2),p} &=&
\cR^2
\Bigg[
 \frac{1}{\beta^2} \Bigg(\frac{4 \pi ^4}{3}\Bigg)
- \frac{24 \pi ^2}{\beta}
+ \Bigg(-32 \pi ^2 \log (2 \beta )-144 \zeta (3)+\frac{8 \pi ^4}{3}-\frac{14
  \pi ^2}{3}
\nonumber\\&&\mbox{}
+94+32 \pi ^2 { l_2}\Bigg)
\Bigg]
+\cA \cR
\Bigg[
 \frac{1}{\beta} \Bigg(\frac{62 \pi ^2}{9}-\frac{44}{3} \pi ^2 \log (2 \beta )\Bigg)
\nonumber\\&&\mbox{}
+ \Bigg(-16 \pi ^2 \log (2 \beta )-96 \zeta (3)+\frac{94 \pi ^2}{3}-\frac{34}{3}-32 \pi ^2 { l_2}\Bigg)
\Bigg]
\nonumber\\&&\mbox{}
+\cR \tf \nl
\Bigg[
 \frac{1}{\beta} \Bigg(\frac{16}{3} \pi ^2 \log (2 \beta )-\frac{40 \pi ^2}{9}\Bigg)
+ \frac{8}{3}
\Bigg]
+\cR \tf \nh
 \Bigg(\frac{344}{3}-\frac{32 \pi ^2}{3}\Bigg)
\,.
\end{eqnarray}
For the $\cR^2/\beta^2$ and the $\{\cA\cR,\cR\tf\nl\}/\beta$ terms of $\Delta^{(2),\delta}$ we find
agreement with Ref.~\cite{Chetyrkin:1997mb}.
At three-loop order we have
\begin{eqnarray}
\Delta^{(3),a} &=&
\nc^3
\Bigg[
 \frac{\pi ^4}{\beta^3} 
+ \frac{1}{\beta^2} \Bigg(-\frac{44}{9} \pi ^4 \log (2 \beta )-\frac{44}{3}
\pi ^2 \log (2 \beta )-\frac{88 \pi ^2 \zeta (3)}{3}+\frac{158 \pi ^4}{27}
\nonumber\\&&\mbox{}
+\frac{356 \pi ^2}{9}\Bigg)
+ \frac{1}{\beta} \Bigg(\frac{484}{9} \pi ^2 \log ^2(2 \beta )-6 \pi ^4 \log
(2 \beta )-\frac{4088}{27} \pi ^2 \log (2 \beta )-\frac{145 \pi ^2 \zeta
  (3)}{6}
\nonumber\\&&\mbox{}
-\frac{\pi ^6}{4}+\frac{653 \pi ^4}{27}+\frac{16181 \pi ^2}{162}-3 \pi ^4 { l_2}\Bigg)
\Bigg]
+\cR^2 \tf \nl
\Bigg[
 \frac{1}{\beta^2} \Bigg(\frac{64}{9} \pi ^4 \log (2 \beta )
\nonumber\\&&\mbox{}
+\frac{64}{3} \pi ^2 \log (2 \beta )+\frac{128 \pi ^2 \zeta (3)}{3}-\frac{352 \pi ^4}{27}-\frac{640 \pi ^2}{9}\Bigg)
+ \frac{1}{\beta} \Bigg(-\frac{80}{3} \pi ^2 \log (2 \beta )
\nonumber\\&&\mbox{}
+32 \pi ^2 \zeta (3)+\frac{454 \pi ^2}{9}\Bigg)
+ \Bigg(\frac{8000 \afour}{9}
-\frac{320}{9} \pi ^2 \log ^2(2 \beta )+\frac{128}{9} \pi ^4 \log (2 \beta
)
\nonumber\\&&\mbox{}
+\frac{6704}{27} \pi ^2 \log (2 \beta )
+\frac{256 \pi ^2 \zeta (3)}{3}+\frac{10780 \zeta (3)}{9}-\frac{9683 \pi
  ^4}{405}-\frac{12100 \pi ^2}{81}+\frac{1912}{27}
\nonumber\\&&\mbox{}
+\frac{1000 { l_2^4}}{27}
-\frac{112}{27} \pi ^2 { l_2^2}-\frac{2296}{27} \pi ^2 { l_2}\Bigg)
\Bigg]
\nonumber\\&&\mbox{}
+\cA \cR \tf \nl
\Bigg[
 \frac{1}{\beta} \Bigg(-\frac{704}{9} \pi ^2 \log ^2(2 \beta )
+\frac{7696}{27} \pi ^2 \log (2 \beta )-\frac{112 \pi ^2 \zeta
  (3)}{3}-\frac{352 \pi ^4}{27}
\nonumber\\&&\mbox{}
-\frac{20348 \pi ^2}{81}\Bigg) 
+ \Bigg(-\frac{6016 \afour}{9}
-\frac{128}{9} \pi ^2 \log ^2(2 \beta )+\frac{2576}{27} \pi ^2 \log (2 \beta
)+\frac{1204 \zeta (3)}{9}
\nonumber\\&&\mbox{}
+\frac{1274 \pi ^4}{81}-\frac{8912 \pi ^2}{81}
+\frac{35792}{81}-\frac{752 { l_2^4}}{27}+\frac{416}{27} \pi ^2 { l_2^2}
\nonumber\\&&\mbox{}
+\frac{3128}{27} \pi ^2 { l_2}\Bigg)\Bigg]
+\cR \tf^2 \nl^2
\Bigg[
 \frac{1}{\beta} \Bigg(\frac{128}{9} \pi ^2 \log ^2(2 \beta )-\frac{1408}{27}
 \pi ^2 \log (2 \beta )+\frac{64 \pi ^4}{27}
\nonumber\\&&\mbox{}
+\frac{3872 \pi ^2}{81}\Bigg)
+ \Bigg(-\frac{4160}{81}-\frac{256 \pi ^2}{27}\Bigg)
\Bigg]
+\cR \tf^2 \nh \nl
 \Bigg(\frac{2560 \pi ^2}{81}-\frac{26560}{81}\Bigg)
\,,\nonumber\\
\Delta^{(3),s} &=&
\nc^3
\Bigg[
 \frac{\pi ^4}{\beta^3}
+ \frac{1}{\beta^2} \Bigg(-\frac{44}{9} \pi ^4 \log (2 \beta )-\frac{44}{3}
\pi ^2 \log (2 \beta )-\frac{88 \pi ^2 \zeta (3)}{3}
\nonumber\\&&\mbox{} 
+\frac{176 \pi ^4}{27}+\frac{374 \pi ^2}{9}\Bigg)
+ \frac{1}{\beta} \Bigg(\frac{484}{9} \pi ^2 \log ^2(2 \beta )
-8 \pi ^4 \log (2 \beta )-\frac{4484}{27} \pi ^2 \log (2 \beta )
\nonumber\\&&\mbox{} 
-\frac{104 \pi ^2 \zeta (3)}{3}-\frac{\pi ^6}{4}+\frac{1375 \pi ^4}{54}+\frac{11434 \pi ^2}{81}-4 \pi ^4 { l_2}\Bigg)
\Bigg]
\nonumber\\&&\mbox{}
+\cR^2 \tf \nl
\Bigg[
 \frac{1}{\beta^2} \Bigg(\frac{64}{9} \pi ^4 \log (2 \beta )+\frac{64}{3} \pi
^2 \log (2 \beta )
\nonumber\\&&\mbox{}
+\frac{128 \pi ^2 \zeta (3)}{3}-\frac{352 \pi ^4}{27}-\frac{640 \pi ^2}{9}\Bigg)
+ \frac{1}{\beta} \Bigg(-\frac{16}{3} \pi ^2 \log (2 \beta )+32 \pi ^2 \zeta
(3)-\frac{202 \pi ^2}{9}\Bigg)
\nonumber\\&&\mbox{}
+ \Bigg(\frac{6400 \afour}{9}-\frac{512}{9} \pi ^2 \log ^2(2 \beta
)+\frac{128}{9} \pi ^4 \log (2 \beta )+\frac{8384}{27} \pi ^2 \log (2 \beta )
\nonumber\\&&\mbox{}
+\frac{256 \pi ^2 \zeta (3)}{3}+\frac{3308 \zeta (3)}{3}+\frac{68 \pi
  ^4}{405}-\frac{24904 \pi ^2}{81}+\frac{1820}{9}+\frac{800 { l_2^4}}{27}
\nonumber\\&&\mbox{}
+\frac{64}{27} \pi ^2 { l_2^2}+\frac{152}{3} \pi ^2 { l_2}\Bigg)
\Bigg]
+\cA \cR \tf \nl
\Bigg[
 \frac{1}{\beta} \Bigg(-\frac{704}{9} \pi ^2 \log ^2(2 \beta )
\nonumber\\&&\mbox{}
+\frac{7696}{27}
 \pi ^2 \log (2 \beta )
-\frac{112 \pi ^2 \zeta (3)}{3}
-\frac{352 \pi ^4}{27}-\frac{20348 \pi ^2}{81}\Bigg)
+ \Bigg(-\frac{5888 \afour}{9}
\nonumber\\&&\mbox{}  
-\frac{128}{9} \pi ^2 \log ^2(2 \beta)
+\frac{2576}{27} \pi ^2 \log (2 \beta )+260 \zeta (3)
+\frac{7364 \pi ^4}{405}-\frac{5968 \pi ^2}{81}-\frac{15968}{81}
\nonumber\\&&\mbox{}  
-\frac{736
  { l_2^4}}{27}
+\frac{448}{27} \pi ^2 { l_2^2}+\frac{584}{9} \pi ^2 { l_2}\Bigg)
\Bigg]
\nonumber\\&&\mbox{}
+\cR \tf^2 \nl^2
\Bigg[
 \frac{1}{\beta} \Bigg(\frac{128}{9} \pi ^2 \log ^2(2 \beta )
-\frac{1408}{27} \pi ^2 \log (2 \beta )+\frac{64 \pi ^4}{27}+\frac{3872 \pi ^2}{81}\Bigg)
\nonumber\\&&\mbox{}
+ \Bigg(\frac{2336}{81}-\frac{128 \pi ^2}{27}\Bigg)
\Bigg]
+ \cR \tf^2 \nh \nl
 \Bigg(\frac{640 \pi ^2}{27}-\frac{20096}{81}\Bigg)
\,,
\nonumber\\
\Delta^{(3),p} &=&
\nc^3
\Bigg[
 \frac{1}{\beta^2} \Bigg(-\frac{44}{9} \pi ^4 \log (2 \beta )-\frac{88 \pi ^2 \zeta (3)}{3}+\frac{8 \pi ^4}{27}\Bigg)
+ \frac{1}{\beta} \Bigg(\frac{484}{9} \pi ^2 \log ^2(2 \beta )
\nonumber\\&&\mbox{}
-16 \pi ^4 \log
(2 \beta )-\frac{788}{27} \pi ^2 \log (2 \beta )-\frac{230 \pi ^2 \zeta
  (3)}{3}-\frac{\pi ^6}{4}+\frac{1483 \pi ^4}{54}+\frac{1942 \pi ^2}{81}
\nonumber\\&&\mbox{}
-8 \pi^4 { l_2}\Bigg) { \Bigg]}
+\cR^2 \tf \nl
\Bigg[
 \frac{1}{\beta^2} \Bigg(\frac{64}{9} \pi ^4 \log (2 \beta )
+\frac{128 \pi ^2 \zeta (3)}{3}-\frac{160 \pi ^4}{27}\Bigg)
\nonumber\\&&\mbox{}
+ \frac{1}{\beta} \Bigg(-48 \pi ^2 \log (2 \beta )
+32 \pi ^2 \zeta (3)+22 \pi ^2\Bigg)
+ \Bigg(\frac{4096 \afour}{3}-\frac{256}{3} \pi ^2 \log ^2(2 \beta
)
\nonumber\\&&\mbox{}
+\frac{128}{9} \pi ^4 \log (2 \beta )+\frac{3008}{9} \pi ^2 \log (2 \beta
)+\frac{256 \pi ^2 \zeta (3)}{3}
+\frac{5828 \zeta (3)}{3}+\frac{32 \pi ^4}{27}
\nonumber\\&&\mbox{}
-\frac{3616 \pi
  ^2}{27}+\frac{1892}{9}+\frac{512 { l_2^4}}{9}
+\frac{256}{9} \pi ^2 { l_2^2}-\frac{296}{9} \pi ^2 { l_2}\Bigg)
\Bigg]
\nonumber\\&&\mbox{}
+\cA \cR \tf \nl
\Bigg[
 \frac{1}{\beta} \Bigg(-\frac{704}{9} \pi ^2 \log ^2(2 \beta )
\nonumber\\&&\mbox{}
+\frac{3472}{27}
 \pi ^2 \log (2 \beta )-\frac{112 \pi ^2 \zeta (3)}{3}
-\frac{352 \pi ^4}{27}
-\frac{3596 \pi ^2}{81}\Bigg)
+ \Bigg(-1280 \afour
\nonumber\\&&\mbox{}
-\frac{128}{3} \pi ^2 \log ^2(2 \beta )
+\frac{1552}{9} \pi
^2 \log (2 \beta )+\frac{668 \zeta (3)}{3}
+\frac{1804 \pi ^4}{45}
-\frac{3988 \pi ^2}{27}+\frac{8992}{27}
\nonumber\\&&\mbox{}
-\frac{160 { l_2^4}}{3}+\frac{64}{3} \pi ^2 { l_2^2}+\frac{1256}{9} \pi ^2
{ l_2}\Bigg)
\Bigg]
\nonumber\\&&\mbox{} 
+\cR \tf^2 \nl^2
\Bigg[
 \frac{1}{\beta} \Bigg(\frac{128}{9} \pi ^2 \log ^2(2 \beta )-\frac{640}{27}
 \pi ^2 \log (2 \beta )
+\frac{64 \pi ^4}{27}
\nonumber\\&&\mbox{}
+\frac{800 \pi ^2}{81}\Bigg)
+ \Bigg(-\frac{1312}{27}-\frac{128 \pi ^2}{9}\Bigg)
\Bigg]
+\cR \tf^2 \nh \nl
 \Bigg(\frac{896 \pi ^2}{27}-\frac{9728}{27}\Bigg)
\,.
\end{eqnarray}
At one- and two-loop order the leading terms are proportional to $C_F/\beta$
and $(C_F/\beta)^2$, respectively. At three-loop order we observe $1/\beta^3$
terms only in the axial-vector and scalar case but not for the vector and
pseudo-scalar currents.  Our findings are in agreement with considerations in
the non-relativistic limit which can be used to predict the leading terms of
order $(\alpha_s/\beta)^n$.  In fact, in this limit $\Delta^\delta$ can be
written as a combination of the Sommerfeld factor and a factor taking into
account $P$-wave scattering~\cite{threshold} (see
also~\cite{Chetyrkin:1997mb}):
\begin{eqnarray}
  \Delta^\delta &=& \frac{y}{1-e^{-y}}\left(1 + P^\delta \frac{y^2}{4\pi^2}\right) +
  \ldots
  \nonumber\\
  &=& 1 + \frac{\alpha_s}{4\pi} C_F \frac{2\pi^2}{\beta}
  + \left(\frac{\alpha_s}{4\pi}\right)^2
  \frac{C_F^2}{\beta^2}\left(\frac{4\pi^4}{3} + P^\delta 4 \pi^2\right)
  + \left(\frac{\alpha_s}{4\pi}\right)^3 \frac{C_F^3}{\beta^3} P^\delta 8 \pi^4
  +
  \ldots \,.
\end{eqnarray}
After the second equality sign $y = C_F\alpha_s\pi/\beta$ has been used and
$P^\delta$ is zero for $S$-wave and unity for $P$-wave processes, i.e., we
have $P^v=P^p=0$ and $P^a=P^s=1$.  The ellipses in the above equations
represent subleading terms.


\subsection{\label{sub::cusp}$\Gamma_{\rm cusp}$ and checks}

An interesting feature of the (renormalized) form factors is the presence of
infrared poles that can be described by a universal function, the cusp
anomalous dimension $\Gamma_{\rm
  cusp}$~\cite{Polyakov:1980ca,Brandt:1981kf,Korchemsky:1987wg},
which is independent of the external current.
This means that we can write
\begin{eqnarray}
  F = Z F^{f}\,,
  \label{eq::F_Ff}
\end{eqnarray}
where $F$ is any of our (six) scalar form factors and $F^{f}$ is the corresponding
(ultraviolet and intrared) finite version.
The factor $Z$, which is defined in the $\overline{\rm MS}$ scheme and
thus only contains poles in $\epsilon$, absorbs the infrared divergences and
$F^{f}$ is finite.  The single $\epsilon$ pole of the $n$-loop corrections of 
$Z$ contains the $n$-loop expression of $\Gamma_{\rm cusp}$. Using the notation
\begin{eqnarray}
  \Gamma_{\rm cusp} &=& \sum_{i \ge 1} \Gamma_{\rm cusp}^{(i)}
  \left(\frac{\alpha_s^{(n_l)}}{\pi}\right)^i \,,
  \nonumber\\
  Z &=& 1 + \sum_{ 1 \le j \le i } 
  \frac{z_{i,j}}{\eps^j} \left(\frac{\alpha_s^{(n_l)}}{\pi}\right)^i
  \,,
\end{eqnarray}
where $\alpha_s^{(n_l)}$ is the strong coupling constant with decoupled heavy
quark, we have
\begin{eqnarray}
  z_{1,1} &=& -\frac{1}{2} \Gamma^{(1)}_{\rm cusp}\,, \nonumber\\
  z_{2,1} &=& -\frac{1}{4} \Gamma^{(2)}_{\rm cusp} \,,\nonumber\\
  z_{3,1} &=& -\frac{1}{6} \Gamma^{(3)}_{\rm cusp} \,.
  \label{eq::Z_Gcusp}
\end{eqnarray}

We have used Eq.~(\ref{eq::F_Ff}) for the four form factors $F_1^v$, $F_1^a$,
$F^s$ and $F^p$, and have determined the corresponding $Z$ factor to three
loops. This requires to use decoupling relations for $\alpha_s$ up to two
loops (including higher order $\epsilon$ terms~\cite{Grozin:2007fh}) since 
the calculation of the form factors described above has been peformed
in the full theory with $\alpha_s\equiv \alpha_s^{(n_f)}$.
Afterwards, one-, two- and three-loop corrections to $\Gamma_{\rm
  cusp}$ are extracted with the help of Eq.~(\ref{eq::Z_Gcusp}) where at three
loops we have to restrict ourselves to the complete $n_l$ and the $N_c^3$ terms
of the remainder.  For all four currents we have obtained the same result for
$\Gamma^{(i)}_{\rm cusp}$ ($i=1,2,3$) which constitutes a strong cross check
of our calculations. Furthermore, our results agree with
Refs.~\cite{Polyakov:1980ca,Korchemsky:1987wg,Grozin:2014hna,Grozin:2015kna}
where dedicated calculations of $\Gamma_{\rm cusp}$ have been performed.

Besides the correct infrared structure there are several other checks which
our analytic expressions fulfill:
\begin{itemize}
\item As mentioned in Section~\ref{sub::high} we
observe that the results for the vector (scalar) and axial-vector
(pseudo-scalar) form factors agree in the high-energy limit.
\item We have
furthermore performed the calculation for general gauge parameter which drops
out for the renormalized form factors. Note that the cancellation is
non-trivial and only occurs in the proper interplay between bare three-loop
expression and wave function and quark mass counterterm contributions.  
\item We
have also numerically cross checked all three-loop master integrals up to the
finite term in $\epsilon$ using the program {\tt
  FIESTA}~\cite{Smirnov:2015mct}.
\item 
As a further check, our results fulfill the axial Ward identity which is
given by
\begin{eqnarray}
  q^\mu \Gamma_\mu^a &=& 2 i \Gamma^p
  \,.
\end{eqnarray}
Using Eq.~(\ref{eq::Gamma}) this leads to
\begin{eqnarray}
  F_1^a + \frac{q^2}{4m^2} F_2^a &=& F^p
  \,,
\end{eqnarray}
which is satisfied by our explicit results up to three loops after
transforming the mass parameter $m$ in Eq.~(\ref{eq::currents}) into the
on-shell scheme.
\end{itemize}



\section{\label{sec::num}Numerical results}

In this section we evaluate both the exact result and the approximations in
the various limits numerically. For illustration we plot for each of the six
form factors the $\epsilon^0$ term both as a function of $x\in[-1,1]$ and
$\phi\in[0,\pi]$. This means we cover the whole real $q^2$ axis. In the plots
we restrict ourselves to the real part of the form factors. Using the results
from~\cite{progdata} it is straightforward to obtain plots for the imaginary
parts, as well.

\begin{figure}[t] 
  \begin{center}
    \begin{tabular}{ccc}
      \includegraphics[width=0.3\textwidth]{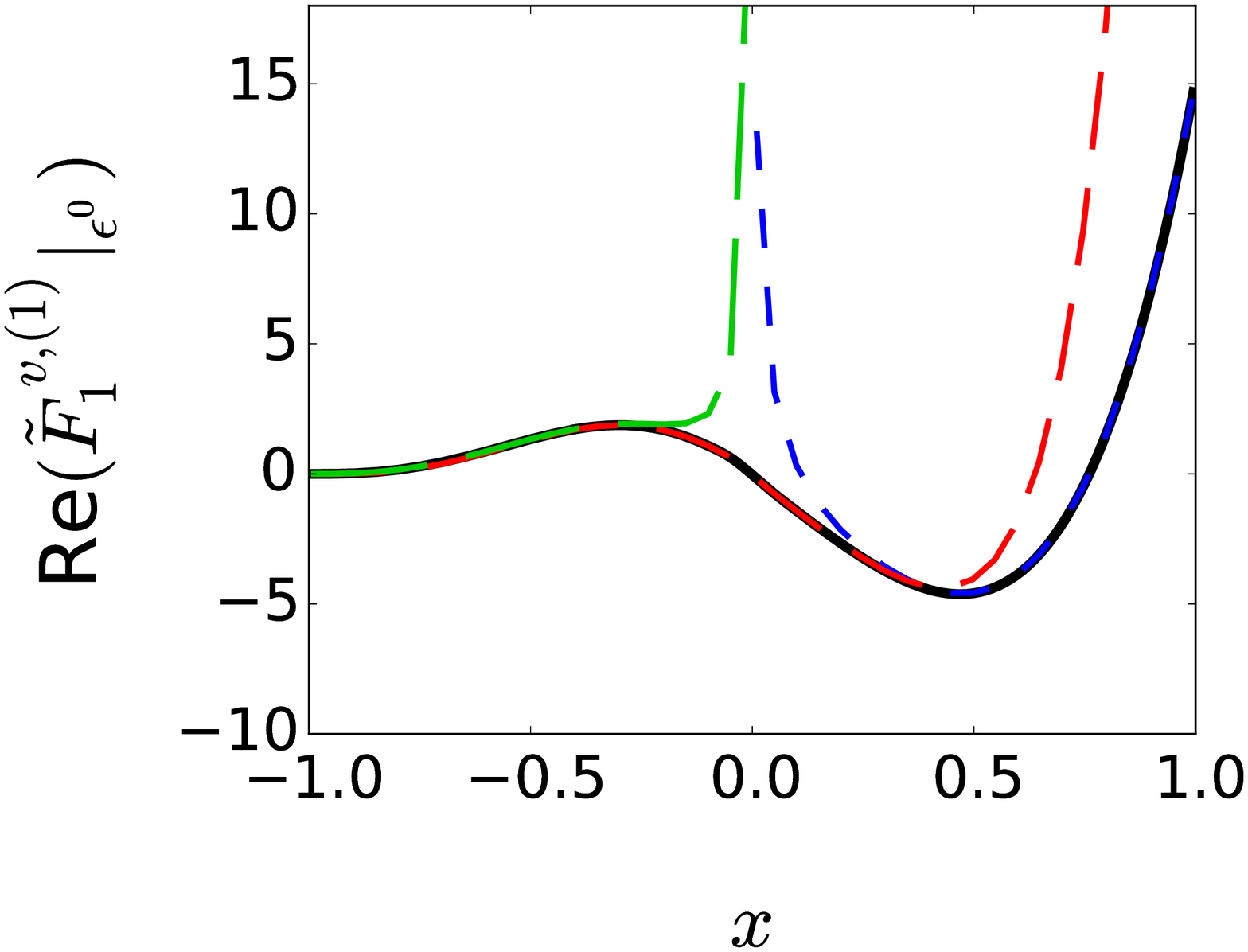} &
      \includegraphics[width=0.3\textwidth]{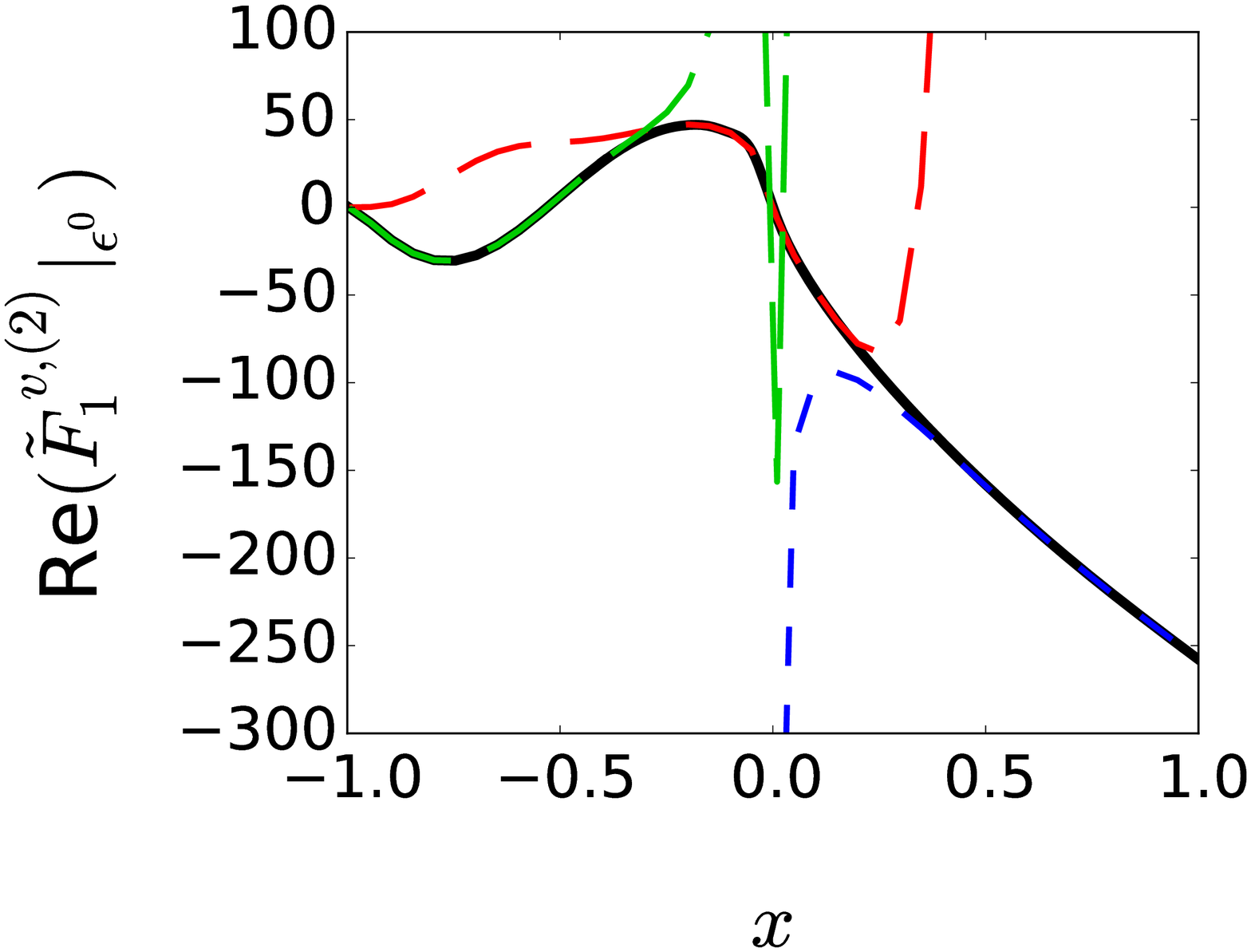} &
      \includegraphics[width=0.3\textwidth]{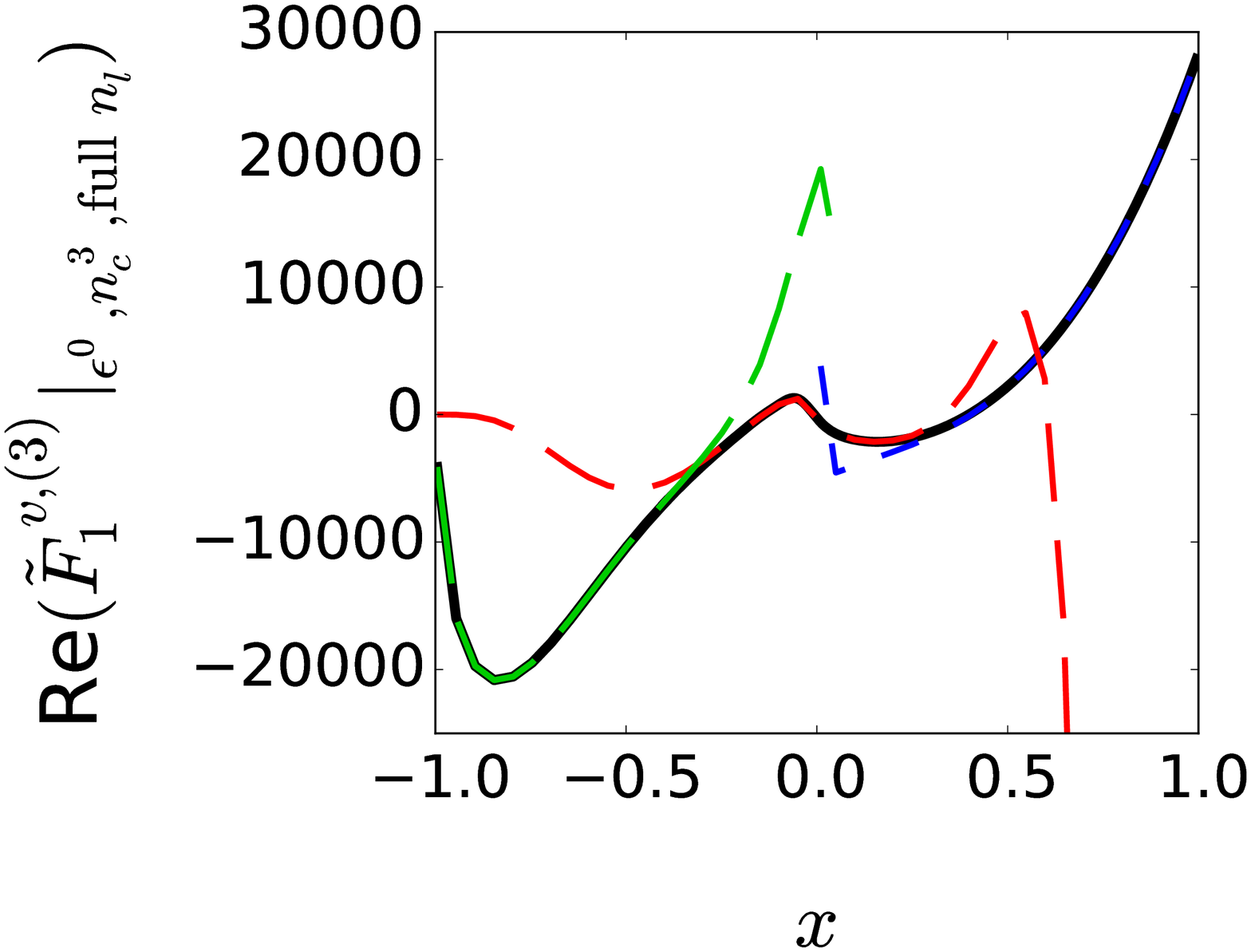} 
      \\
      \includegraphics[width=0.3\textwidth]{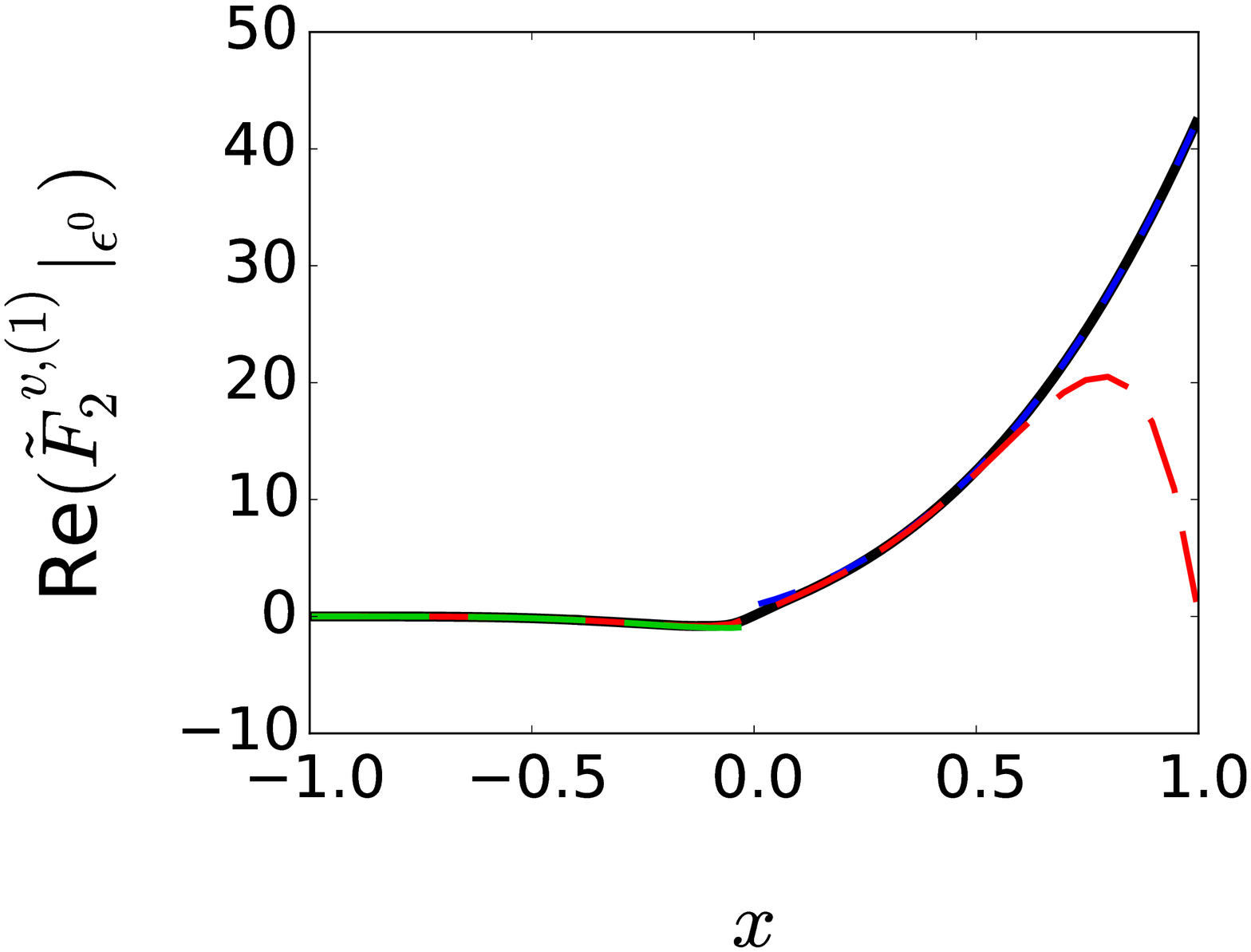} &
      \includegraphics[width=0.3\textwidth]{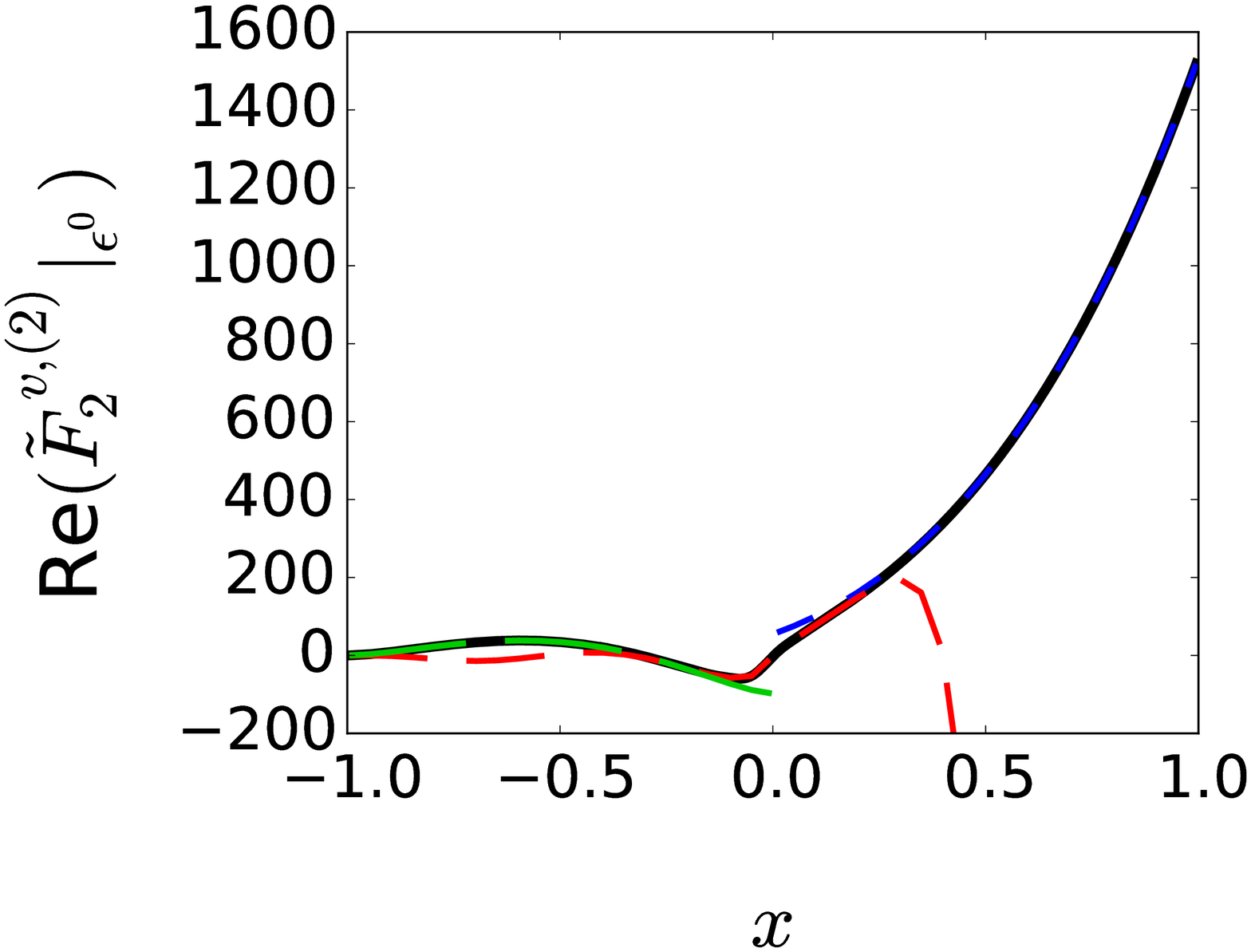} &
      \includegraphics[width=0.3\textwidth]{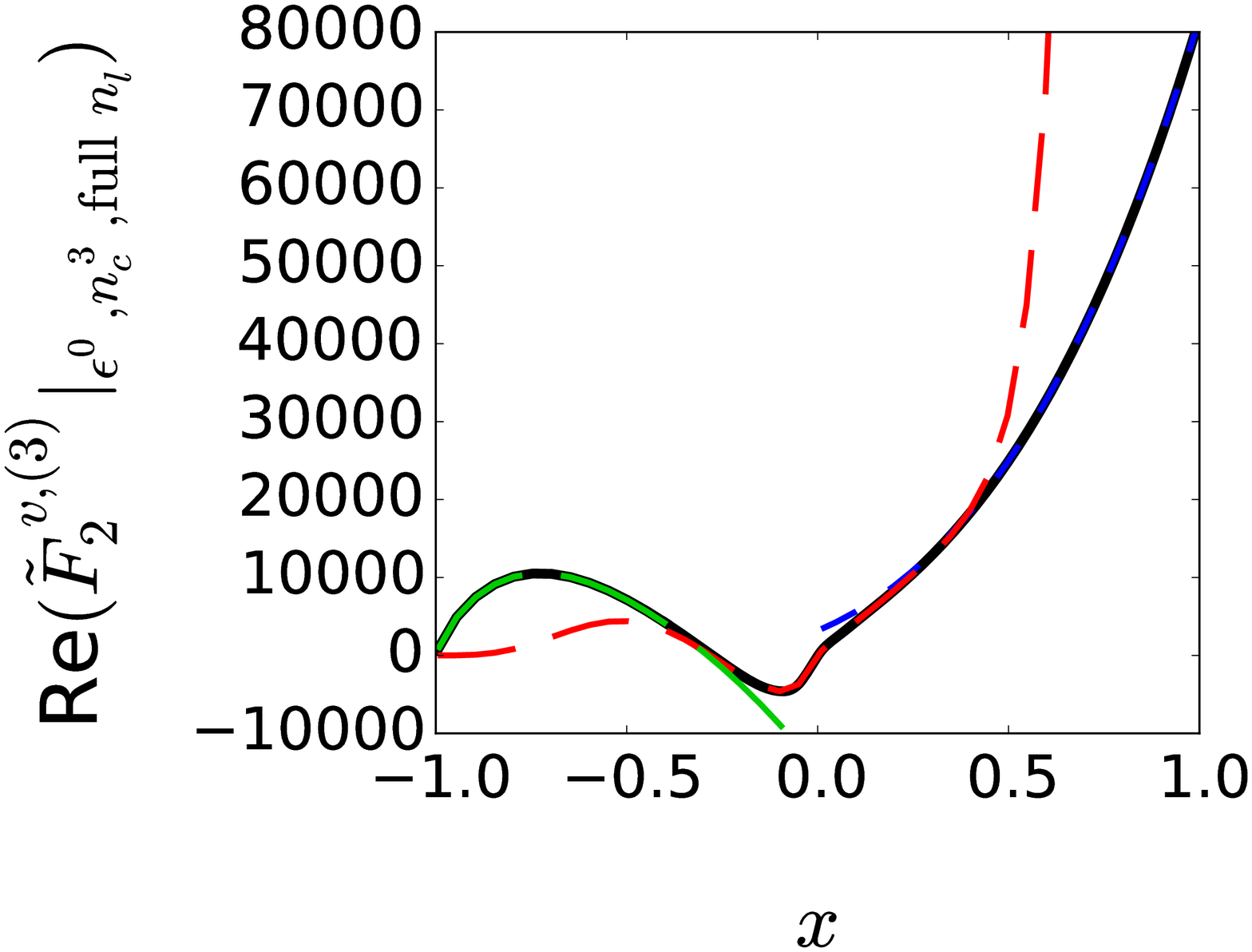} 
      \\
      \includegraphics[width=0.3\textwidth]{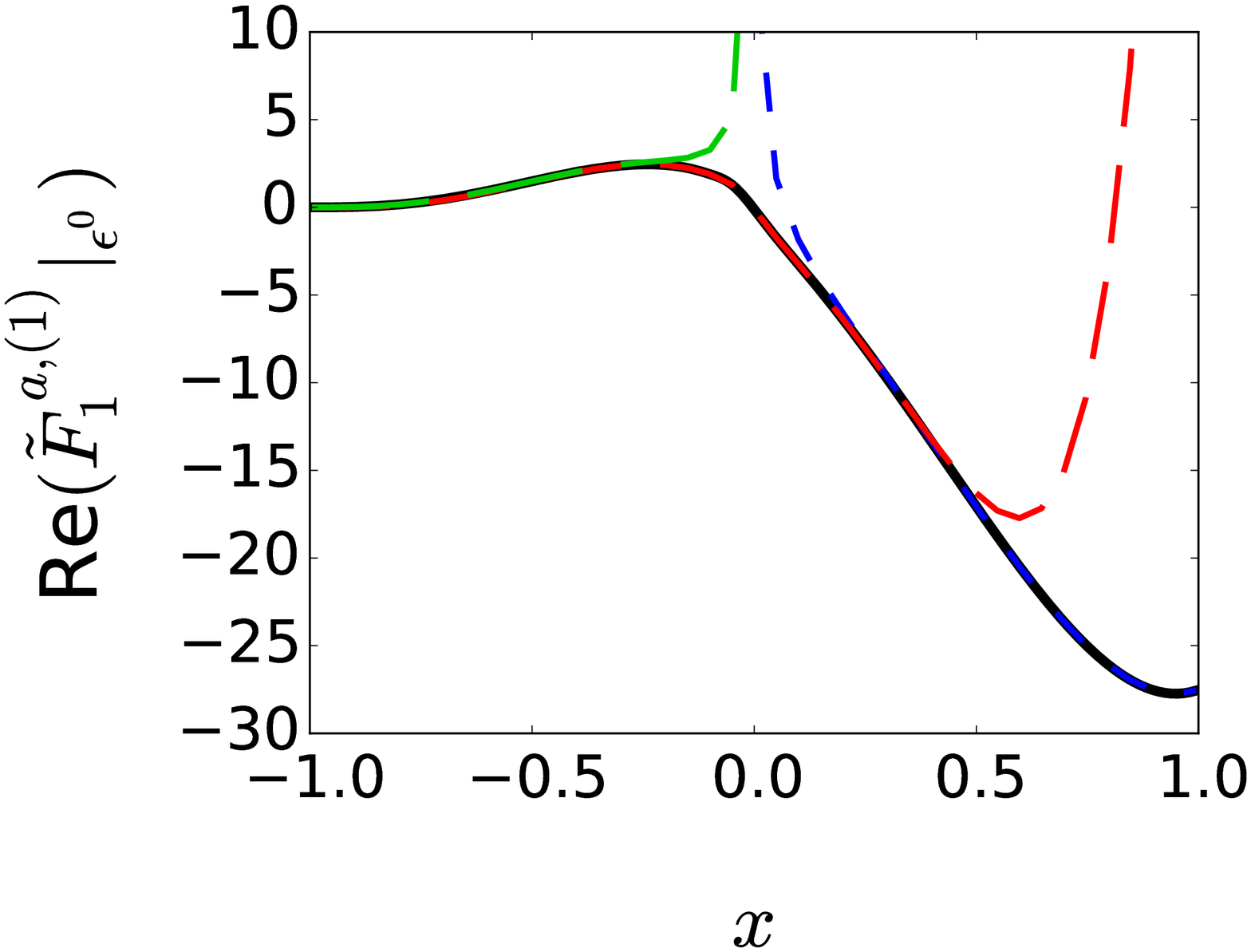} &
      \includegraphics[width=0.3\textwidth]{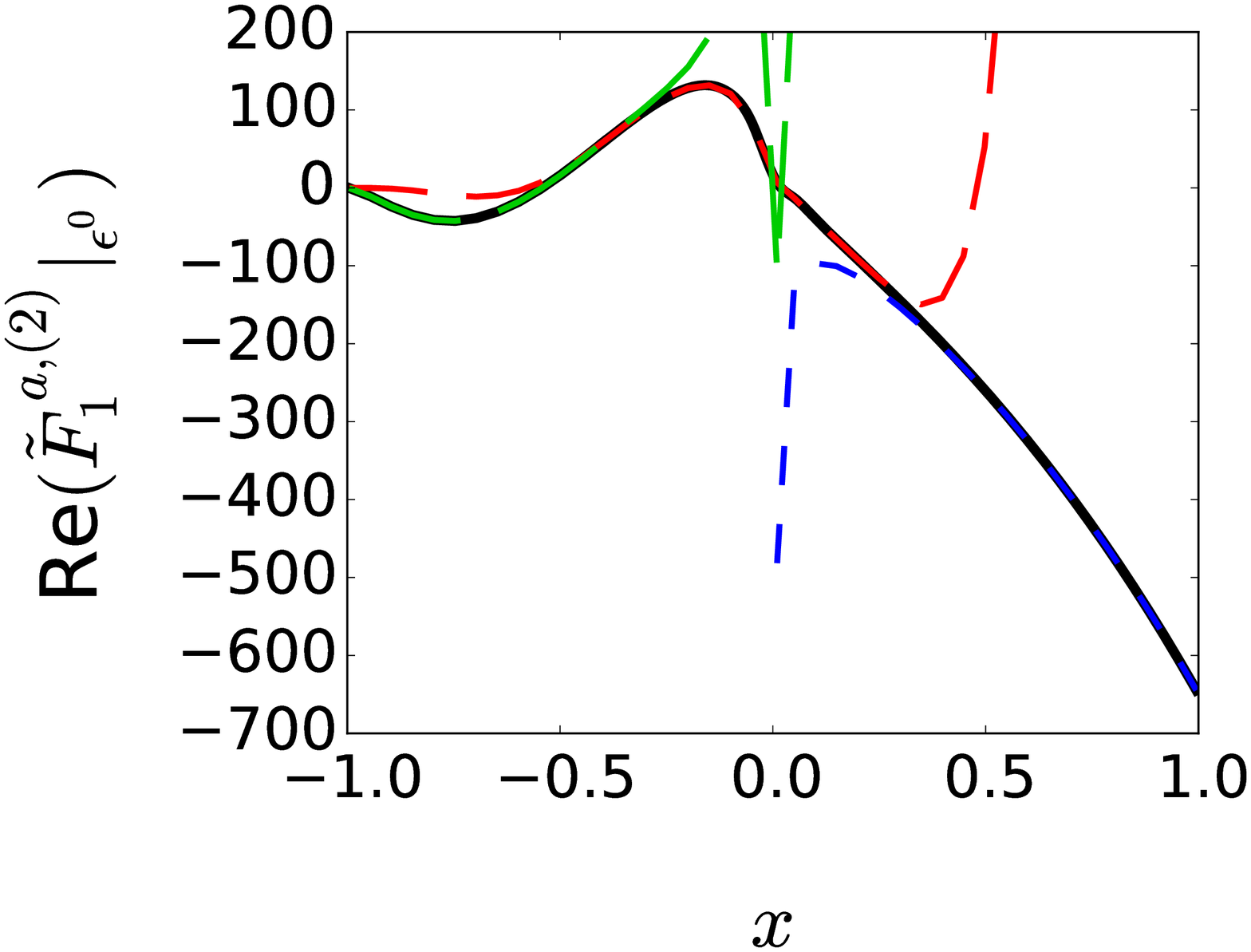} &
      \includegraphics[width=0.3\textwidth]{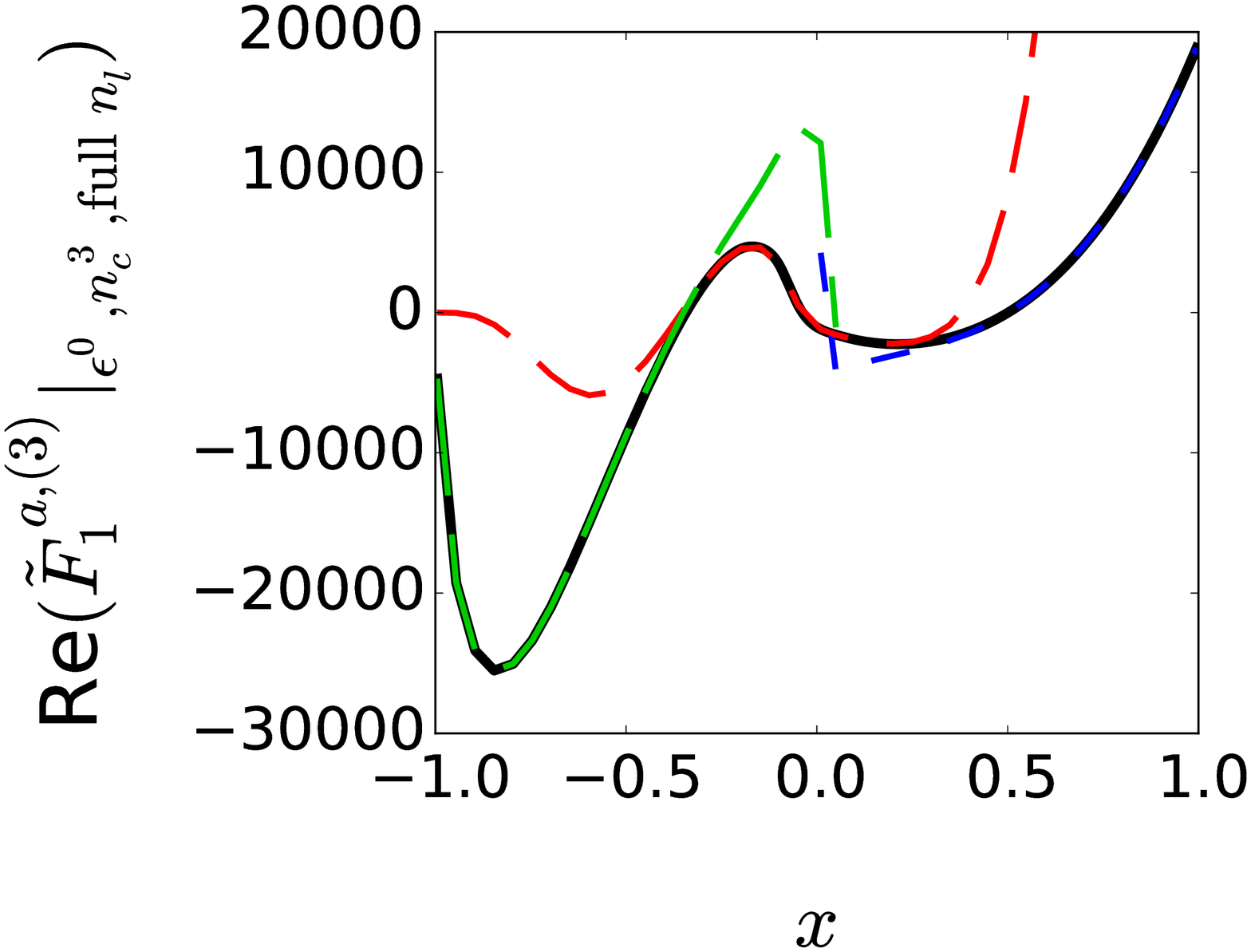} 
      \\
      \includegraphics[width=0.3\textwidth]{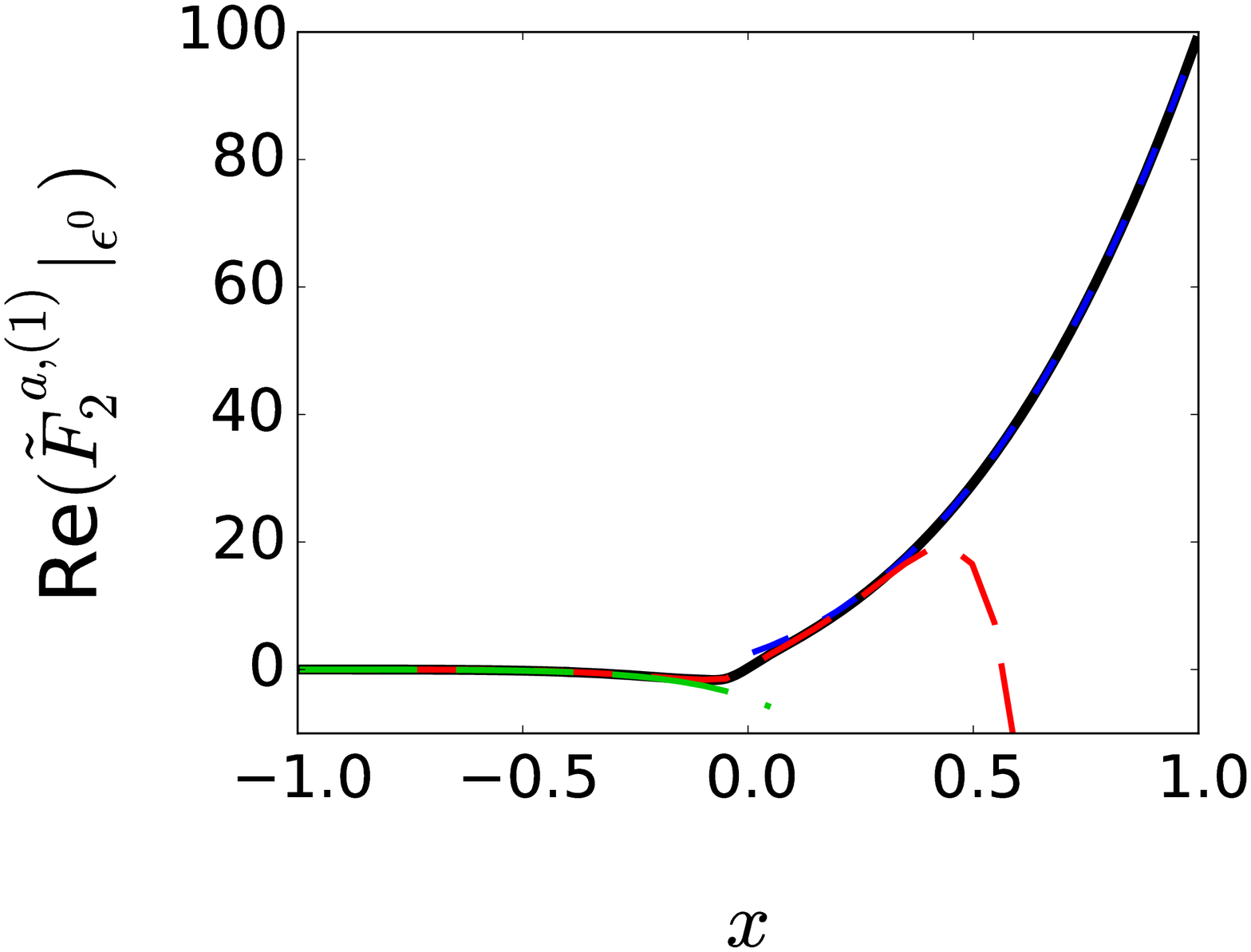} &
      \includegraphics[width=0.3\textwidth]{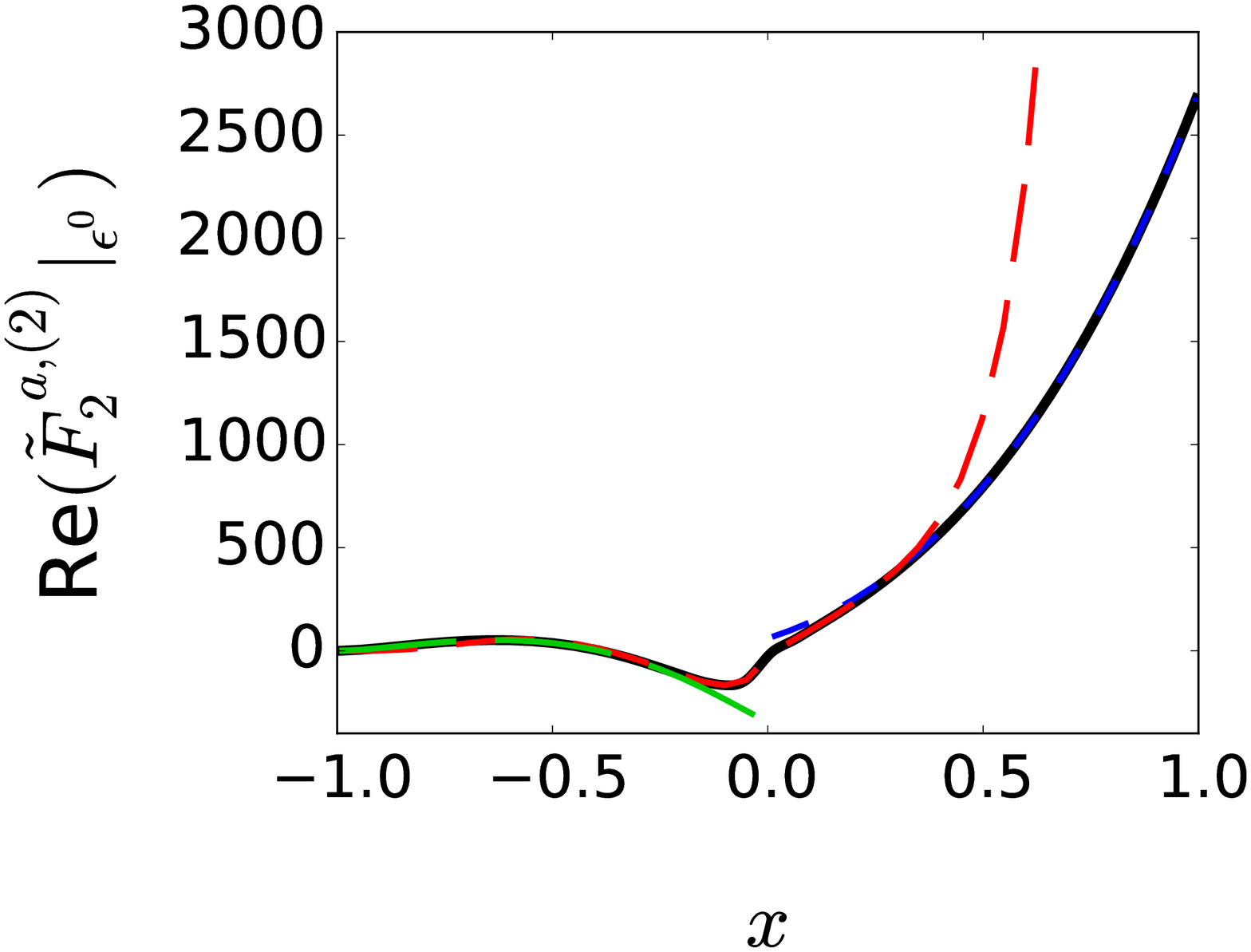} &
      \includegraphics[width=0.3\textwidth]{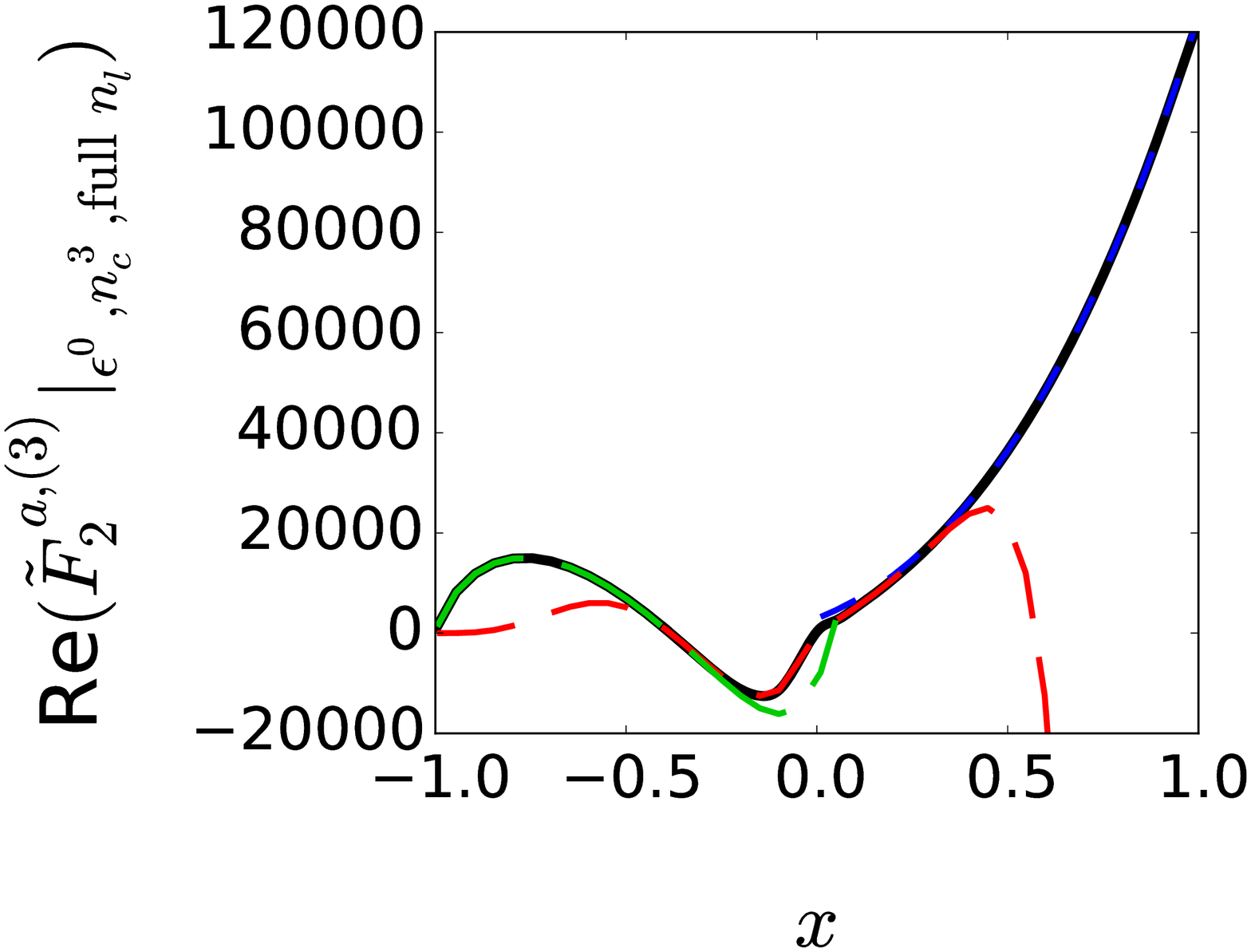} 
    \end{tabular}
    \caption{\label{fig::x_re_va}Real part of the $\epsilon^0$ term of the
      vector and axial-vector form factors as a function of $x$. Exact results
      and approximations are shown as solid and dashed lines, respectively. At
      three-loop order we add the complete light-fermion part for $n_l=5$ and
      the $N_c^3$ contribution. Short- (blue), medium- (red) and long- (green)
      dashed lines correspond to the low-energy, high-energy and threshold
      approximation, respectively}
  \end{center}
\end{figure}

\begin{figure}[t] 
  \begin{center}
    \begin{tabular}{ccc}
      \includegraphics[width=0.3\textwidth]{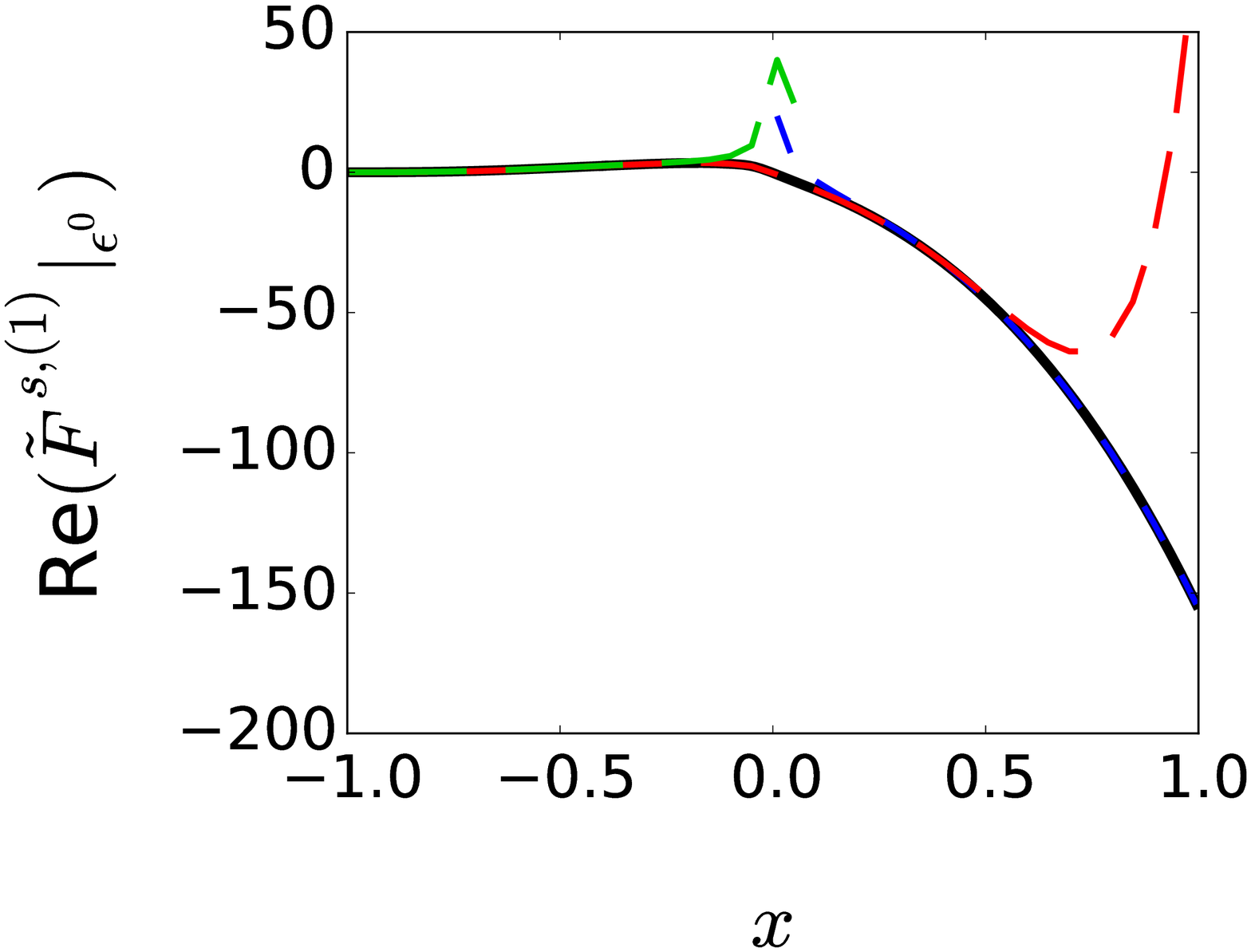} &
      \includegraphics[width=0.3\textwidth]{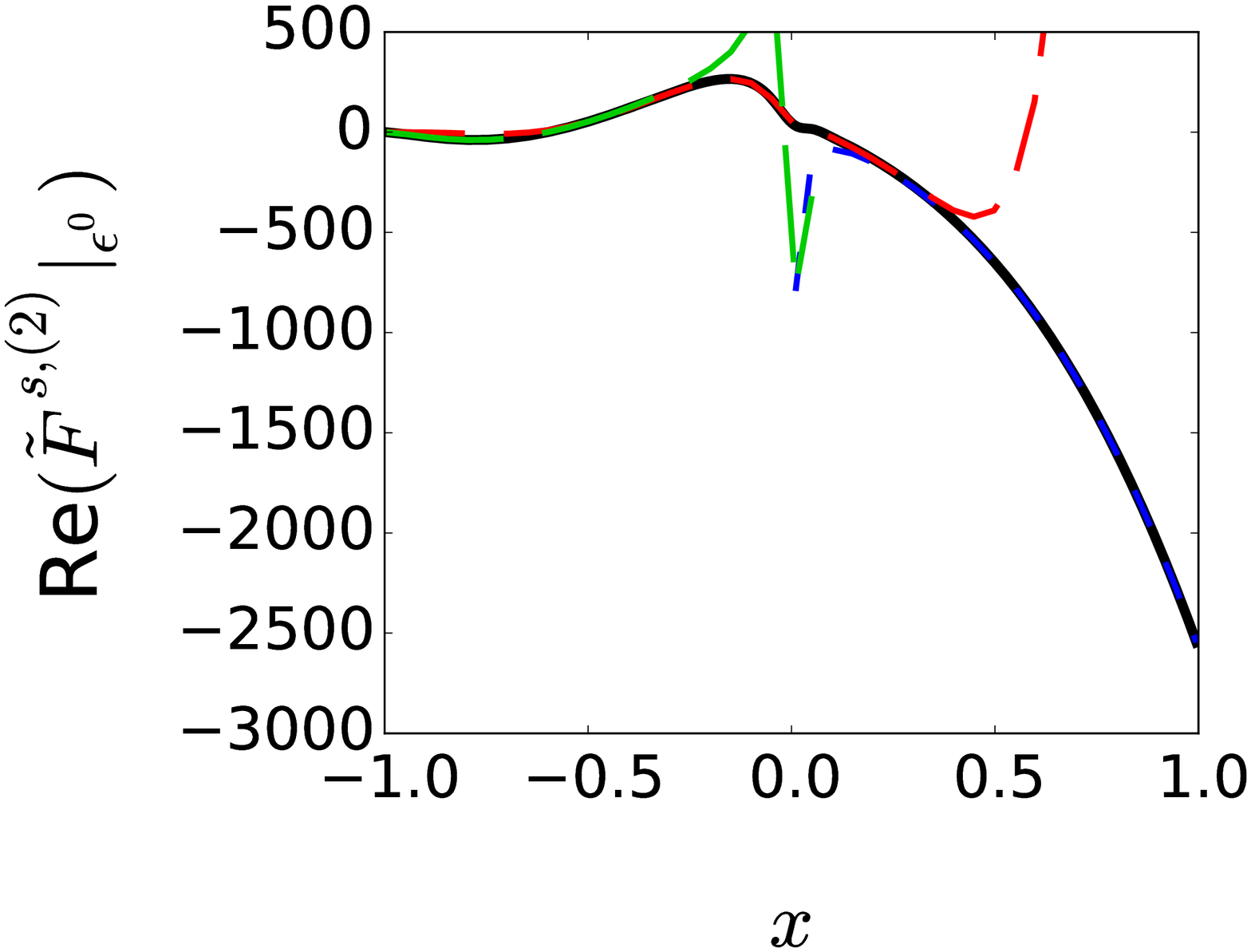} &
      \includegraphics[width=0.3\textwidth]{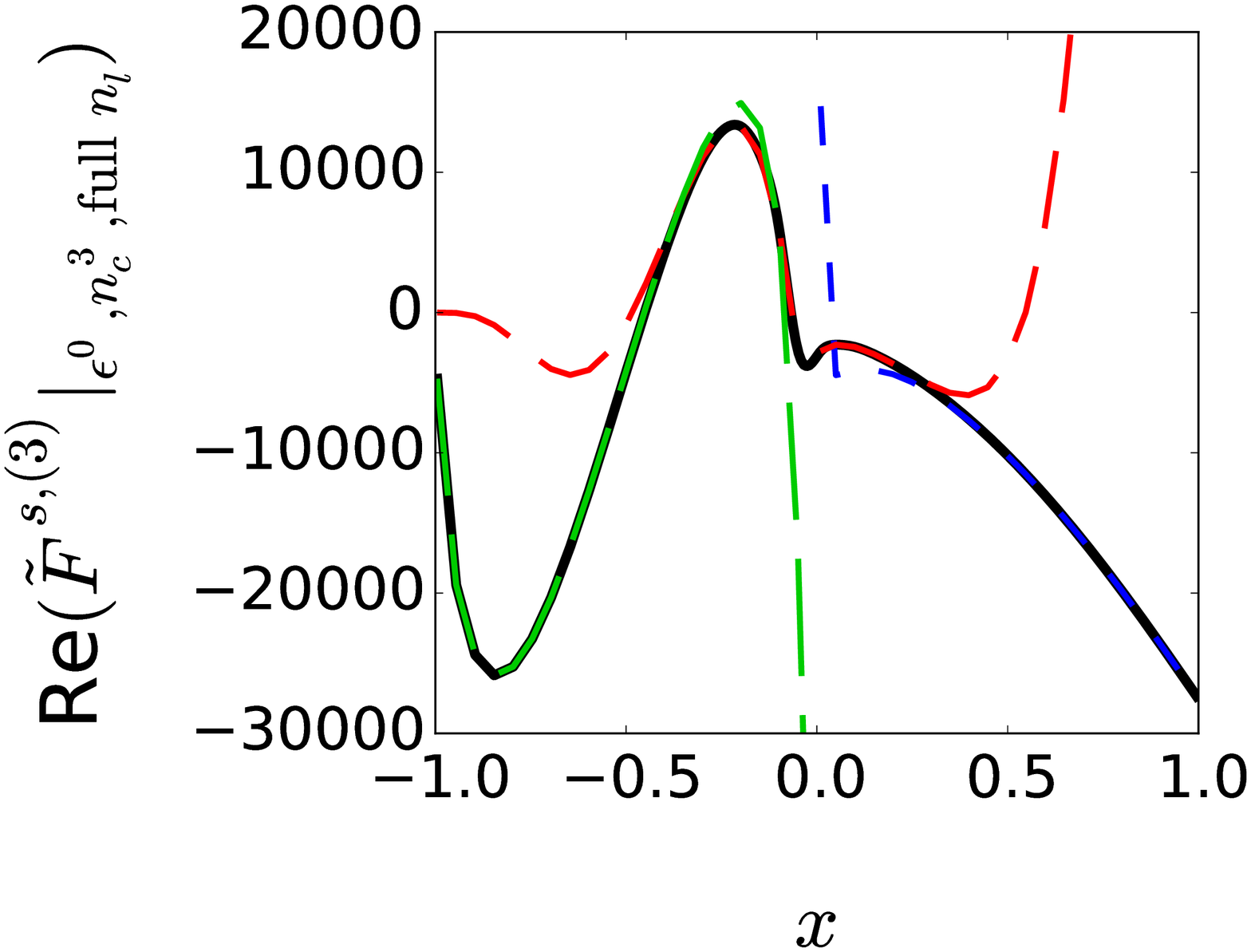} 
      \\
      \includegraphics[width=0.3\textwidth]{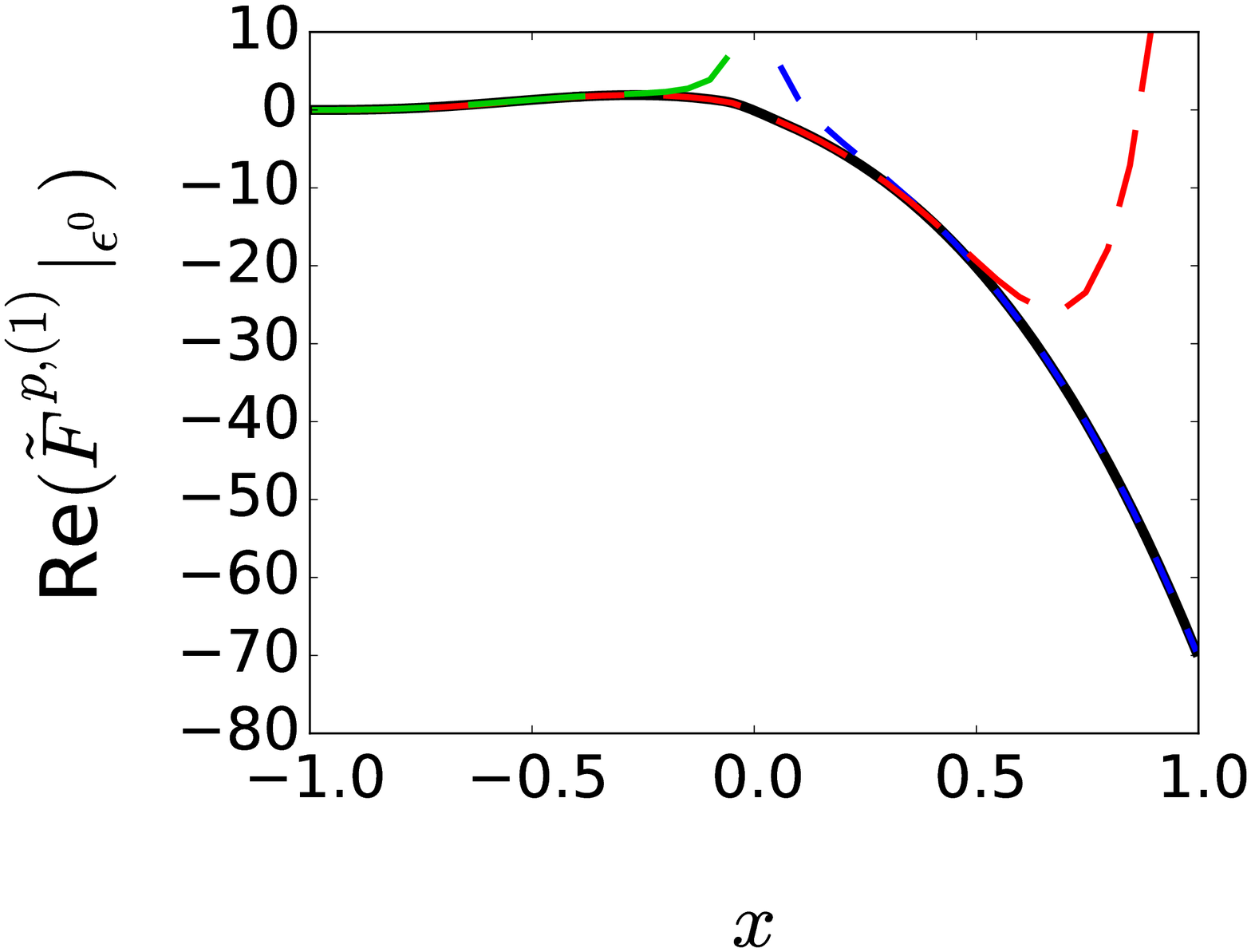} &
      \includegraphics[width=0.3\textwidth]{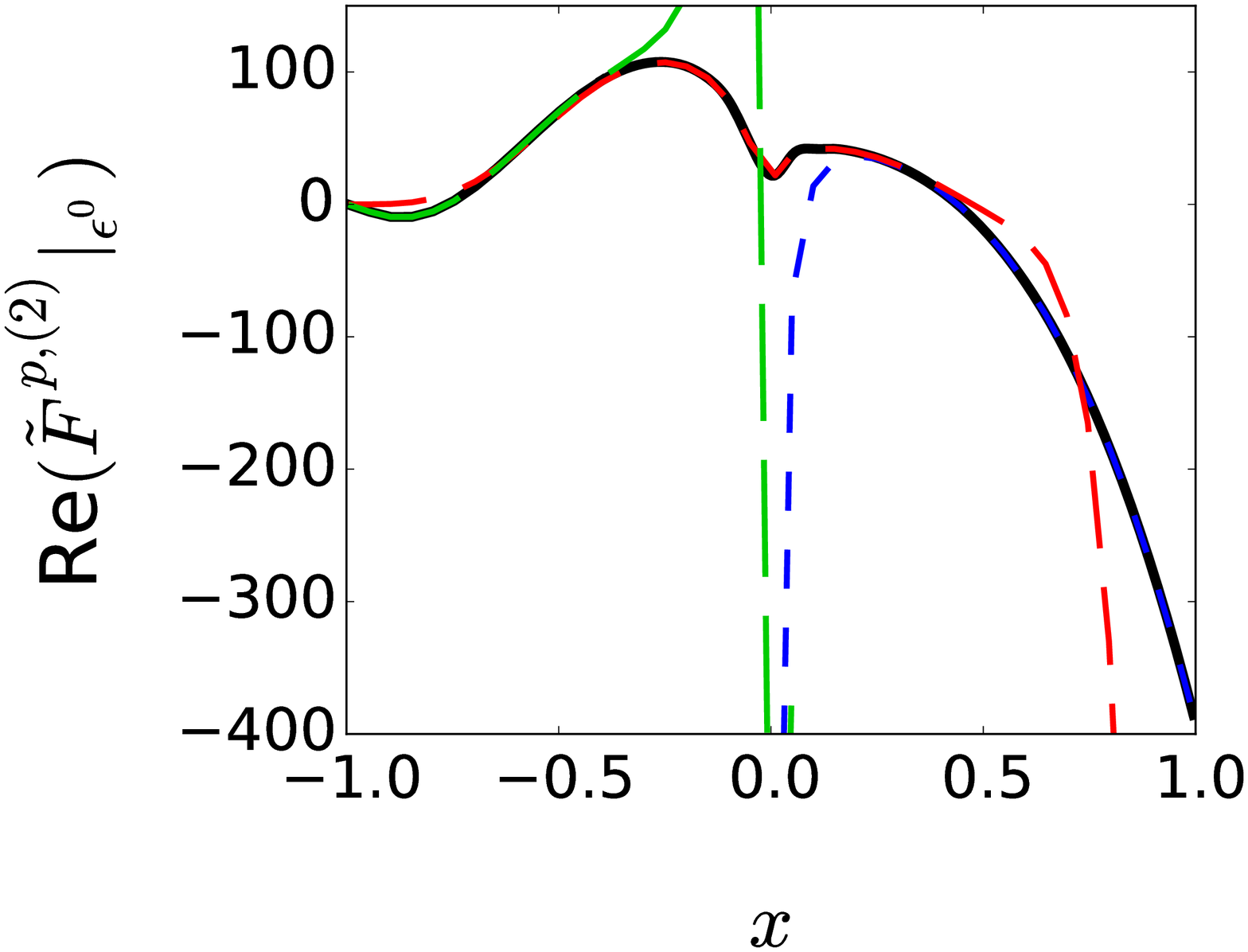} &
      \includegraphics[width=0.3\textwidth]{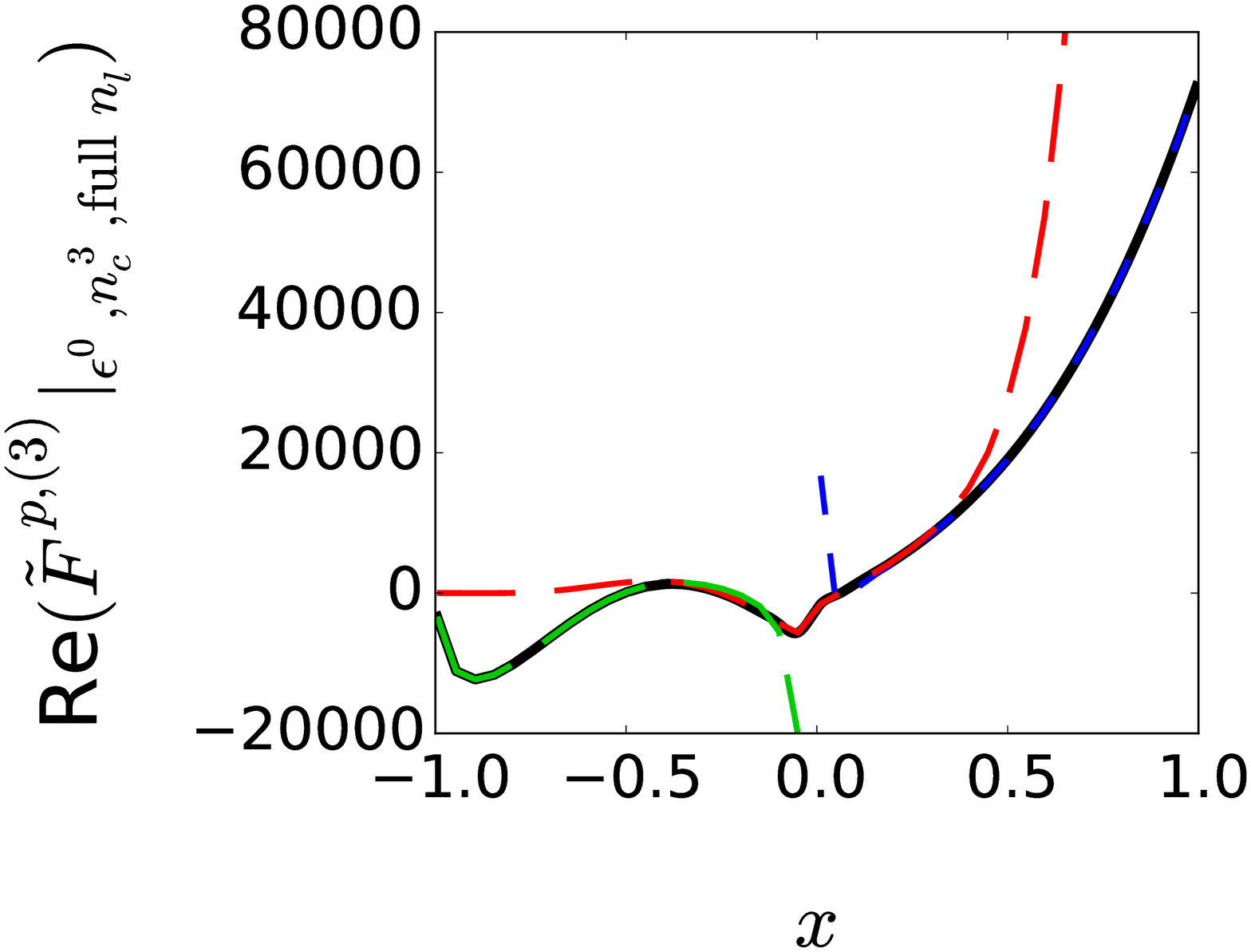} 
    \end{tabular}
    \caption{\label{fig::x_re_sp}Same as Fig.~\ref{fig::x_re_va} but for the
      the scalar and pseudo-scalar currents.}
  \end{center}
\end{figure}

At one- and two-loop order we plot the complete (non-singlet) results and set
$N_c=3, n_l=5$ and $n_h=1$. These values are also used at three loops where
the sum of the complete $n_l$ and the $N_c^3$ terms are
shown. For the numerical evaluation of the GPLs we use the program {\tt
  ginac}~\cite{Bauer:2000cp,Vollinga:2004sn} which is straightforward for real
values of $x$. GPLs with complex arguments (in our case for $x=e^{i\phi}$ with
$\phi\in[0,\pi]$) are evaluated with the help of
transformation rules given in Ref.~\cite{Vollinga:2004sn}. Some of the
GPLs involving $r_1$ require extraordinary long run times. In some
cases the results are even unstable. For this reason we generate in a first step
for each GPL, which is present in our anayltic result, a data base for
$\phi\in[0,\pi]$ and construct an interpolation function. Afterwards the
numerical evaluation of the form factors is fast and stable.

The approximations shown in the plots contain terms up to order $x^6$ and
$(1-x)^6$ in the high- and low-energy expansion, respectively. At threshold we
only include terms up to order $\beta^3$ although higher order terms are
available~\cite{progdata}. However, for the $N_c^3$ term we observe a bad
convegence behaviour which is the reason that we drop $\beta^4$ and higher
terms.

In the high-energy limit the form factors exhibit logarithmic singularities.
Thus, for the plots in the range $x\in[-1,1]$ we subtract the leading
high-energy behaviour, i.e., all terms which are not power-suppressed by $x$,
in order to ensure a smooth behaviour for $x\to0$. Furthermore, we multiply by
$(1+x)^4$ to ensure that at threshold (i.e. for $x=-1$) the one-, two- and
three-loop expressions become zero. This leads to a numerical enhancement for
$x=1$, however, also in this limit finite results are obtained. Thus, the
function we use for the plots reads
\begin{eqnarray}
  \tilde{F}(q^2) = (1+x)^4 \left[ F(q^2) - F(q^2)\Big|_{q^2\to\infty} \right]
  \,.
\end{eqnarray}
Our results for the six scalar form factors are shown in
Figs.~\ref{fig::x_re_va} and~\ref{fig::x_re_sp}.  The exact result is shown as
solid (black) curve and the approximations are plotted as dashed lines.  Note
that in all cases the whole range $x\in[-1,1]$ can be covered by the
approximations, i.e., for each $x$-value there is at least one dashed curve on
top of the (black) solid line.

\begin{figure}[t] 
  \begin{center}
    \begin{tabular}{ccc}
      \includegraphics[width=0.3\textwidth]{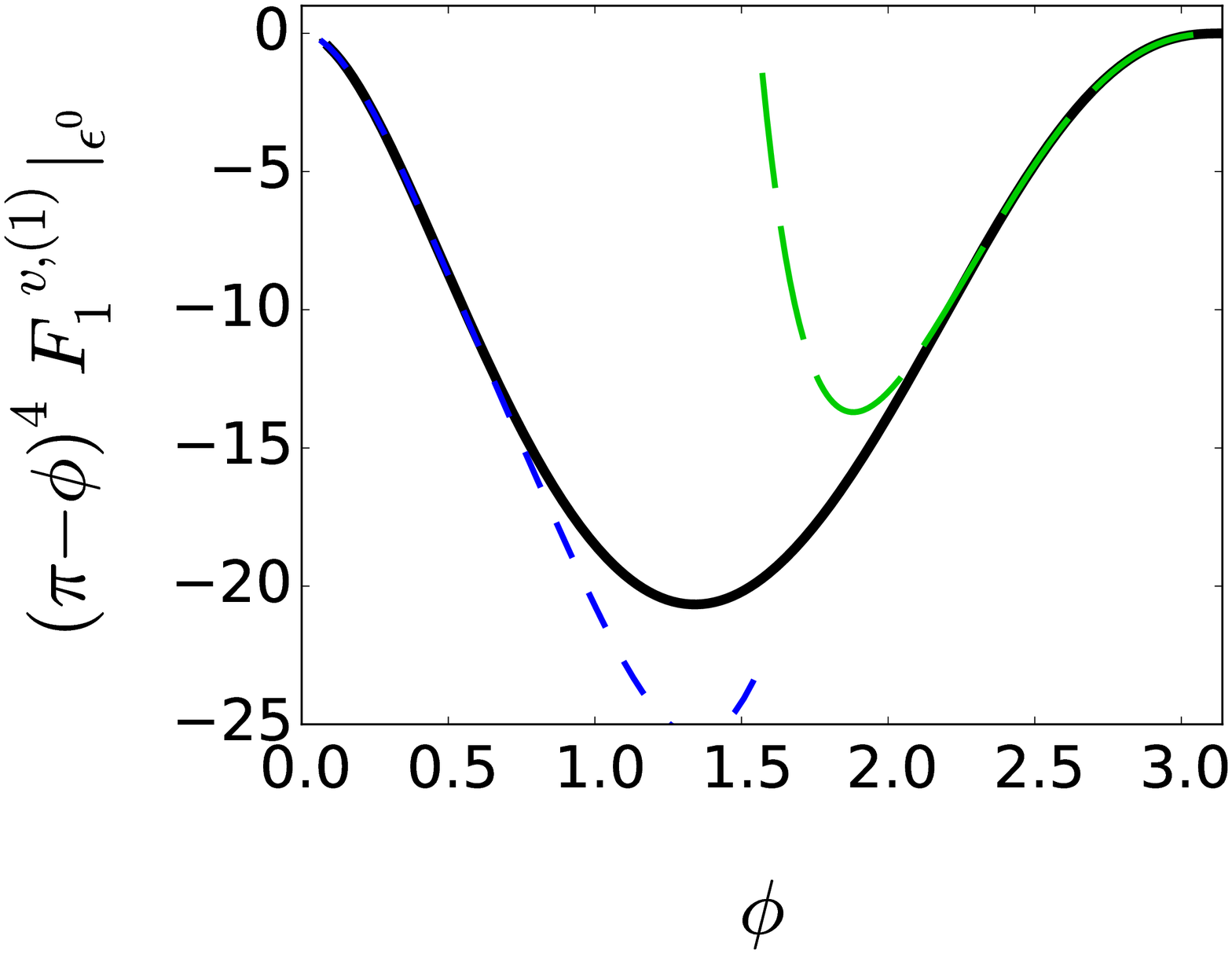} &
      \includegraphics[width=0.3\textwidth]{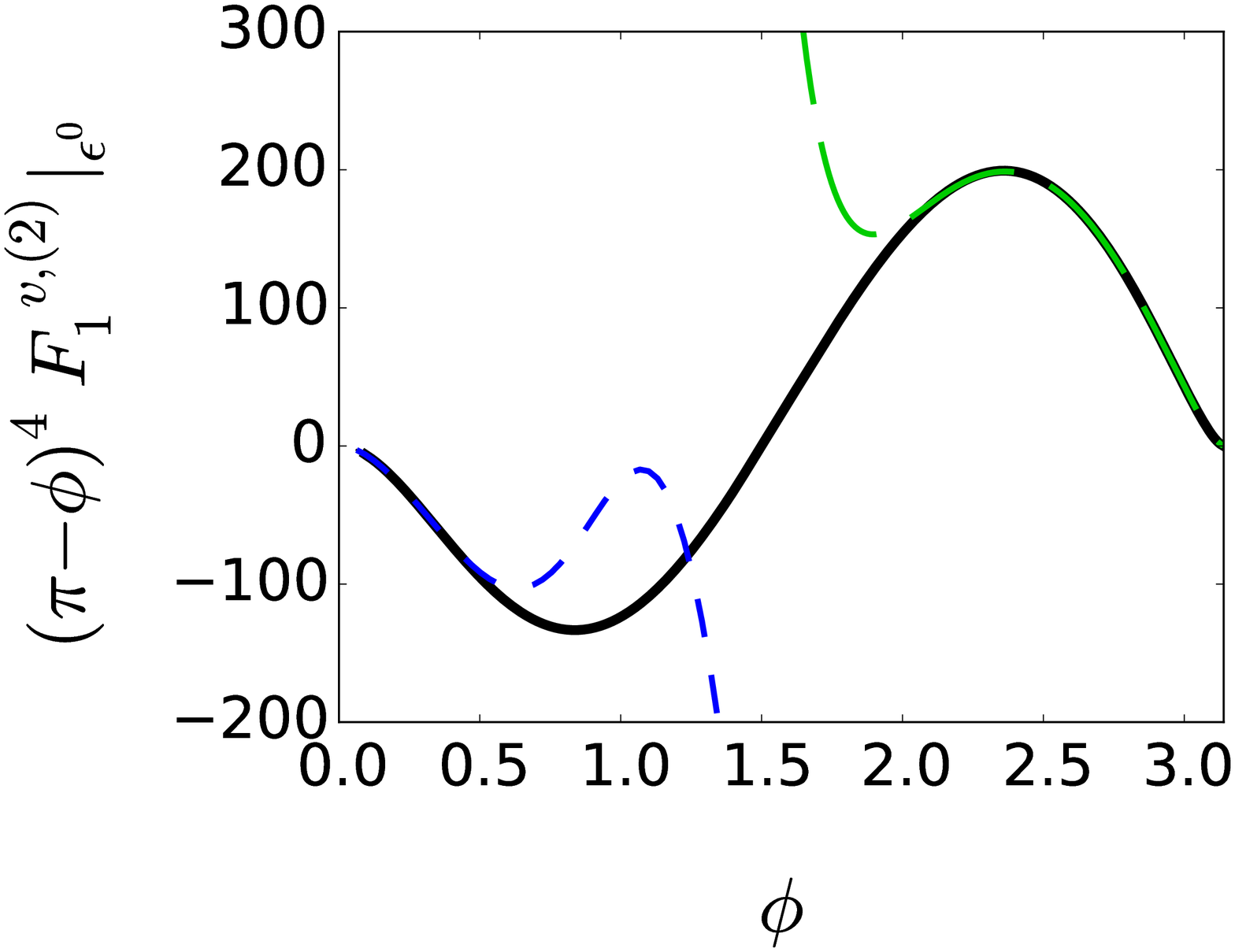} &
      \includegraphics[width=0.3\textwidth]{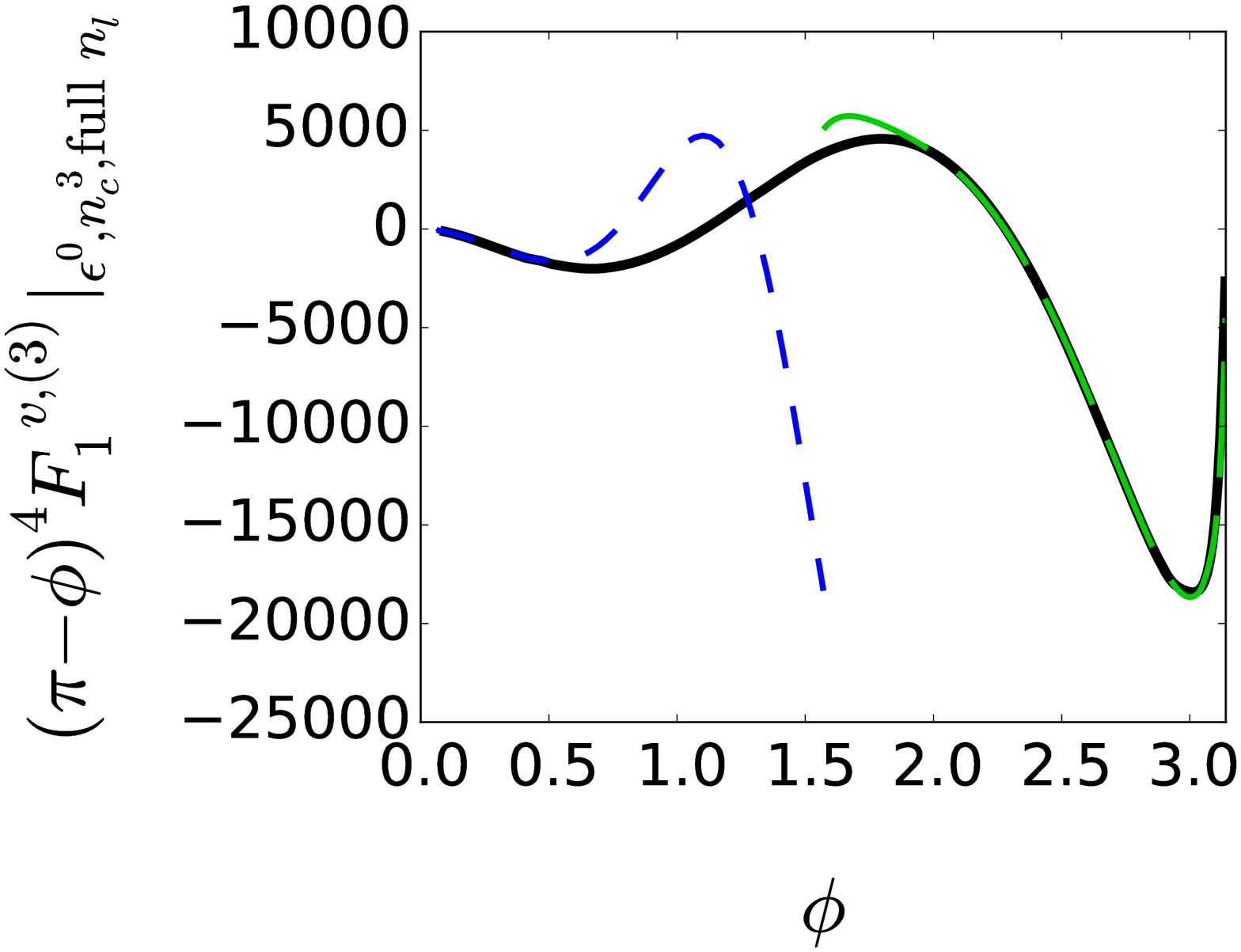} 
      \\
      \includegraphics[width=0.3\textwidth]{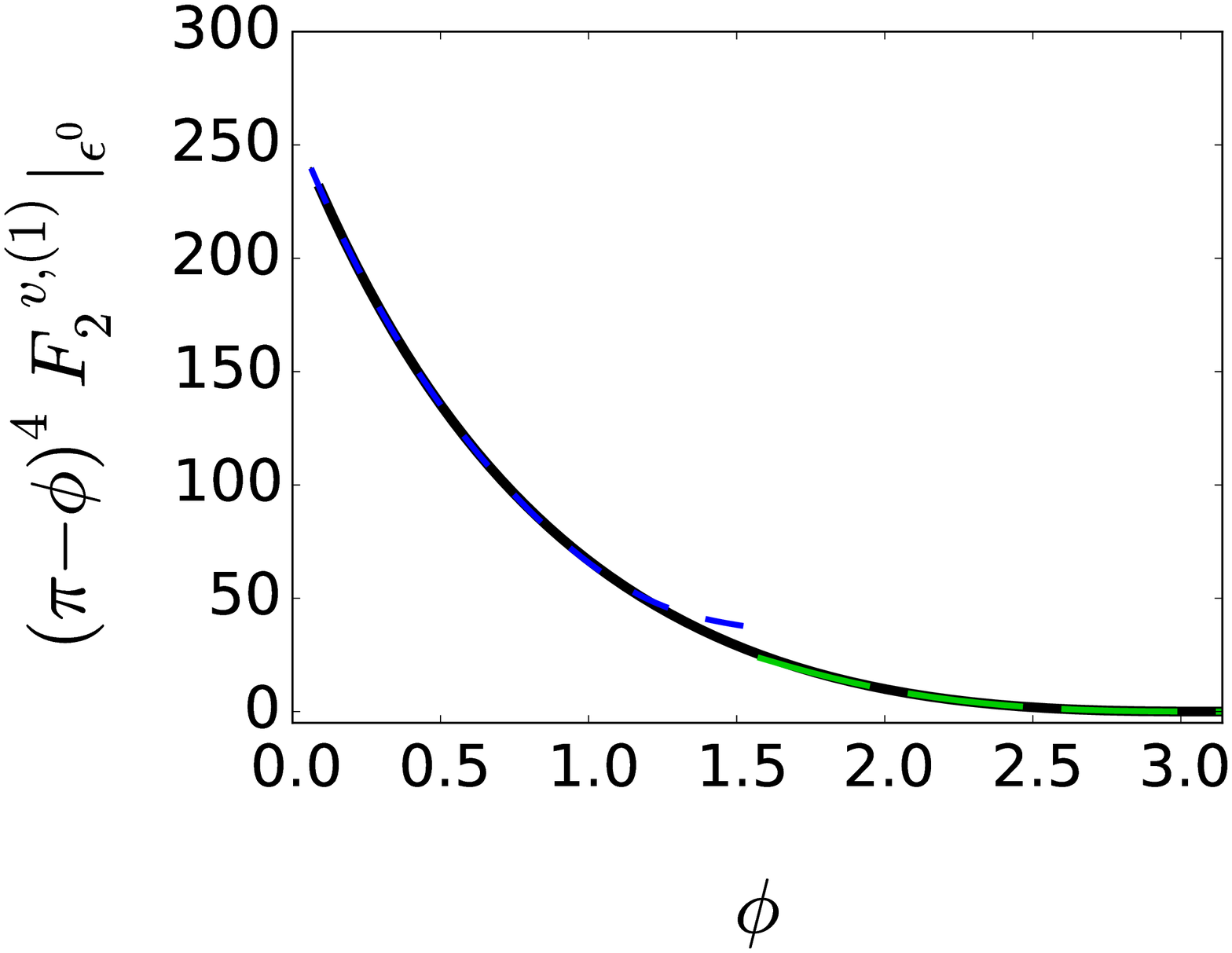} &
      \includegraphics[width=0.3\textwidth]{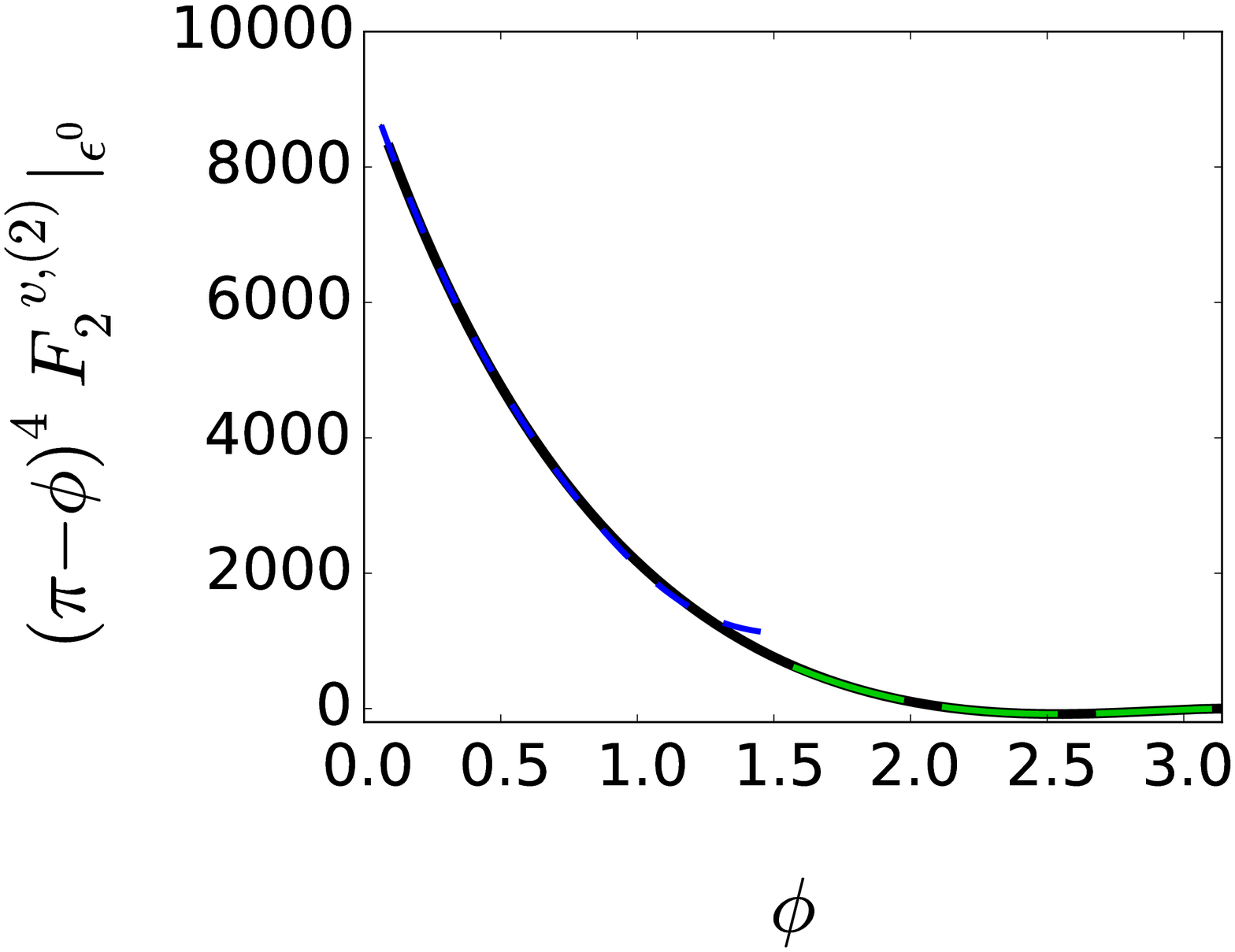} &
      \includegraphics[width=0.3\textwidth]{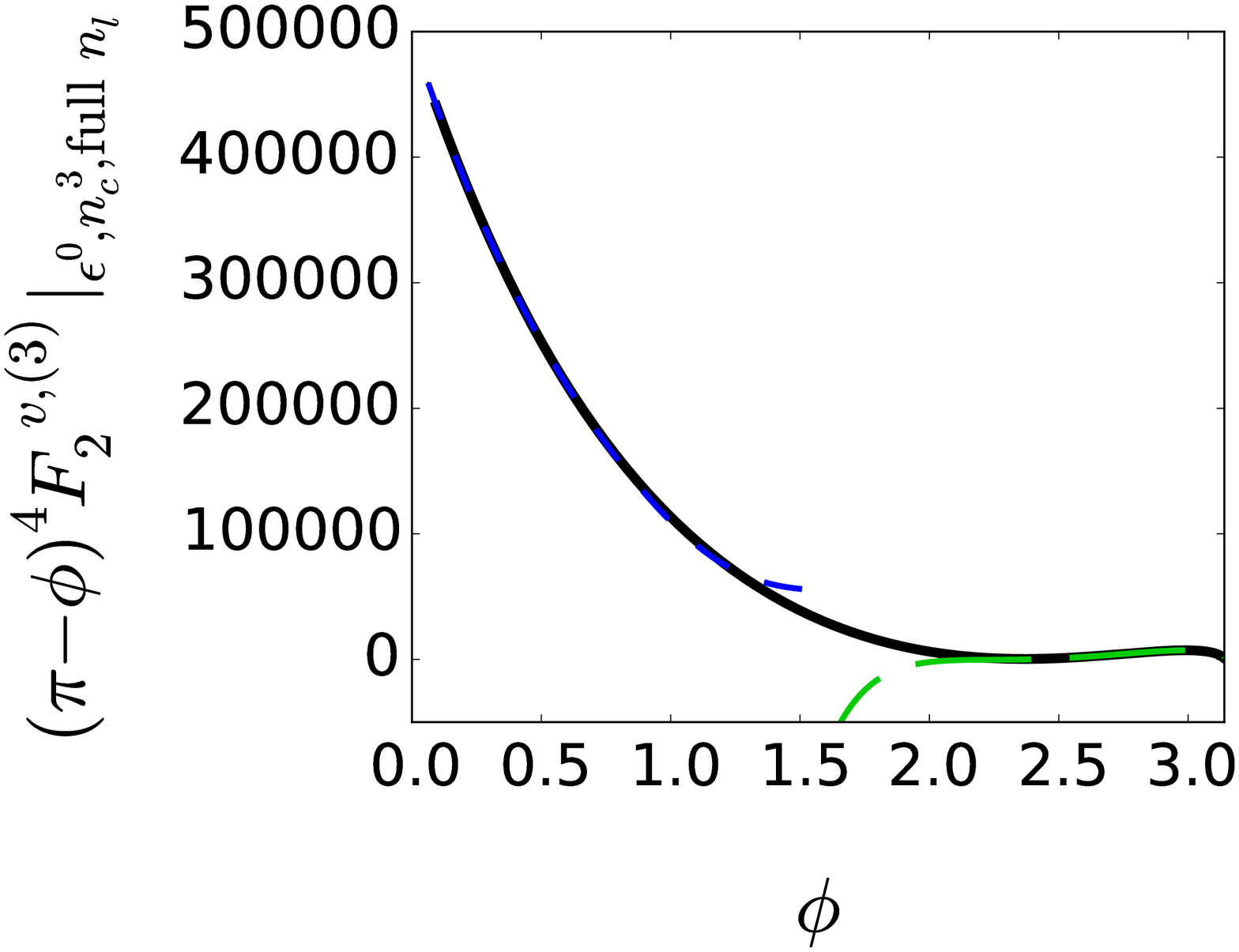} 
      \\
      \includegraphics[width=0.3\textwidth]{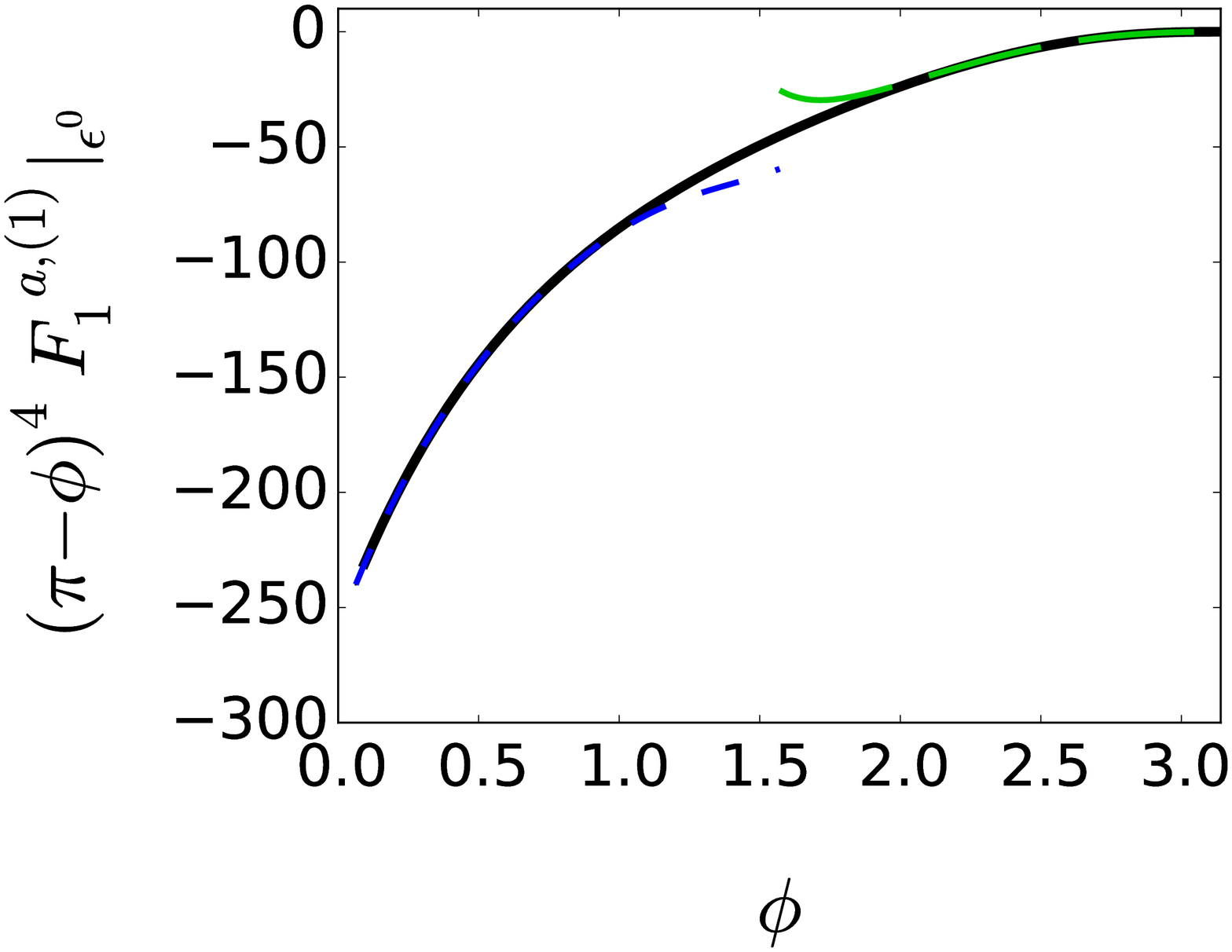} &
      \includegraphics[width=0.3\textwidth]{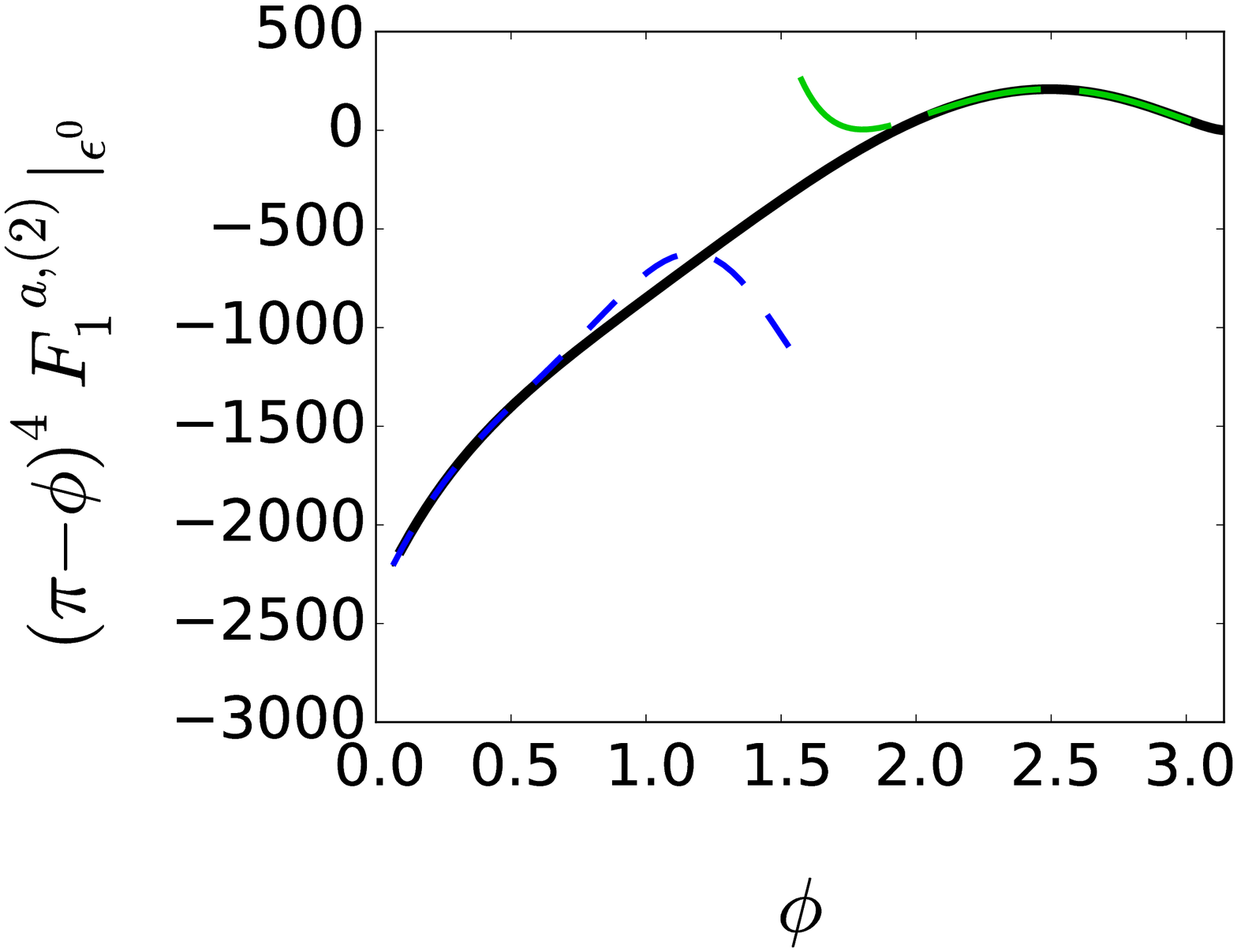} &
      \includegraphics[width=0.3\textwidth]{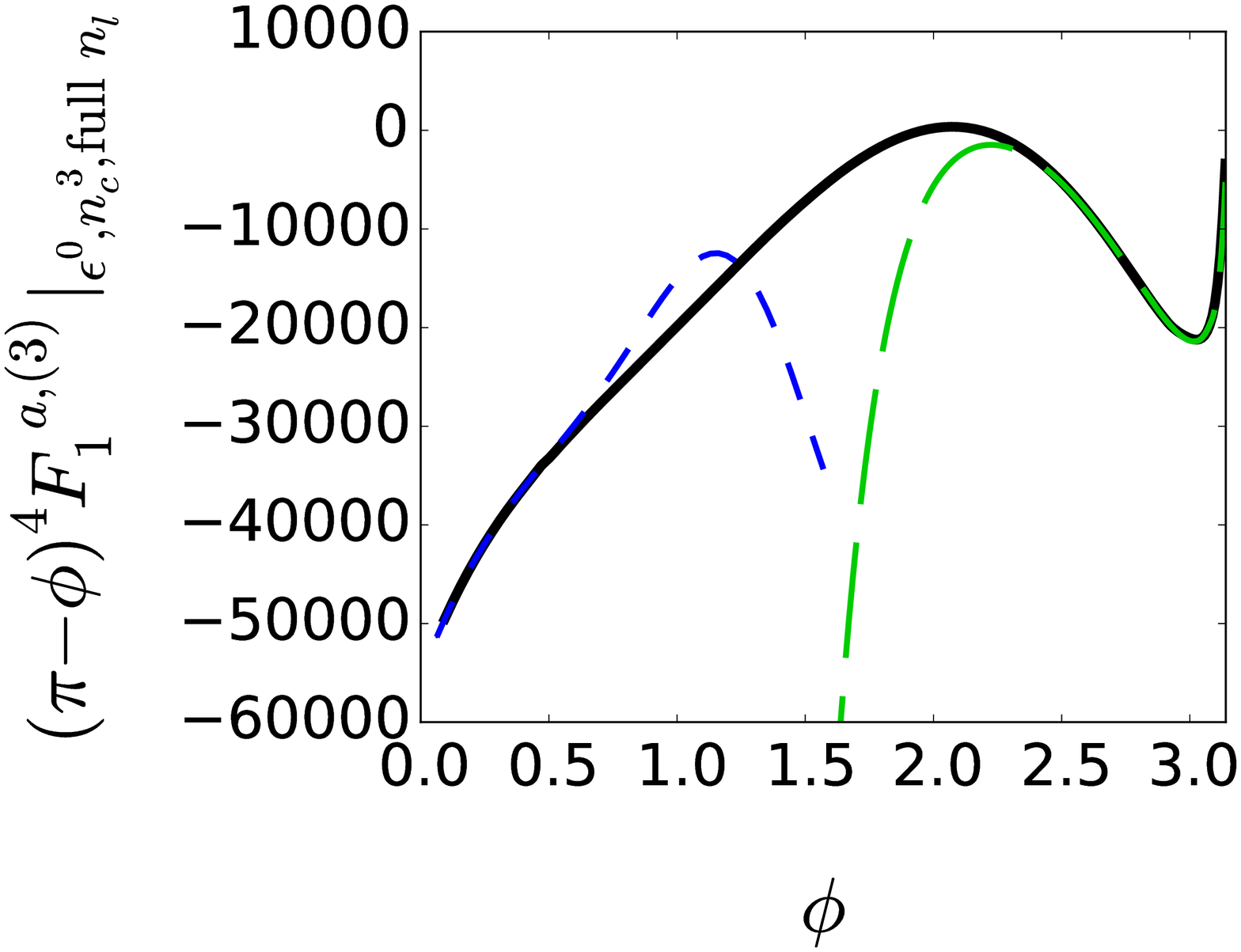} 
      \\
      \includegraphics[width=0.3\textwidth]{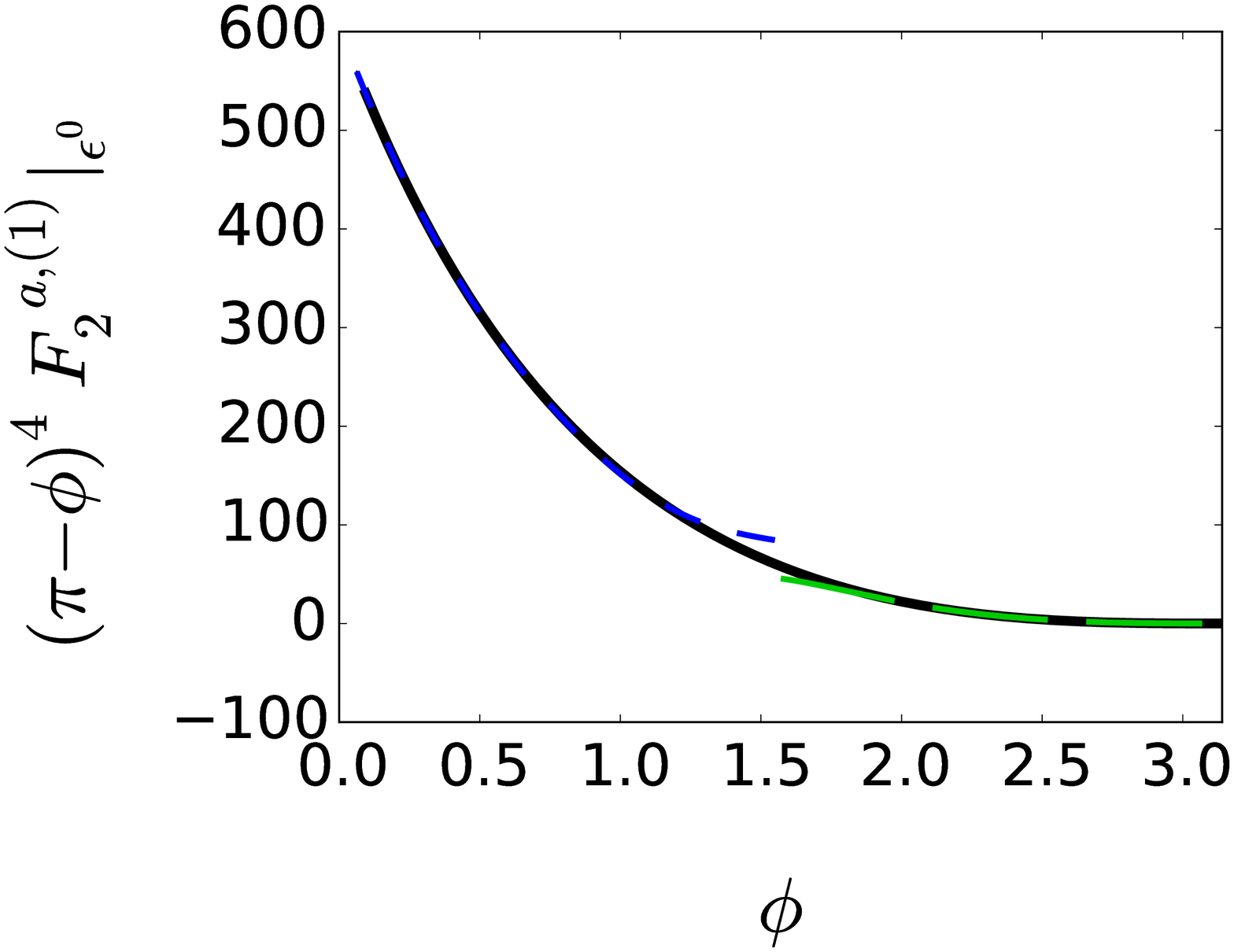} &
      \includegraphics[width=0.3\textwidth]{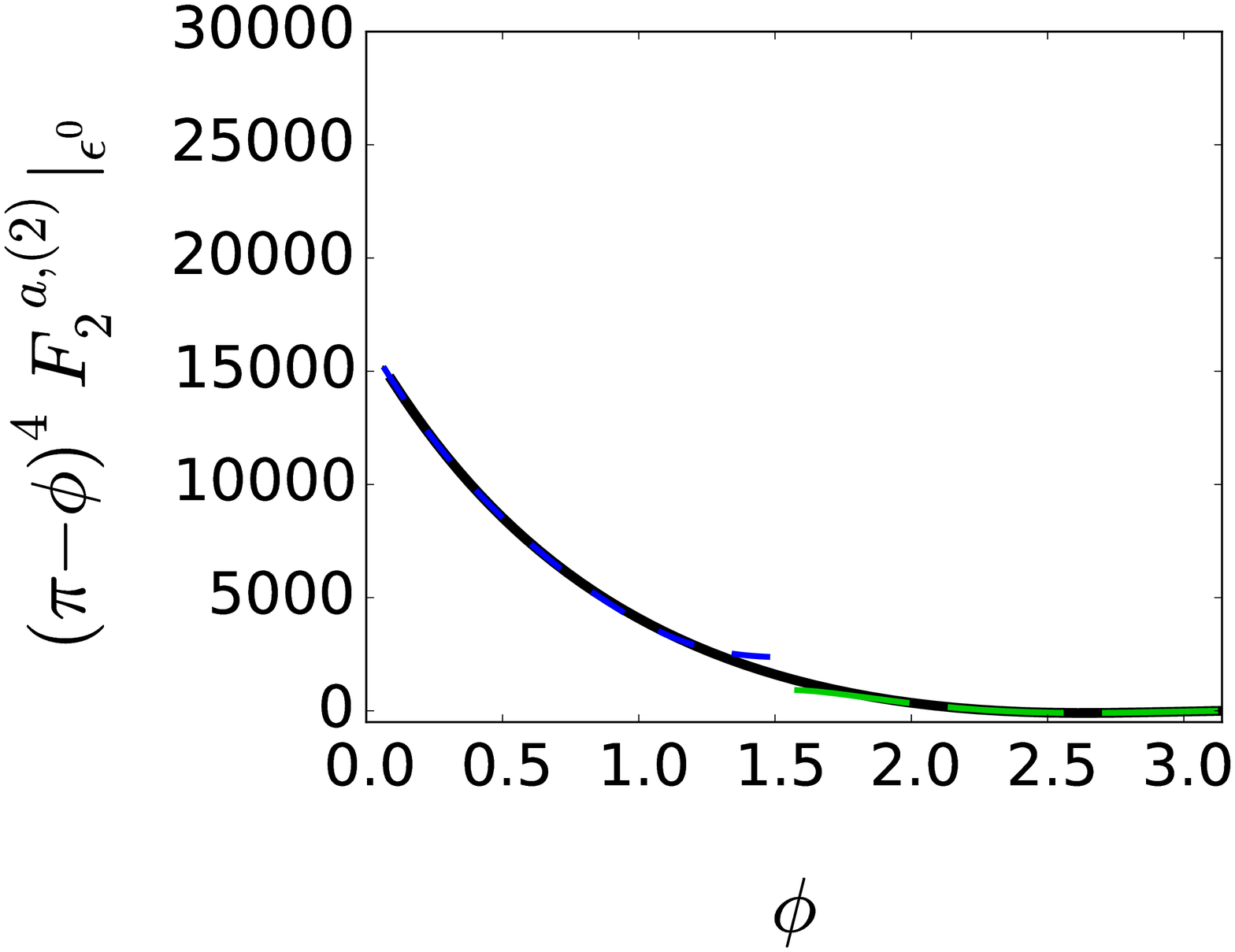} &
      \includegraphics[width=0.3\textwidth]{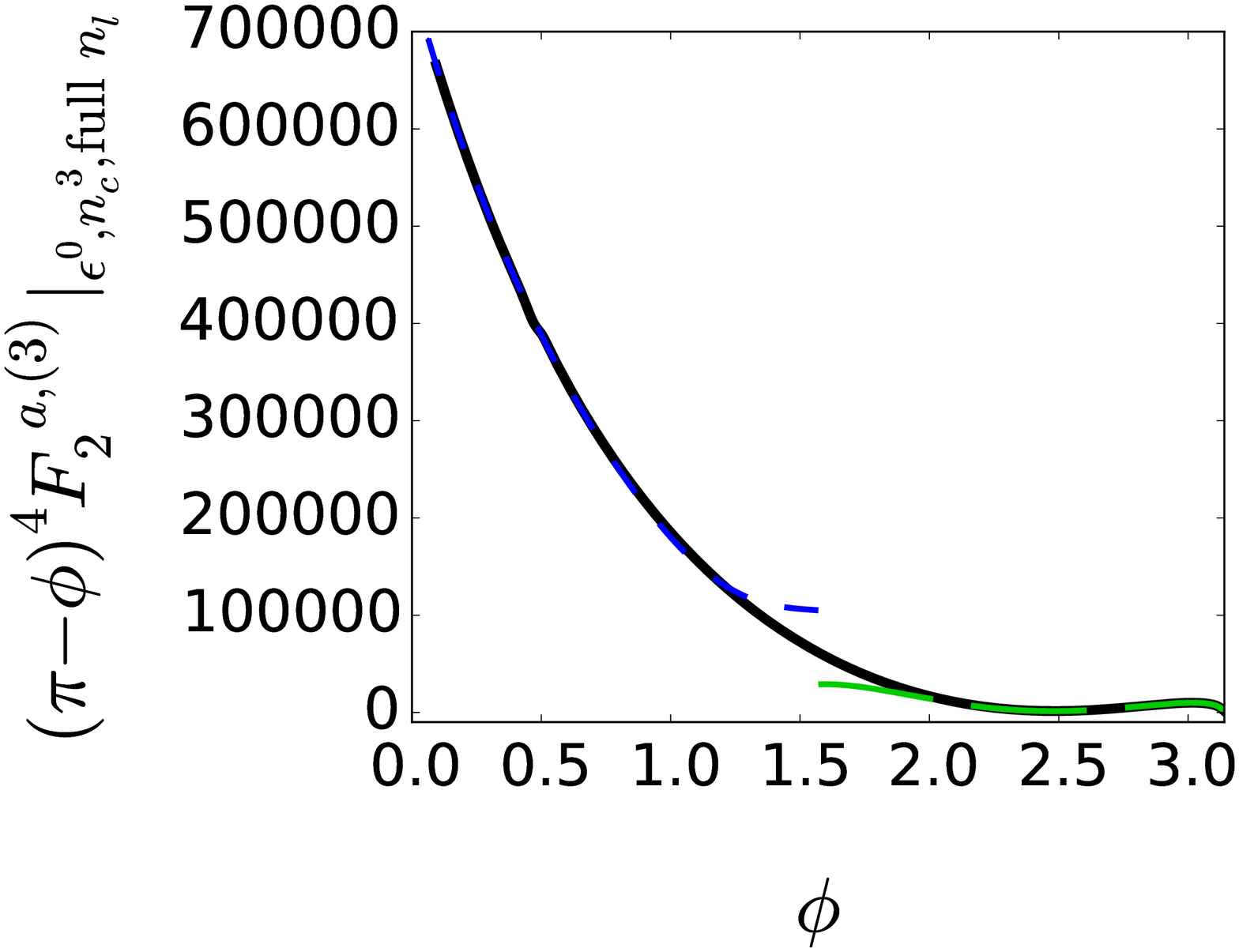} 
    \end{tabular}
    \caption{\label{fig::phi_va}$\epsilon^0$ term of the vector and axial-vector
      form factors as a function of $\phi$.  Exact results and approximations
      are shown as solid and dashed lines, respectively. At three-loop order
      we add the complete light-fermion part for $n_l=5$ and the $N_c^3$
      contribution. Note that for $\phi\in[0,\pi]$ the form factors are real.
      Short- (blue) and long- (green)
      dashed lines correspond to the low-energy and threshold
      approximation, respectively}
  \end{center}
\end{figure}

\begin{figure}[t] 
  \begin{center}
    \begin{tabular}{ccc}
      \includegraphics[width=0.3\textwidth]{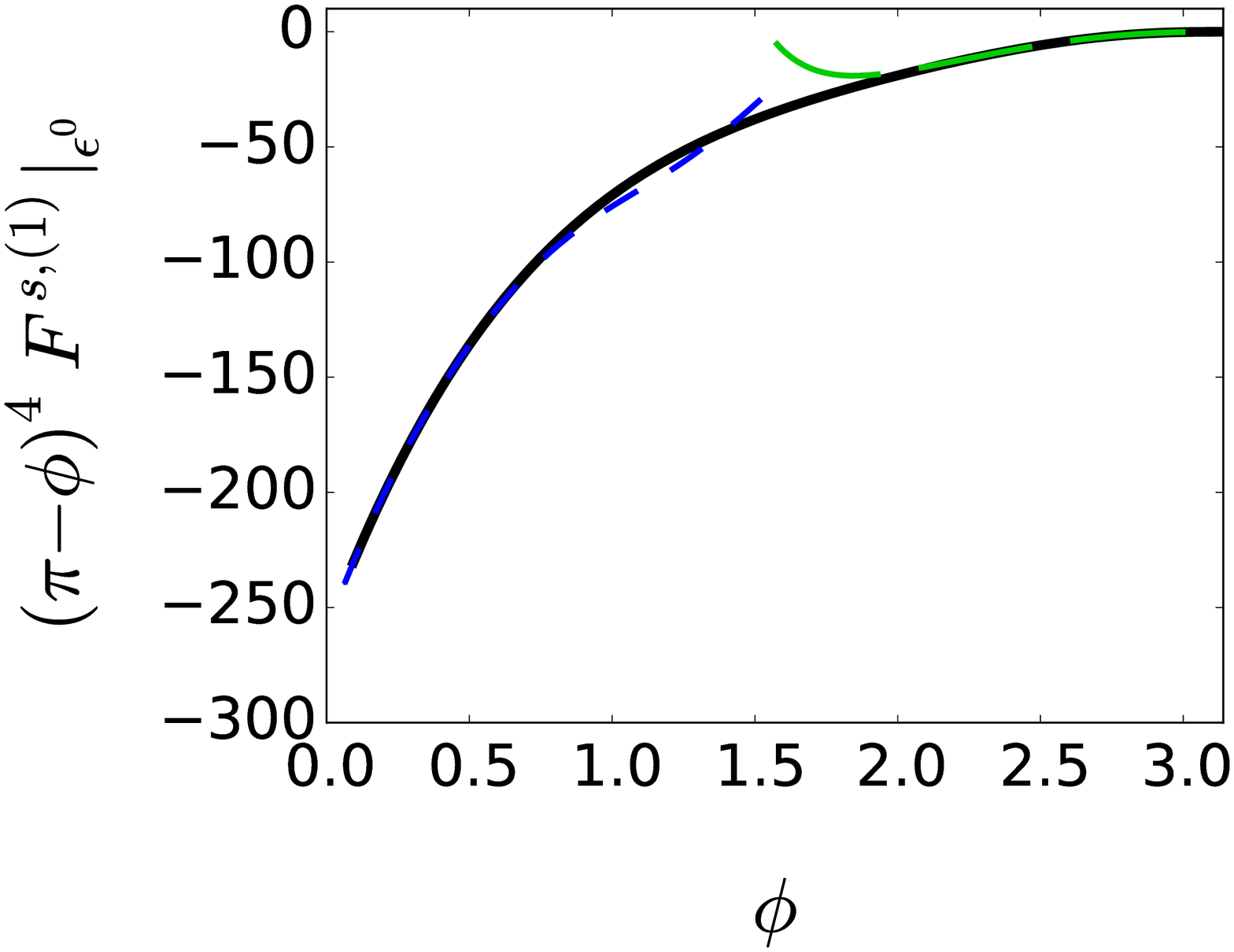} &
      \includegraphics[width=0.3\textwidth]{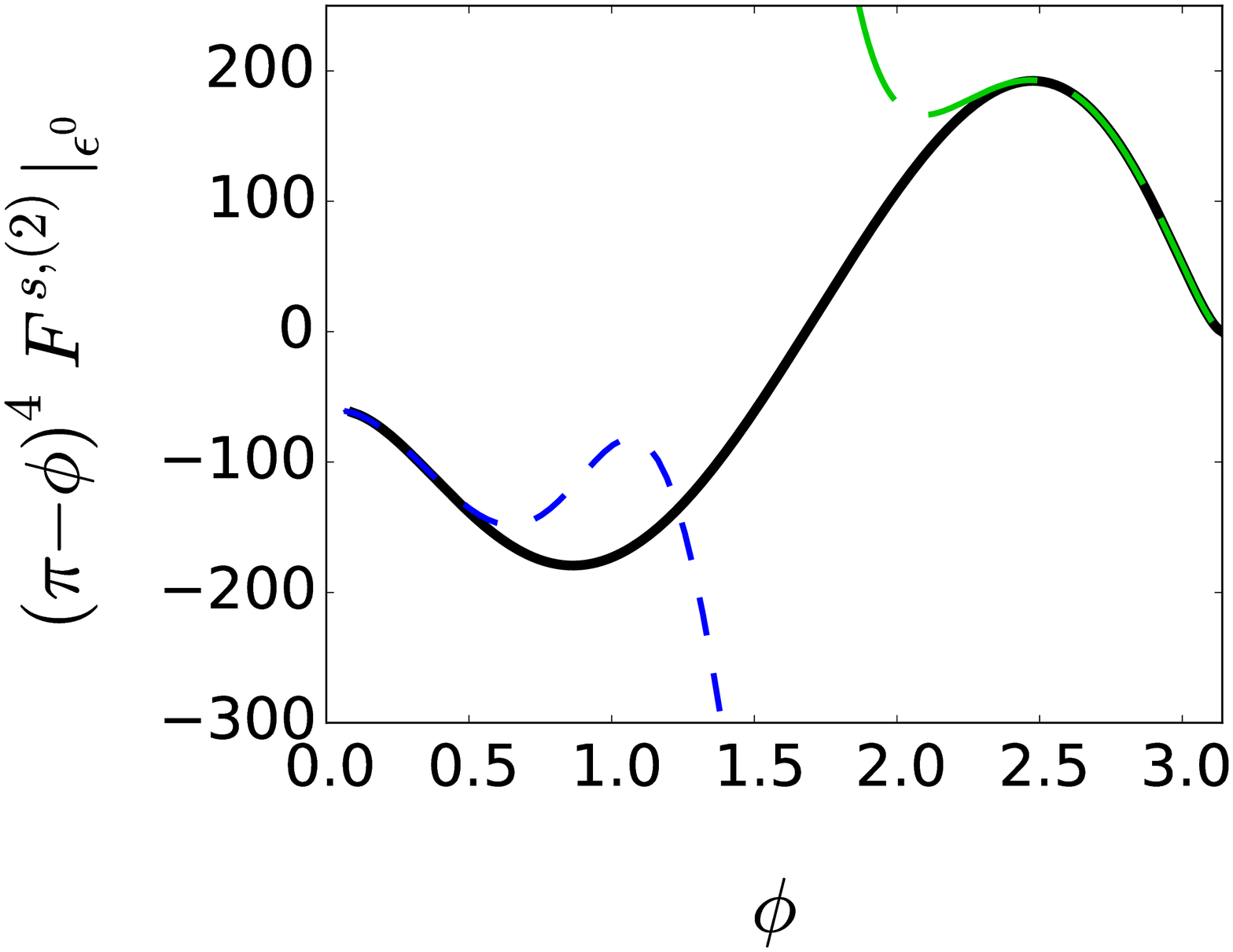} &
      \includegraphics[width=0.3\textwidth]{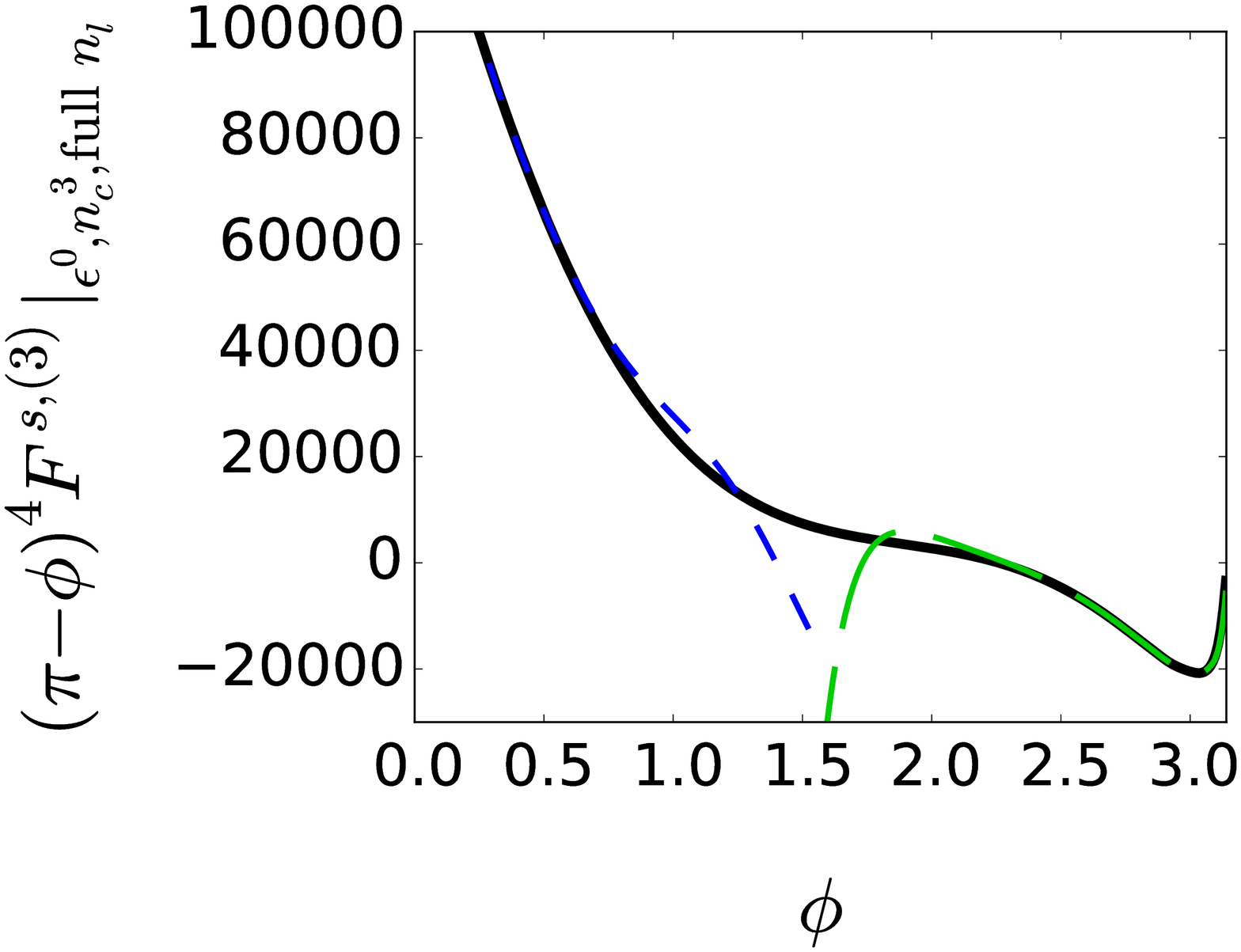} 
      \\
      \includegraphics[width=0.3\textwidth]{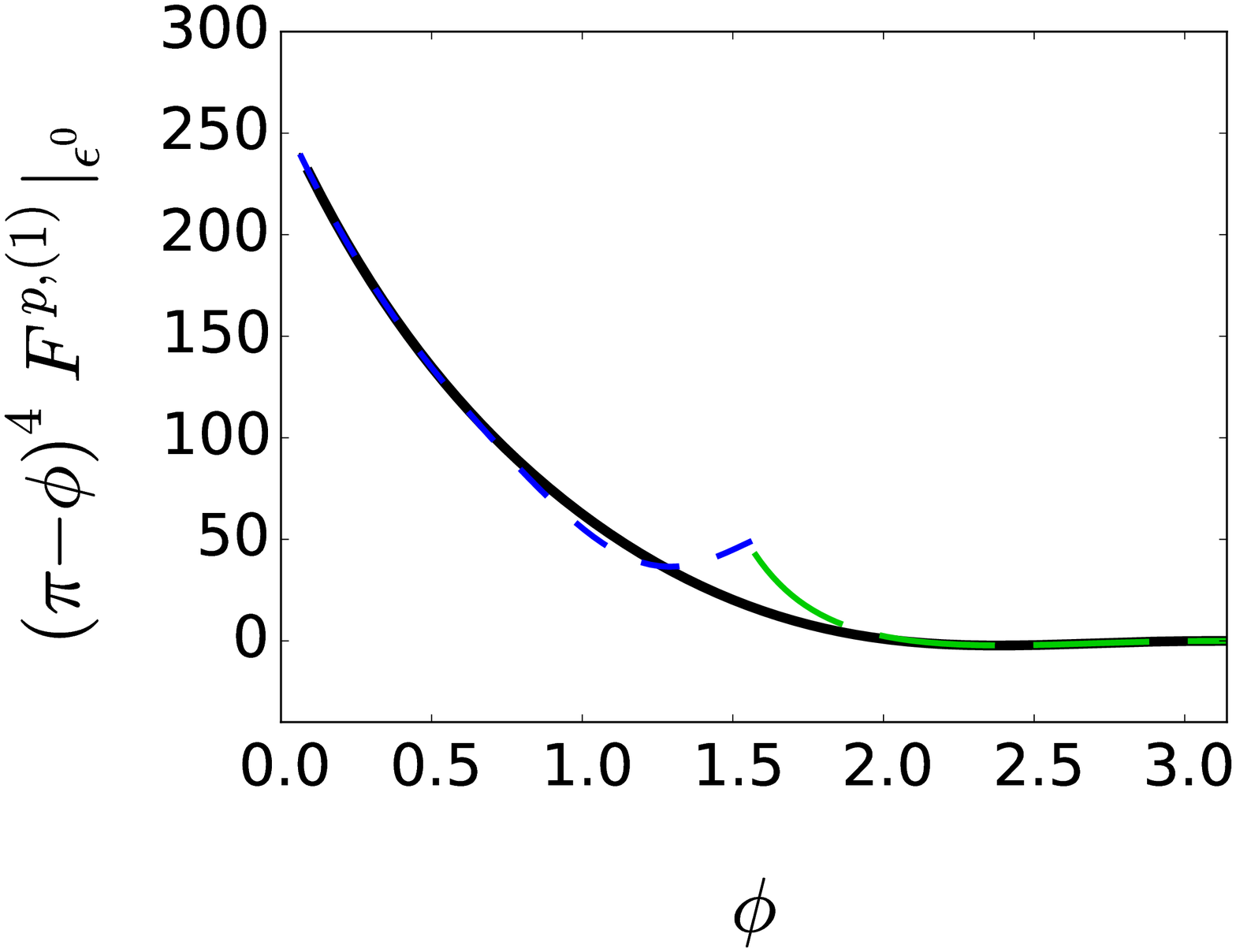} &
      \includegraphics[width=0.3\textwidth]{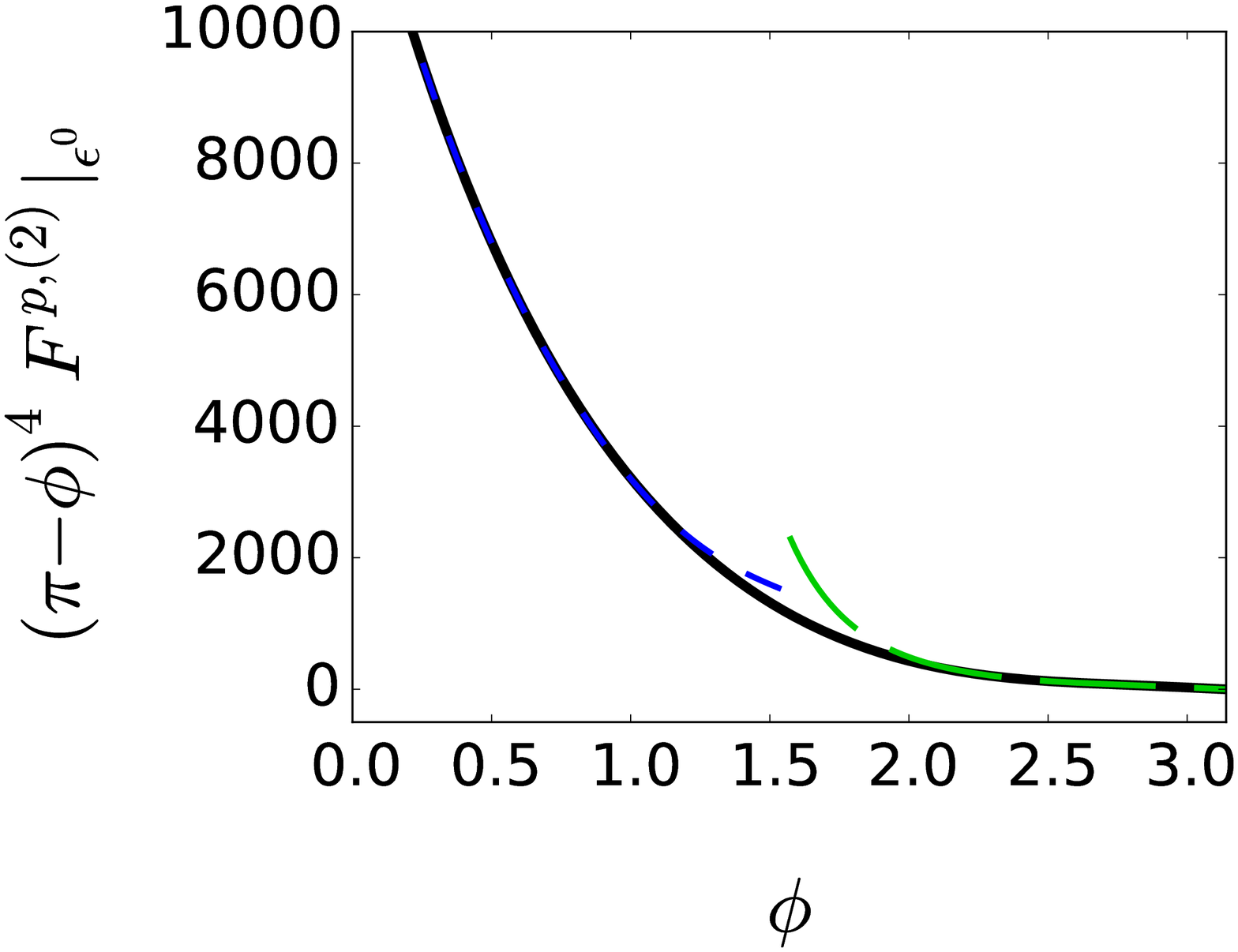} &
      \includegraphics[width=0.3\textwidth]{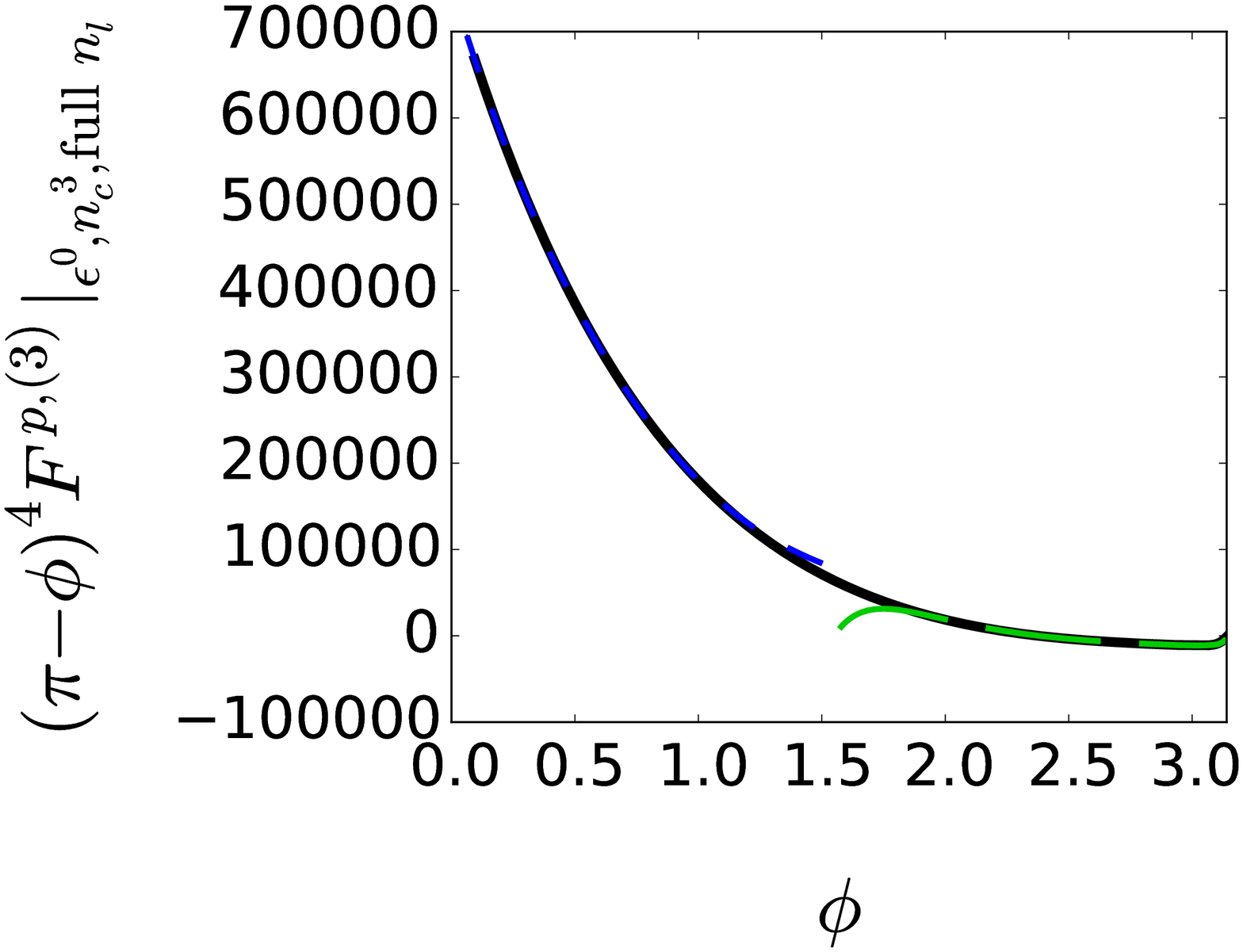} 
    \end{tabular}
    \caption{\label{fig::phi_sp}Same as Fig.~\ref{fig::phi_va} but for the
      scalar and pseudo-scalar currents.}
  \end{center}
\end{figure}

It is interesting to mention that some of the plots show a peculiar bump-like
structure at or close to $x=0$. This is not a numerical artifact but probably
related to the way the high-energy limit is subtracted. Note that the exact
results and the high-energy expansions (red dashed curve), which is simple to
evaluate numerically, perfectly agree with each other.

Figures~\ref{fig::phi_va} and~\ref{fig::phi_sp} show the results for the one-,
two- and three-loop form factors for $\phi\in[0,\pi]$ where $x=e^{i\phi}$. For
these values of $x$ the form factors have to be real which we checked
numerically.  To suppress the threshold singularities we plot $(\pi-\phi)^4
F$ and thus ensure that also at three loops the plotted functions are
zero at threshold, i.e. for $x=1$. 
One observes that the approximations agree with the exact result
for $\phi\lesssim 0.5$ and $\phi\gtrsim 2.0$ which corresponds to 
$q^2/m^2 \lesssim 0.25$ and $q^2/m^2\gtrsim 2.8$, respectively.


\section{\label{sec::con}Conclusions}

We have considered the vertex form factors induced by
vector, axial-vector, scalar and pseudo-scalar heavy quark currents,
which play an important role both in the Standard Model but also
in extensions. The form factors are parametrized by six scalar functions,
which we have computed up to three-loop order. Our results are
expressed in terms of GPLs with letters $\{-1,0,1,r_1=e^{i\pi/3}\}$
and the argument $x$ defined via the relation $q^2/m^2 = -(1-x)^2/x$.
After expanding the GPLs for small and large $q^2$ and around
the threshold given by $q^2=4m^2$ we obtain compact and easy to evaluate
expansions in the corresponding kinematical regions. We have discussed the
convergence properties by comparing to the exact expressions.
On the way to our three-loop result we have obtained the two-loop 
form factors including order $\epsilon^2$ terms.
This work extends the considerations of Refs.~\cite{Henn:2016tyf}
and~\cite{Lee:2018nxa} to axial-vector, scalar and pseudo-scalar currents.
Obvious next steps towards the full result are singlet contributions 
and the subset of Feynman diagrams containing closed massive fermion loops.
However, it can be expected that even these sub-classes show a more involved
mathematical structure and it is likely that not all pieces of the final result can
be expressed in terms of GPLs.



\section*{\label{sec::ack}Acknowledgments}

V.S. is thankful to Claude Duhr for permanent help in manipulations with GPLs.
This work is supported by RFBR, grant 17-02-00175A, and by the Deutsche
Forschungsgemeinschaft through the project ``Infrared and threshold effects in
QCD''.  R.L.  acknowledges support from the ``Basis'' foundation for
theoretical physics and mathematics.  The Feynman diagrams were drawn with the
help of {\tt Axodraw}~\cite{Vermaseren:1994je} and {\tt
  JaxoDraw}~\cite{Binosi:2003yf}.

\vspace*{1em}

{\bf Note added:}
\\
While finishing the write-up of the paper we became aware of
Ref.~\cite{zeuthen} where the large-$N_c$ result of the form factors
considered in this paper have been computed. Complete agreement has been
found. We would like to thank the authors of Ref.~\cite{zeuthen} for
the comparison of the results prior to publication.


\end{document}